\documentclass[11pt]{article}
\pdfoutput=1
\def\be{\begin{equation}}
\def\ee{\end{equation}}
\def\bea{\begin{eqnarray}}
\def\eea{\end{eqnarray}}
\usepackage{graphicx}
\usepackage{amssymb,amsmath}
\usepackage{cite,color,url}
\usepackage[colorlinks=true
,urlcolor=blue
,anchorcolor=blue
,citecolor=blue
,filecolor=blue
,linkcolor=blue
,menucolor=blue
,linktocpage=true
,pdfproducer=medialab
,pdfa=true
]{hyperref}

\catcode`\@=11
\def\lsim{\mathrel{\mathpalette\@versim<}}
\def\gsim{\mathrel{\mathpalette\@versim>}}
\def\@versim#1#2{\vcenter{\offinterlineskip
\ialign{$\m@th#1\hfil##\hfil$\crcr#2\crcr\sim\crcr } }}
\catcode`\@=12
\catcode`@=12
\begin{document}
\thispagestyle{empty}
\begin{flushright}
\end{flushright}
\vspace{0.3in}
\begin{center}
{\Large \bf Decays of $H^0/A^0$ in supersymmetric scenarios \\ with heavy sfermions \\} 
\vspace{1.0in}
{\bf Ernesto Arganda$^a$, J. Lorenzo Diaz-Cruz$^b$ and Alejandro Szynkman$^a$}
\vspace{0.2in} \\
{\sl $^a$ IFLP, CONICET - Dpto. de F\'{\i}sica, Universidad Nacional de La Plata, \\ 
C.C. 67, 1900 La Plata, Argentina} \\
{\sl $^b$ Facultad de Ciencias F\'{\i}sico-Matem\'aticas, \\
Benem\'erita Universidad Aut\'onoma de Puebla, Puebla, M\'exico }
\end{center}
\vspace{1.0in}
\begin{abstract}
The recent discovery of a new boson at the LHC, which resembles a SM-like Higgs 
boson with $m_h=125$ GeV, is starting to provide strong guidelines into
SUSY model building. For instance, the identification of such a state with the 
lightest CP-even Higgs boson of the MSSM ($h^0$), requires large values of 
$\tan\beta$ and/or heavy sfermions. One outcome of this result is the possibility  
to solve the SUSY flavor and CP problems by decoupling, which points towards some 
realization of Split-inspired SUSY scenarios, in which scalars are much 
heavier than gauginos and higgsinos. 
However, we argue here that the remaining Higgs bosons of the MSSM ($H^0$, $A^0$, $H^{\pm}$) 
do not have to be as heavy as the sfermions, and having them with masses near the EW scale  
does not pose any conflict with current MSSM constraints. 
We discuss then some SUSY scenarios with heavy sfermions, from a bottom-up approach,
which contain the full Higgs sector, as well as a possible dark matter candidate, with masses
near the EW scale, and identify  distinctive signals from these scenarios that
could be searched at the LHC.

\end{abstract}

\vspace*{30mm}
\noindent {\footnotesize E-mail: {\tt \href{mailto:ernesto.arganda@fisica.unlp.edu.ar}{ernesto.arganda@fisica.unlp.edu.ar}, \href{mailto:jldiaz@fcfm.buap.mx}{jldiaz@fcfm.buap.mx}, \href{mailto:szynkman@fisica.unlp.edu.ar}{szynkman@fisica.unlp.edu.ar}}}

\newpage

\section{Introduction}
\label{intro}

 After so many years of expectation, the LHC has now tested in very significant ways the
mechanism of Electro-Weak symmetry breaking (EWSB). In fact, the recent LHC 
results~\cite{:2012gk,:2012gu} indicate that a new particle with Standard Model (SM)-like Higgs properties 
and a mass around $m_{h_\text{SM}} \simeq 125$ GeV has been detected. 
Furthermore, this mass  value agrees quite  well with the range preferred by the analysis 
of electroweak precision tests~\cite{Flacher:2008zq,Erler:2010wa}, which can be seen as another 
confirmation of the SM success.
It has been suggested that further studies of the Higgs couplings predicted within the SM 
are required in order to confirm its nature \cite{Espinosa:2012im,Giardino:2012ww}. In fact, one expects that some deviation
could appear when one considers that the SM should not be the final theory of fundamental
particles, as its many open aspects seem to indicate, and some new theory will eventually
replace the SM. In fact, the LHC has already provided important bounds on the scale
of new physics beyond the SM \cite{Ellis:2010wx}. 

In particular, supersymmetry (SUSY) has been one of the most popular extensions of the
SM ~\cite{Haber:1984rc, Aitchison:2005cf}, with motivations that include:
a) a solution to the hierarchy problem, b) viable unification of SM gauge couplings,
c) a Dark Matter (DM) candidate and d) possible connection with String theory.
The Minimal Supersymmetric extension of the SM (MSSM) \cite{Chung:2003fi} has been widely studied in order to test
the realization of SUSY at the electroweak scale. The model predicts the existence of
superpartners for each SM particle: squarks/sleptons, gauginos and higgsinos are the partners
of quarks/leptons, gauge and Higgs bosons, respectively. The Higgs sector contains two scalars
doublets, with a spectrum that includes three neutral Higgs bosons ($h^0$, $H^0$, $A^0$) and one charged 
Higgs pair ($H^{\pm}$) \cite{Carena:2002es}. Thus, one expects that the recent Higgs mass result 
$m_{h_\text{SM}} \simeq 125$ GeV should be  useful to constrain SUSY.

The MSSM predicts the tree-level value $m_{h^0} \simeq m_Z$, while radiative corrections 
involving the top/stop system are needed in order to bring $m_{h^0}$ above the LEP bound, 
$m_{h^0} > 115$ GeV \cite{mssmhix}. In fact, to make the MSSM light Higgs boson to reach a mass in the 
range 125$-$126 GeV, one needs to include  stop masses of order TeV and/or large values 
of $\tan\beta$.  Similarly, the direct search for squarks and gluinos at LHC is actually lifting
the sfermions mass limits to a multi-TeV
range~\cite{ATLAS-CONF-2012-103,ATLAS-CONF-2012-104,ATLAS-CONF-2012-105,ATLAS-CONF-2012-109,CMS-SUS11016,CMS-SUS12016,CMS-SUS12017}.
On the other hand, this pattern of heavy sfermions has a positive side, namely the possibility
to solve the SUSY flavor and CP problems \cite{Gabbiani:1996hi} by decoupling \cite{ArkaniHamed:1997ab,DiazCruz:2005qz}. 

Thus, current searches for Higgs and SUSY at LHC suggest that 
the surviving MSSM should have the following features:
\begin{itemize}
\item It contains a SM-like Higgs boson with $m_{h^0} \simeq$ 125 GeV.
\item It contains heavy sfermions of the third generation (with $m = {\cal O}$(TeV)), 
to account for the Higgs mass value.
\item It includes heavy sfermions of first and second generations in order to solve
the  SUSY flavor and CP problems by decoupling.
\end{itemize}
Furthermore, such a model should contain a DM candidate with $m_{\tilde \chi_1^0}={\cal O}$(100 GeV) \cite{Baer:2011ab}
and the masses of all the MSSM particles must also agree with all bounds from 
collider and low-energy frontiers \cite{Arbey:2011aa}.

This suggests that some scenario with Split SUSY~\cite{ArkaniHamed:2004fb,Giudice:2004tc,ArkaniHamed:2004yi}
could be emerging from current data. The original Split SUSY, which has been widely studied 
lately~\cite{Zhu:2004ei,Kilian:2004uj,Hewett:2004nw,Masiero:2004ft,Gambino:2005eh,Kilian:2005kr,Gupta:2005fq,Wang:2005kf,
Kersting:2008qn,Cao:2011ht,Alves:2011ug,Giudice:2011cg,Hall:2011jd,Dhuria:2012bc,Arvanitaki:2012ps,Hall:2012zp,Unwin:2012fj},
assumed that, except for the light SM-like Higgs boson ($h^0$), all scalars are in the multi-TeV range, 
while gauginos and higgsinos would have lower masses and could be 
at the reach of the LHC.
Alternative models based on pure gravity mediation give place to similar spectra with heavy scalar masses~\cite{Ibe:2011aa,Ibe:2012hu,Bhattacherjee:2012ed}.
However, when one considers the role played by the full heavy Higgs spectrum ($H^0$, $A^0$ and $H^{\pm}$),
one notices that having them with masses near the EW scale does not pose in principle any phenomenological conflict. 
For instance, the approximate degeneracy between the heavy Higgs bosons facilitates the agreement with EW precision tests (EWPT); 
similar conclusion holds for the implications of the Yukawa couplings for low-energy flavor observables and 
collider results \cite{DiazCruz:2008ry}.
In fact, these Higgs scalars could be searched at the LHC and provide the first signature of SUSY at the EW scale, 
together with a DM candidate.

The aim of this paper is to discuss the possible realization of a Split-inspired non-universal Higgs
scenario, i.e. a scenario with Higgs masses near the EW
scale and heavy sfermions.
The possibility of having heavy Higgs boson masses at the EW scale in Split SUSY models
was already mentioned in~\cite{Delgado:2005ek},
where the meaning of the fine-tuning which Split philosophy implies was clarified. In this work it is   
pointed out that, in order to have both Higgs doublets near the EW scale, the
imposition of a second fine-tuning is required, besides the one needed to have a light SM-like 
Higgs boson at the EW scale.
Split SUSY with a Higgs sector near the EW scale was also discussed in~\cite{Diaz:2006ee,Diaz:2009gf,Diaz:2011pc}, 
within the context of R-parity breaking models and with focus on the generation of neutrino 
masses\footnote{Even if Split SUSY scenarios with a Higgs sector near the EW scale may
seem to be similar to the Split-inspired non-universal Higgs scenarios
proposed here, in terms of the mass spectrum, it is important to
emphasize a conceptual difference among them. Whereas in the former ones all the heavy states
are decoupled, as Split SUSY imposes, we start from a bottom-up
approach and work within a generic MSSM where only the sfermion of the
first and second generations are very heavy.}.
Thus, we shall consider the MSSM Higgs sector with $m_{h^0} \simeq 125$ GeV and heavy Higgs states 
($H^0$, $A^0$) with mass as low as the LHC admits, i.e. $m_{H^0}$, $m_{A^0} =$ 200$-$600 GeV.
Although the charged Higgs pair ($H^{\pm}$) should be heavier than about 
350 GeV, in order to satisfy the bounds from B-physics, here one would need to include stop-chargino
contributions, which may weaken these bounds. Therefore we will only consider direct search bounds,
thus we shall also assume $m_{H^{\pm}}= $ 200$-$600 GeV.
In some sense we are studying a type of two Higgs doublet model (2HDM) with MSSM parameters
and additional states that include a dark matter candidate, which we assume to be the lightest 
supersymmetric particle (LSP), the neutralino ($\tilde \chi_1^0$),
with MSSM parameters chosen such that $m_{\tilde \chi_1^0} = {\cal O}$(100 GeV).

The paper is organized as follows: in Section~\ref{higgssector}
we analyze the most relevant parameters that provide a light Higgs boson with
$m_{h^0} \simeq 125$ GeV and present the specific
SUSY scenarios with heavy sfermions we shall be working with. Section~\ref{binoLSP} is devoted to the study of
Higgs boson decays in the bino LSP scenario, in which there is only a bino-like
neutralino at low energies. We dedicate Section~\ref{winoLSP} and Section~\ref{higgsinoLSP}
to examine the Higgs boson signatures in wino LSP and
higgsino LSP scenarios, respectively, in which we have two SUSY particles
(one wino-like neutralino and one wino-like chargino) or three SUSY particles
(two higgsino-like neutralinos and one higgsino-like chargino) near the EW scale.
Finally, a discussion of results, perspectives and conclusions is presented in Section~\ref{conclusions}.

\section{SUSY scenarios with heavy sfermions and the Higgs mass}
\label{higgssector}

Now, our first goal is to determine the parameters of the MSSM that provide a light Higgs boson with $m_{h^0} \simeq 125$ GeV.
After considering the results, including  statistical and systematic uncertainties reported by
ATLAS~\cite{:2012gk} and CMS~\cite{:2012gu}, we accept a value of $m_{h^0}$ in our numerical analysis if it lies within the range [124 GeV, 127 GeV].
In order to perform a general study of the Higgs sector within the Split-inspired SUSY scenarios,
we assume that all of the soft masses of squarks and sleptons of the first and second generations are given by only one parameter,
$M_S$, which governs the decoupling scale. We also consider only a common soft mass for the third generation of sfermions,
$m_s$, which is defined as the boundary condition for the renormalization group equations (RGEs).
Our numerical analysis was performed with the package {\tt SUSY-HIT}~\cite{Djouadi:2006bz}
(which includes {\tt SuSpect}~\cite{Djouadi:2002ze}, {\tt SDECAY}~\cite{Muhlleitner:2003vg} and
{\tt HDECAY}~\cite{Djouadi:1997yw}) and the spectrum results were cross-checked with {\tt SOFTSUSY}~\cite{Allanach:2001kg},
with the other relevant MSSM parameters defined as follows:

\begin{itemize}
 
\item $1 < \tan\beta < 60$.

\item $-$3 TeV $< M_1$, $M_2$, $\mu < 3$ TeV\footnote{Although these parameters are varied within this range, we only select points that satisfy current direct bounds on SUSY masses~\cite{Beringer:1900zz}.}.

\item  1 TeV $< M_3 < 3$ TeV.

\item 200 GeV $< m_{A^0} < 600$ GeV.

\item 10 TeV $< M_S <$ 100 TeV.

\item 1 TeV $< m_s <$ 7.5 TeV.

\end{itemize}

The range of values of $M_S$ shown above
were chosen in order to solve the SUSY flavor and CP problems by decoupling~\cite{Gabbiani:1996hi,ArkaniHamed:1997ab}.
The third generation of sfermions is free from this condition and their corresponding masses
could be as low as experimental data permit, whenever they lead to the correct value of $m_{h^0}$.
It is important to note that values of $m_{A^0}$ near the EW scale together with
heavy soft masses of the third generation could be problematic. This situation could drive us
to imaginary tree-level values of $m_{A^0}$ (meaning that the electroweak Higgs minimum is essentially unstable)
or it could induce tachyonic masses for the stops, sbottoms and staus (breaking color and electric charge)~\cite{Ibarra:2005vb}.
In~\cite{Bernal:2007uv} the full one-loop RGEs of Split SUSY  were taken into account for
a generic MSSM with heavy scalars and universal Higgs masses of ${\cal O}$($10^4$ GeV), 
showing that this problem arises in this case too.
These problems can be avoided if we increase $m_{A^0}$ or decrease $m_s$.
We have checked that scenarios with values of $m_s$ below 8 TeV do not present this kind of difficulty.
Consequently, according to our choice of parameters we do not have to deal with this issue.

\begin{figure}[t!]
\begin{center}
\begin{tabular}{cc}
\includegraphics[width=75mm]{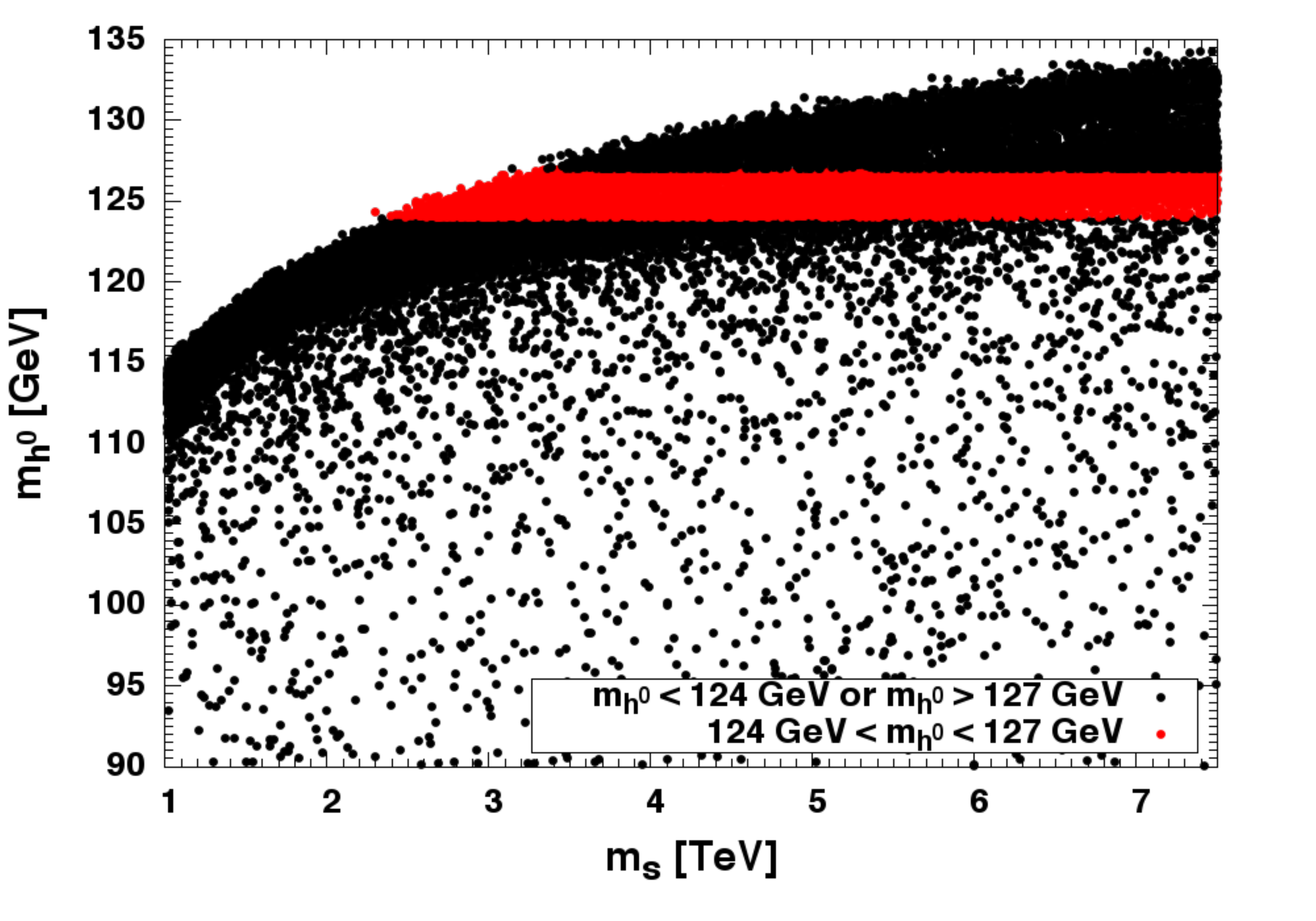} &
\includegraphics[width=75mm]{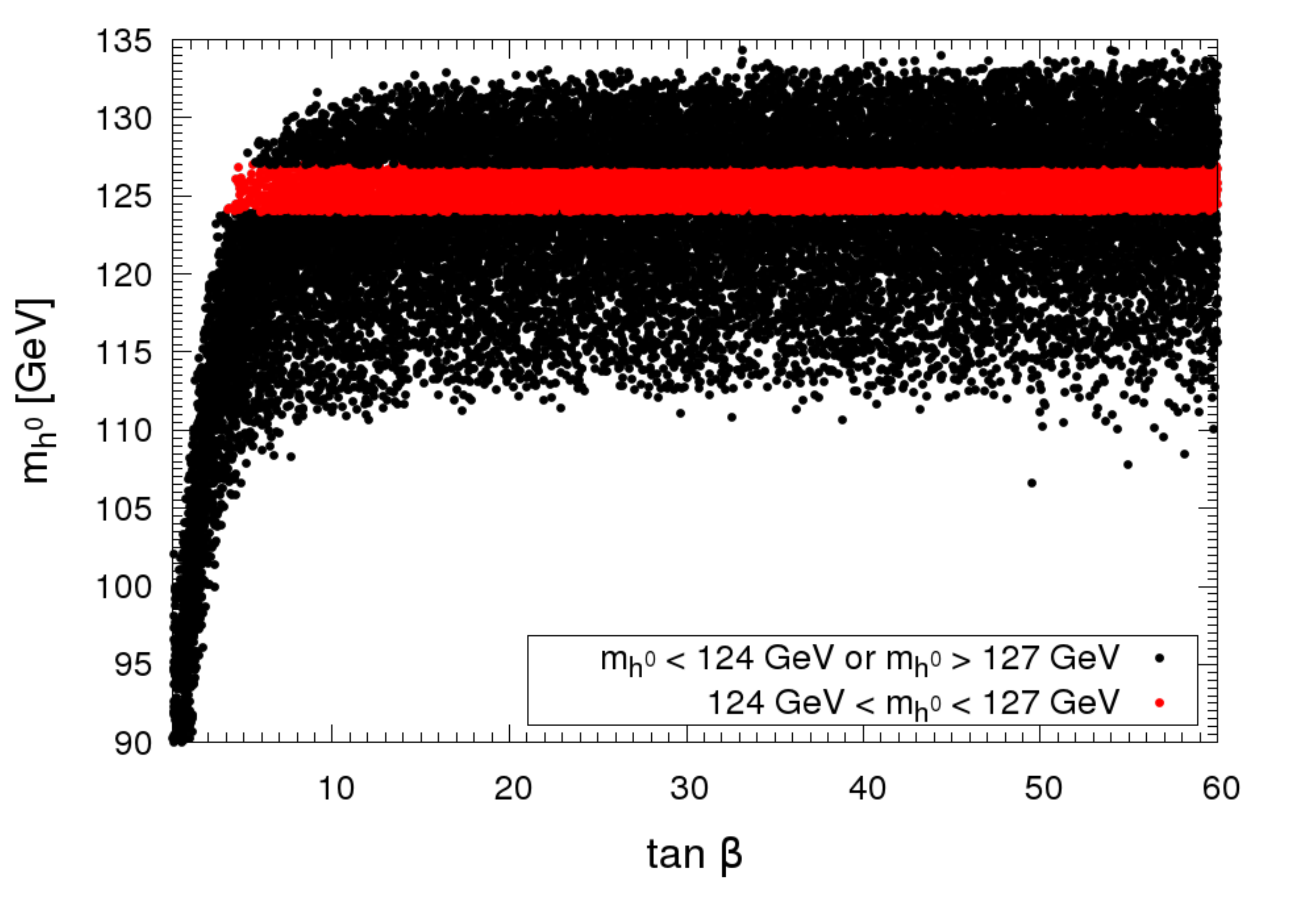} \\
\includegraphics[width=75mm]{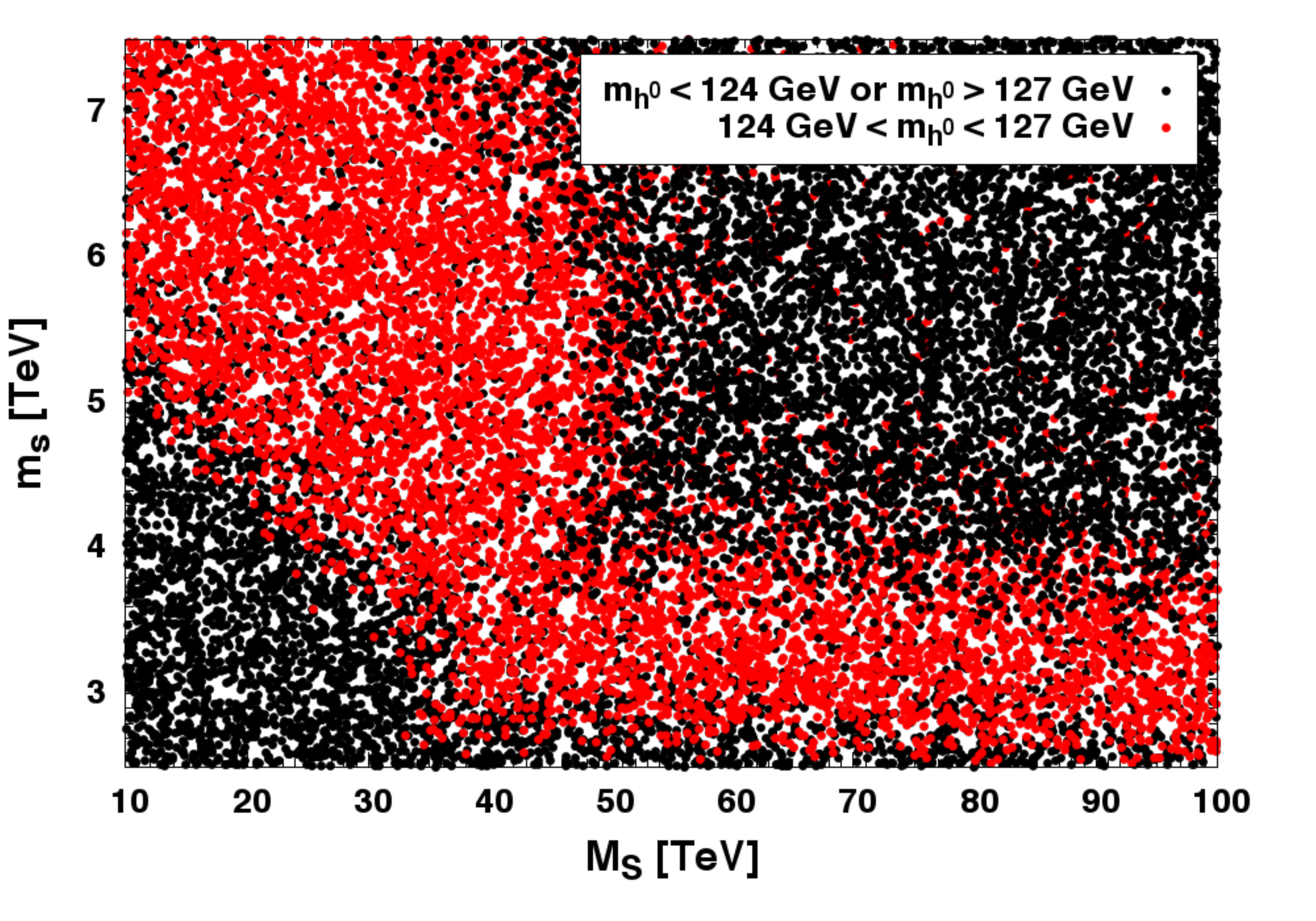} &
\includegraphics[width=75mm]{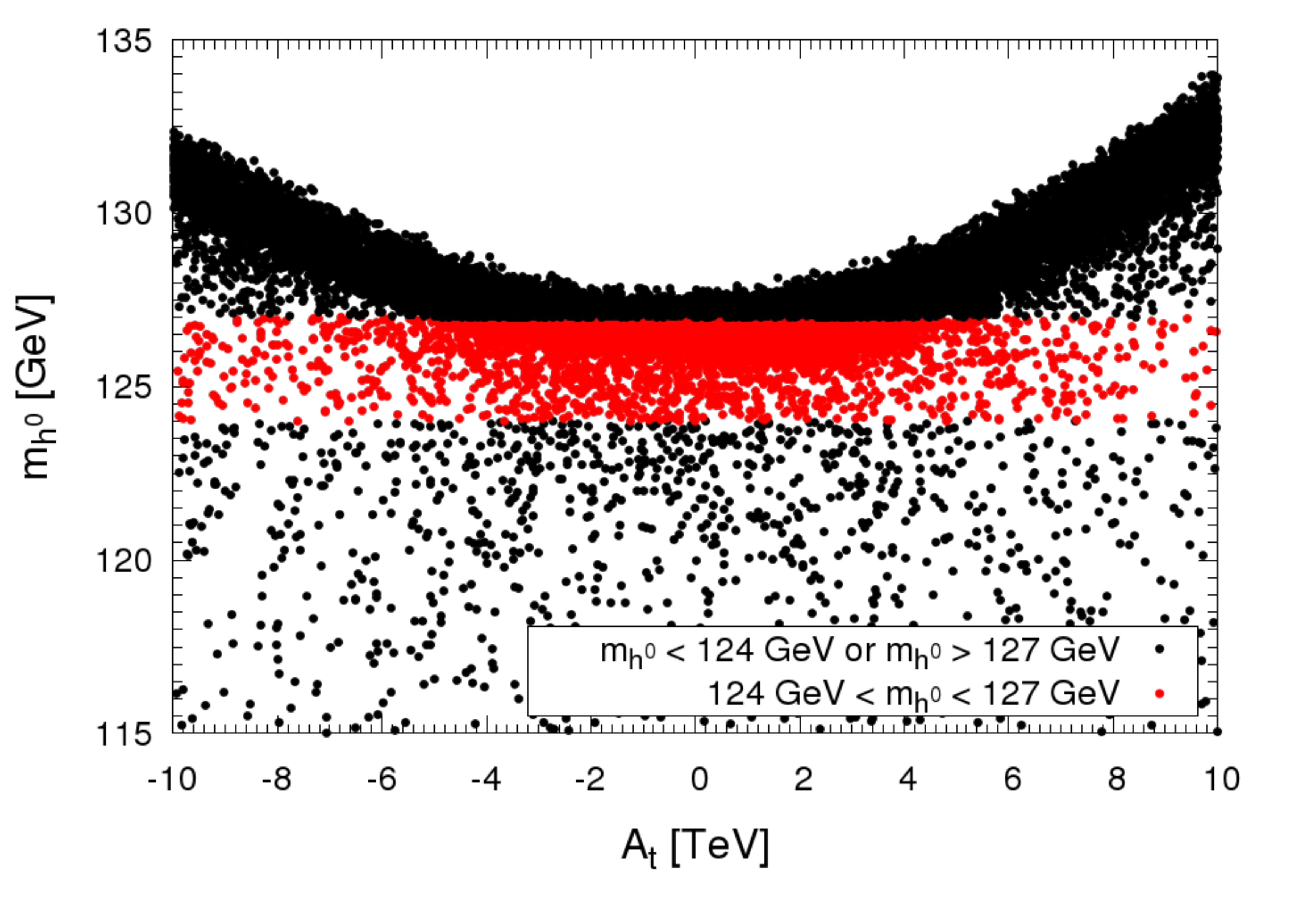}
\end{tabular}
\caption{Scatter plots of the allowed regions in parameter space for $m_{h^0}$.
In all the plots, red dots are for 124 GeV $< m_{h^0} <$ 127 GeV and black dots represent values of $m_{h^0}$
smaller than 124 GeV or larger than 127 GeV. Values for the rest of the parameters were varied randomly.
Upper left panel: $m_{h^0}$ as a function of $m_s$ ($A_t = 0$).
Upper right panel: $m_{h^0}$ as a function of $\tan \beta$ ($A_t = 0$).
Lower left panel: $m_{h^0}$ in the plane $m_s$$-$$M_S$ ($A_t = 0$).
Lower right panel: $m_{h^0}$ as a function of $A_t$ with $M_S =$ 40 TeV and $m_s =$ 7 TeV.}\label{fig:mh0}
\end{center}
\end{figure}

We have generated scatter plots included  in Figure~\ref{fig:mh0} that show the
different regions in parameter space where $m_{h^0}$ lies between 124 GeV and 127 GeV (red dots)
or falls outside this range (black dots). All the points were obtained by setting real random numbers
for the MSSM parameter within the regions defined above. The stop trilinear coupling $A_t$ was set equal
to zero in all the plots except for the lower right panel. From the upper plots, we can see the behavior
of $m_{h^0}$ as a function of $m_s \simeq $ $\sqrt{m_{\tilde t_1} m_{\tilde t_2}}$ and $\tan\beta$.
As expected, in order to obtain a light Higgs boson mass around 125 GeV it is necessary to have stops with
masses above 2$-$2.5 TeV, as shown in the upper left panel. It is clear from this plot that,
as $m_s$ grows up, $m_{h^0}$ also raises and it could reach too large values, exceeding the experimental value.
On the upper right panel we see the known strong 
dependence of $m_{h^0}$ on $\tan\beta$, concluding that we need values
of $\tan\beta$ larger than 5 if we want to accommodate these scenarios to
the present data reported by ATLAS and CMS, with the values of
$M_S$ and $m_s$ within the ranges used in this work.

On the lower left panel of Figure~\ref{fig:mh0} the behavior of $m_{h^0}$ with $M_S$
and $m_s$ is displayed.
Here we have to note that $m_{h^0}$ is sensitive to the squarks of the
the first and second generations due to the fact that we do not decouple these heavy states.
Therefore, despite their Yukawa associated couplings are smaller than the third generation ones,
their masses are very large and could produce important radiative corrections to $m_{h^0}$, although less relevant than
the corrections induced by stops and sbottoms.
For low values of $M_S$, close to 10 TeV, values of $m_s$
smaller than 5 TeV do not allow to get values of $m_{h^0}$ in the valid range. As $M_S$ increases,
the range of $m_s$ that generates correct values of $m_{h^0}$ become larger. For values of $M_S$
between 40 and 50 TeV, stop masses in the range [2.5 TeV, 7.5 TeV] drive us to 124 GeV $< m_{h^0} <$ 127 GeV.
From $M_S \simeq$ 50 TeV, this windows starts to close,
since the combined radiative corrections to $m_{h^0}$ of $m_s$ and $M_S$
exceed the upper value of 127 GeV,
and only low values of $m_s$, between
2.5 TeV and 4 TeV, result in proper values of $m_{h^0}$.

We have also studied the dependence of $m_{h^0}$ on $A_t$, which is allowed to
take values in the range [$-$10 TeV, 10 TeV]. For that, we have set $M_S =$ 40 TeV, $m_s =$ 7 TeV
and varied the other parameters randomly. The results are presented in the lower right panel of Figure~\ref{fig:mh0},
where we can check that it is possible to get a light Higgs boson around 125 GeV for any value of $A_t$. Nevertheless,
we note that large values of $|A_t|$ generate too large radiative corrections to $m_{h^0}$ that surpass the experimental data.
Therefore, we can conclude that we can set $A_t = 0$ without any loss of generality for the phenomenological analysis of the scenarios.

\begin{figure}[t!]
\begin{center}
\begin{tabular}{cc}
\includegraphics[width=75mm]{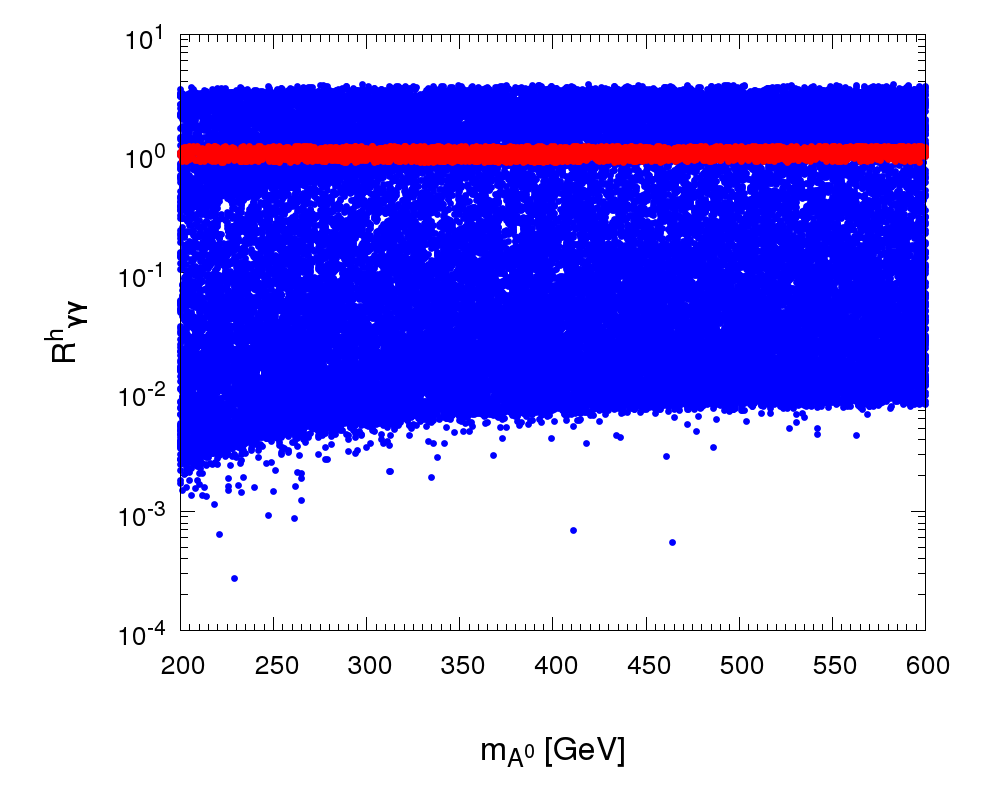} &
\includegraphics[width=75mm]{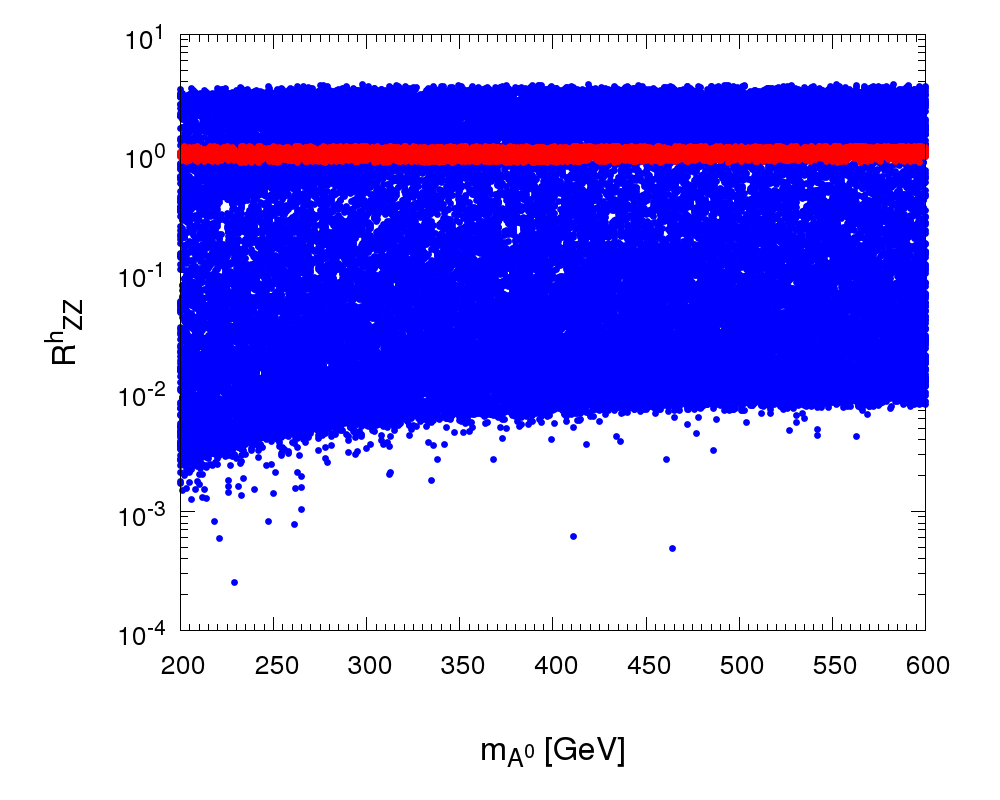}
\end{tabular}
\caption{Scatter plots of the ratios $R^h_{\gamma\gamma}$ (left panel) and $R^h_{ZZ}$ (right panel) as a function of $m_{A^0}$.
All the parameters are randomly generated within the same ranges as in Figure~\ref{fig:mh0}.
Red dots represent values of $R^h_{XX}$ ($X=\gamma, \, Z$) within the range [0.85, 1.15] and blue dots are for values of $R^h_{XX}$
falling outside this range.}\label{fig:Rs}
\end{center}
\end{figure}

The possibility to reproduce the signal rate for the SM-like Higgs signals with $m_h\simeq 125$ GeV
observed at the LHC~\cite{ATLAS-CONF-2012-127,CMS-PAS-HIG-12-045}, within our scenario with non-universal Higgs masses,
is illustrated in Figure~\ref{fig:Rs}. Here we have plotted the ratios:
\begin{equation}
 R^h_{XX} = \frac{\sigma(gg\to h^0)}{\sigma(gg\to h_\text{SM})} \, \frac{\text{BR}(h^0 \to XX)}{\text{BR}(h_\text{SM} \to XX)} 
\end{equation}
for $X=\gamma, \, Z$ (for the calculation of these ratios we have used the code
{\tt FeynHiggs}~\cite{Heinemeyer:1998yj,Heinemeyer:1998np,Degrassi:2002fi,Frank:2006yh}).
From these plots we can see that there are plenty of points where these ratios
are close to 1, as the LHC data indicates.

For completeness, we have analyzed the behavior of $m_{H^0}$ and $m_{H^\pm}$
as a function of $m_{A^0}$. Both masses present a closely linear dependence with $m_{A^0}$
which follows from the approximated relation $m_{H^0}, \, m_{H^\pm} \simeq m_{A^0}$.
In almost all the parameter space, we have found $m_{H^0}$ and $m_{H^\pm}$ slightly larger than $m_{A^0}$,
what converts the pseudoscalar to the lightest of the heavier Higgs bosons.

In the following sections we shall discuss Higgs properties for specific scenarios,
which will be defined according to the nature of the LSP and the number of neutralinos/charginos
which will have masses similar to the Higgs bosons, and thus could be reachable at the LHC.
These scenarios, with {\it ad hoc} choices of the parameters $M_1$, $M_2$ and $\mu$, are defined as follows:

\begin{itemize}
 \item {\bf Bino LSP Scenario:} Only one bino-like neutralino at the EW scale, thus we will have in this case $|M_1| \ll |M_2| , \, |\mu|$.

\item {\bf Wino LSP Scenario:} One wino-like neutralino and one wino-like chargino, degenerate in mass,
which occurs for  $|M_2| \ll |M_1| \,, |\mu|$. 

 \item {\bf Higgsino LSP Scenario:} Two higgsino-like neutralinos and one higgsino-like chargino, 
 in this case: $|\mu| \ll |M_1| \,, |M_2|$.
\end{itemize}

Then, we shall evaluate the branching ratios (BR) for the decays of the heavy Higgs bosons $H^0$ and $A^0$
in each of these three scenarios\footnote{A study of constraints imposed by the relic abundance is needed to see whether the LSP in each of these scenarios is an acceptable DM candidate. This analysis is beyond the scope of this work, however, we point out the results from~\cite{Hooper:2011tk} where these constraints are studied in a similar region of the parameter space we scan here. It is found that it is possible to have LSP masses of ${\cal O}$(100 GeV) consistent with experiments in any of these three scenarios.}.
If possible, we shall try to identify signatures that could have a chance of being detected at the LHC. 
Furthermore, we also would like to find signatures that could allow us to discriminate between the MSSM and the 
2HDM. For instance, we shall evaluate the BR for the invisible decays of $H^0$ and $A^0$,
i.e. the decays $(H^0, \, A^0) \to \tilde \chi^0_1 \, \tilde \chi^0_1$.
In this work we are not considering the charged Higgs boson $H^\pm$ decay channels since we are primarily
interested in studying the neutral resonances, for the mass window of 200$-$600 GeV, which is being studied currently by
the LHC. Secondly, the dominant decay mode, $H^\pm \to t b$, is very difficult to separate from SM backgrounds, 
while the sub-dominant mode,  $H^\pm \to \tau^\pm \nu_\tau$, has more hope of being detectable. In fact, we have 
noticed that both decay modes are almost constant for the three scenarios presented above, with 
BR($H^\pm \to t b$) $\simeq$ 0.9 and BR($H^\pm \to \tau^\pm \nu_\tau$) $\simeq$ 0.1, which is not so different from
the results expected for the general 2HDM.

\section{Bino LSP scenario}
\label{binoLSP}

The content of particles of the bino LSP scenario below 1 TeV scale
is the following: one bino-like neutralino $\tilde \chi_1^0$ (60 GeV $\lesssim m_{\tilde \chi_1^0} \lesssim$ 150 GeV),
one light Higgs boson $h^0$ (124 GeV $< m_{h^0} <$ 127 GeV), one heavy Higgs boson $H^0$,
one pseudoscalar Higgs boson $A^0$ and a charged Higgs pair $H^\pm$
(200 GeV $\lesssim m_{H^0}, \, m_{A^0}, \, m_{H^\pm} \lesssim$ 600 GeV).
The minimum value of $m_{\tilde \chi_1^0}$, governed by the bino soft mass $M_1$, is chosen in order to be conservative and respect the present lower bound~\cite{Beringer:1900zz}.

The case in which only one neutralino is at the EW scale can be realized only for a
bino-like neutralino, and this requires: $|M_1| \ll |M_2|$, $|\mu|$.
However, in this case the invisible decay width of $H^0$ and $A^0$ is very small.
This suppression can be easily understood through the form of the couplings of
neutralinos to neutral Higgs bosons~\cite{Djouadi:2001fa}:
\begin{eqnarray}
G^{L, R}_{\tilde \chi^0_1 \tilde \chi^0_1 H^0} &=& \frac{1}{2 \sin\theta_W} \left( Z_{12}- \tan\theta_W \, Z_{11} \right) 
\left(\cos\alpha \, Z_{13} - \sin\alpha \, Z_{14} \right) \,, \label{Hchi01chi01} \\
G^{L}_{\tilde \chi^0_1 \tilde \chi^0_1 A^0} &=& \frac{1}{2 \sin\theta_W} \left( Z_{12}- \tan\theta_W \, Z_{11} \right) 
\left(-\sin\beta \, Z_{13} + \cos\beta \, Z_{14} \right) \,, \label{Achi01chi01L} \\
G^{R}_{\tilde \chi^0_1 \tilde \chi^0_1 A^0} &=& - G^{L}_{\tilde \chi^0_1 \tilde \chi^0_1 A^0} \,, \label{Achi01chi01R}
\end{eqnarray}
where $Z_{ij}$ are the different entries of the single real matrix $Z$ which diagonalizes
the four-dimensional neutralino matrix in the ($-i \tilde B^0$, $-i \tilde W_3^0$, $\tilde H_1^0$, $\tilde H_2^0$) basis.
It is clear that the coupling $H^0 (A^0)$$-$$\tilde \chi^0_1$$-$$\tilde \chi^0_1 $ only receives
contributions from the Higgs-higgsino-gaugino terms. Thus, if $\tilde \chi^0_1$ was 
pure bino (or pure wino/higgsino), the invisible width of $H^0/A^0$ will vanish.
     
We discuss first the branching ratios of the heavy neutral Higgs bosons of the MSSM, $H^0$ and $A^0$, for a particular 
fixed choice of parameters which lies within the range
60 GeV $< |M_1| <$ 150 GeV and 1 TeV $< |M_2|$, $M_3$, $|\mu| <$ 3 TeV,
with $M_S =$ 40 TeV, $m_s =$ 7 TeV and $A_t = $ 0, as a function of
$m_{H^0 (A^0)}$ and $\tan\beta$.
Then, for the most relevant modes we shall present scatter plots where these parameters are
varied independently.

\begin{figure}[t!]
\begin{center}
\begin{tabular}{cc}
\includegraphics[width=75mm]{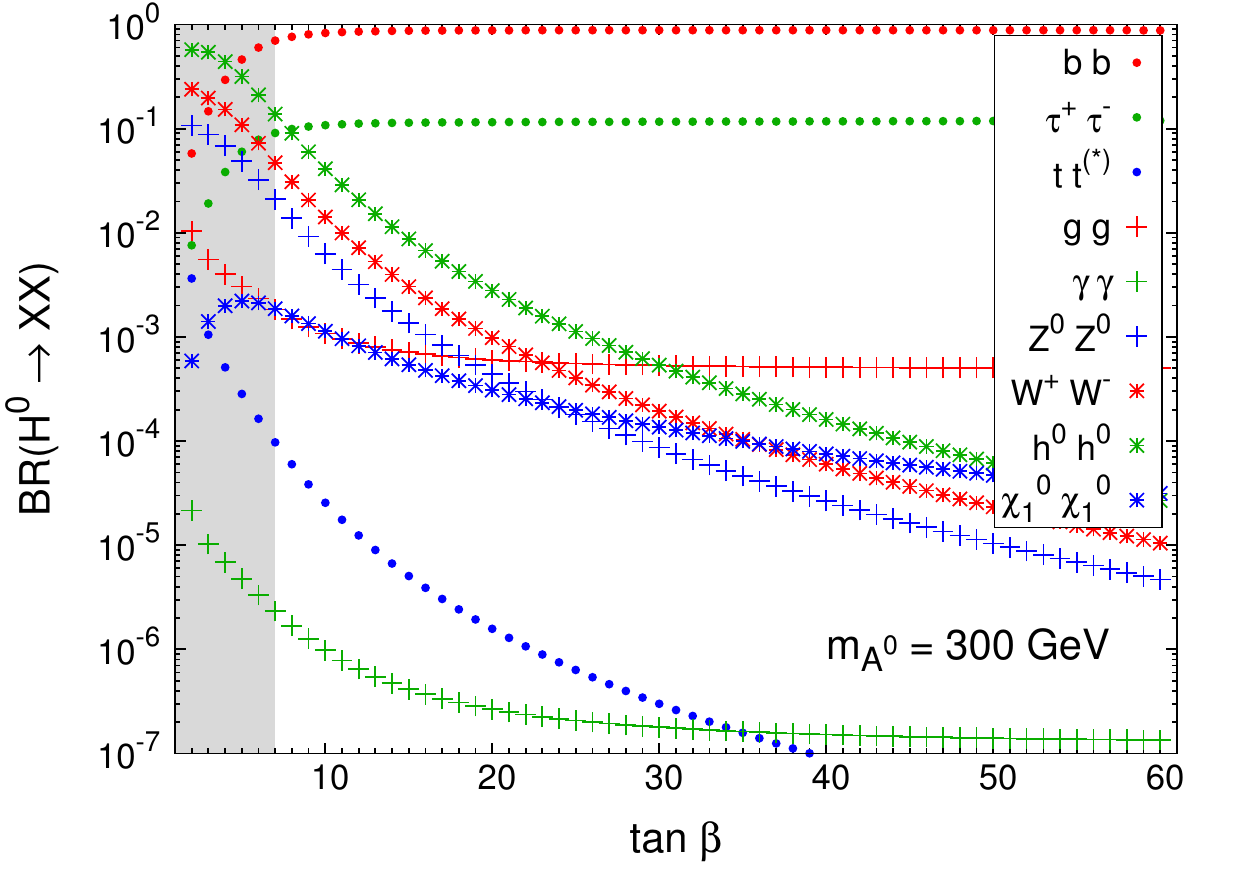} &
\includegraphics[width=75mm]{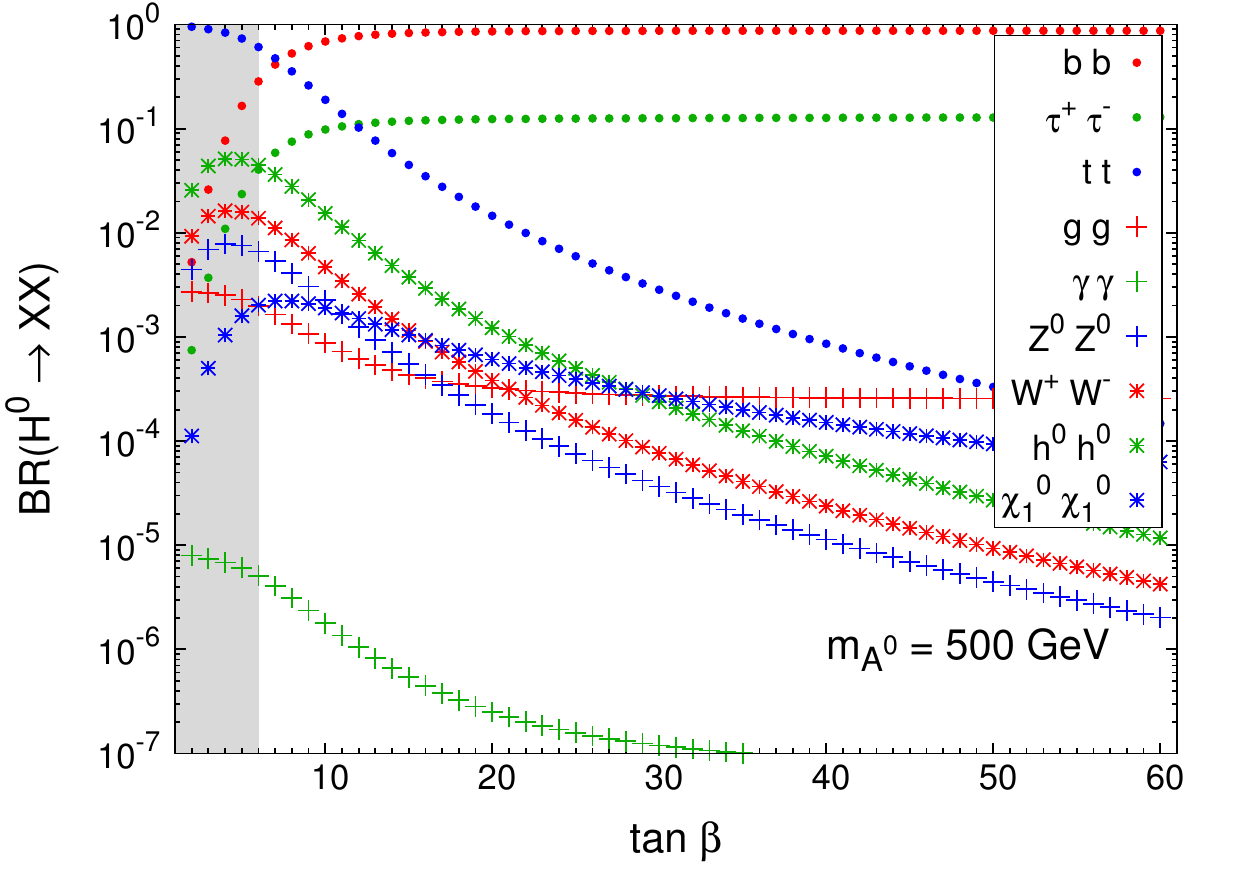} \\
\includegraphics[width=75mm]{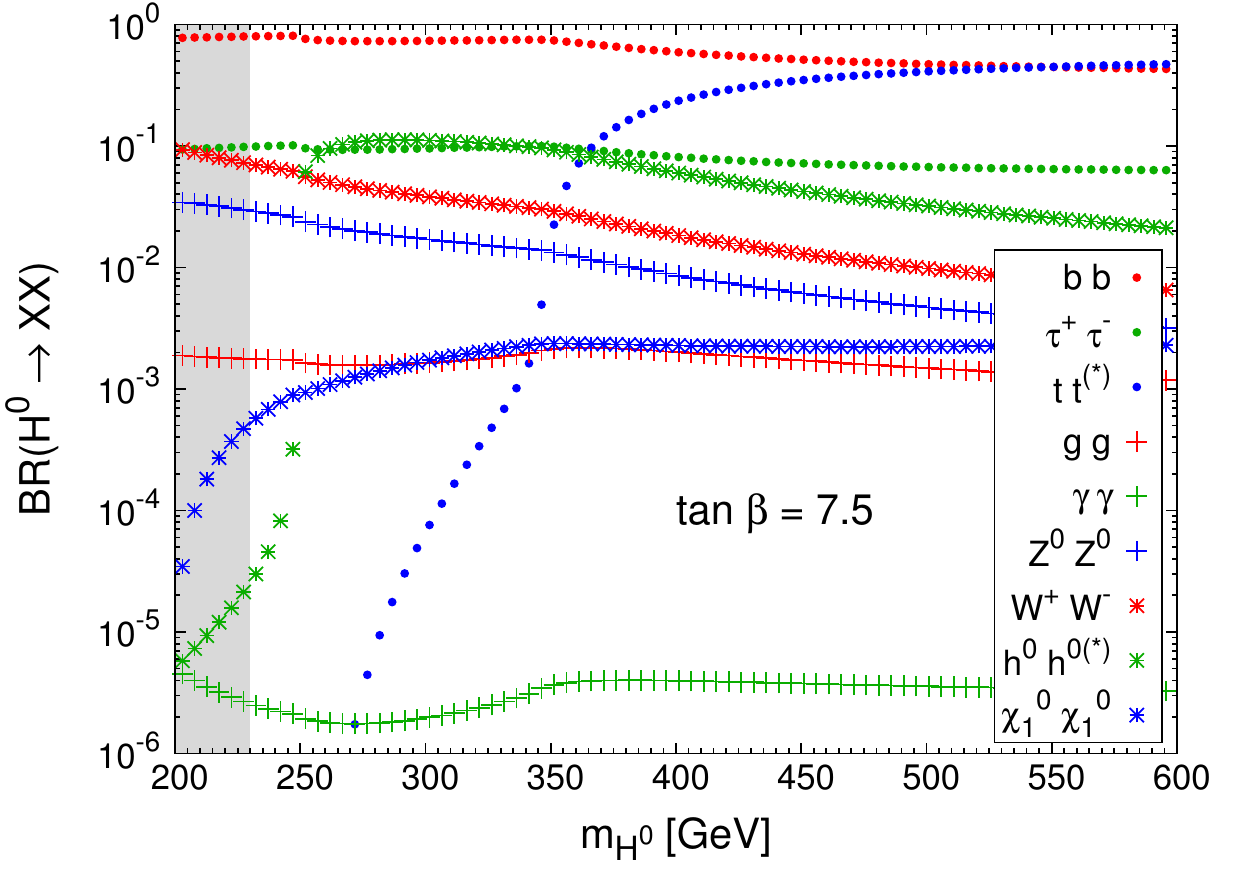} &
\includegraphics[width=75mm]{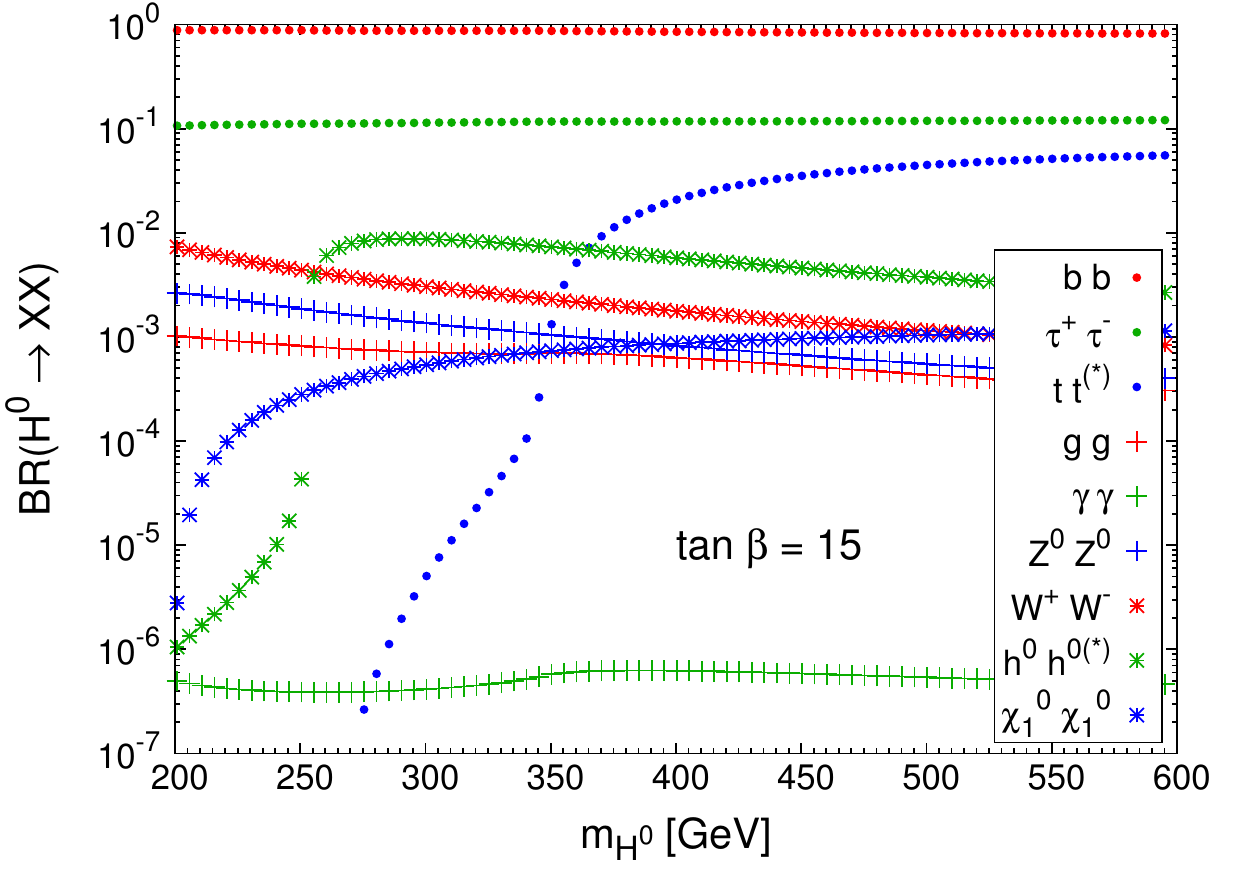}
\end{tabular}
\caption{$H^0$ decay channels in bino LSP scenario for $M_S =$ 40 TeV,
$m_s =$ 7 TeV, $A_t =$ 0, $M_1 =$ 100 GeV, $M_2 = \mu =$ 1 TeV, $M_3 =$ 3 TeV.
Upper left panel: $H^0$ branching ratios as a function of $\tan\beta$ for $m_{A^0} = $ 300 GeV.
Upper right panel: $H^0$ branching ratios as a function of $\tan\beta$ for $m_{A^0} = $ 500 GeV.
Lower left panel: $H^0$ branching ratios as a function of $m_{H^0}$ for $\tan\beta = 7.5$.
Lower right panel: $H^0$ branching ratios as a function of $m_{H^0}$ for $\tan\beta = 15$.
Shaded gray area represents values of $m_{h^0} <$ 124 GeV. The discontinuities around
$m_{H^0} \simeq 250$, 350 GeV in lower panels are due to the fact that $h^0h^0$ and $t \bar t$ channels
start to be kinematically allowed, respectively.}
\label{fig:BRH0decays_scenario1}
\end{center}
\end{figure}

\begin{figure}[t!]
\begin{center}
\begin{tabular}{cc}
\includegraphics[width=75mm]{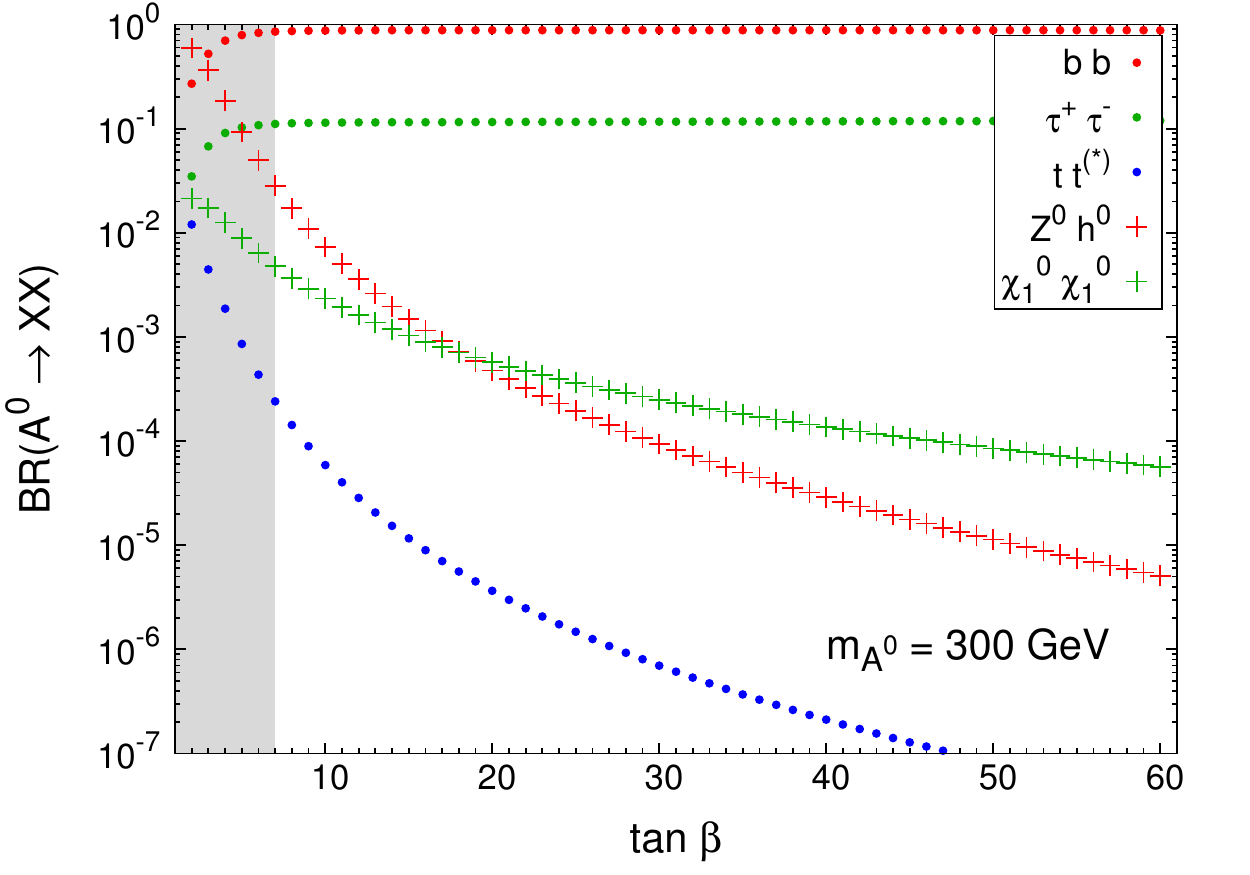} &
\includegraphics[width=75mm]{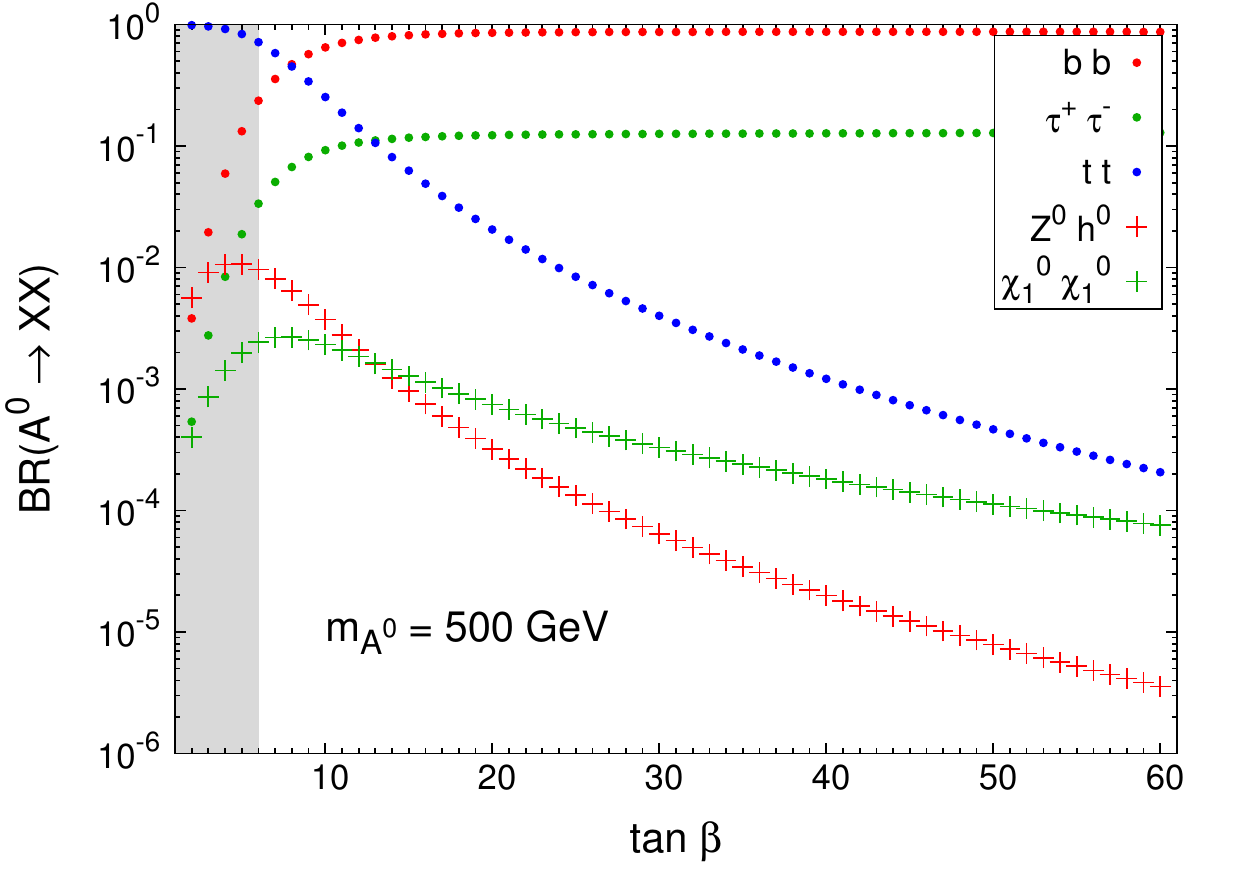} \\
\includegraphics[width=75mm]{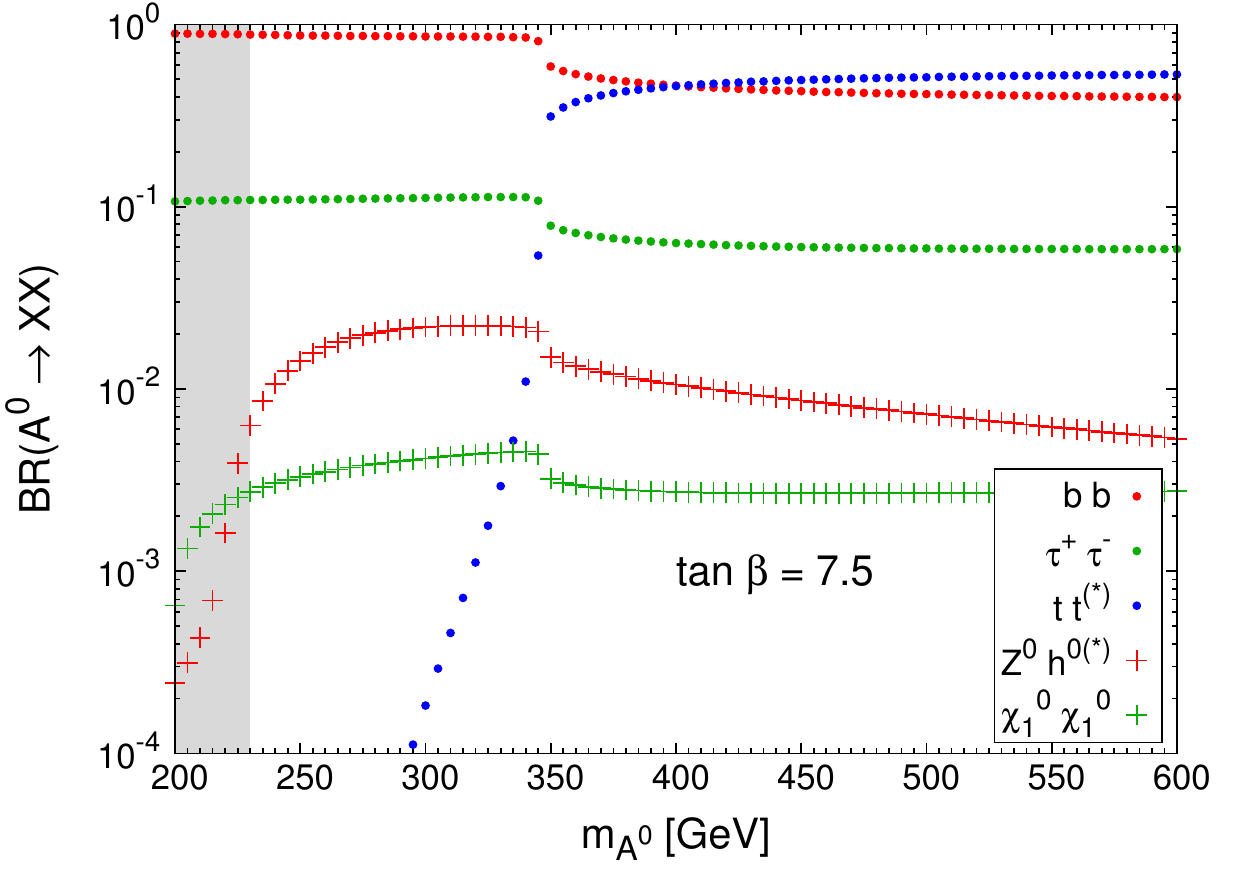} &
\includegraphics[width=75mm]{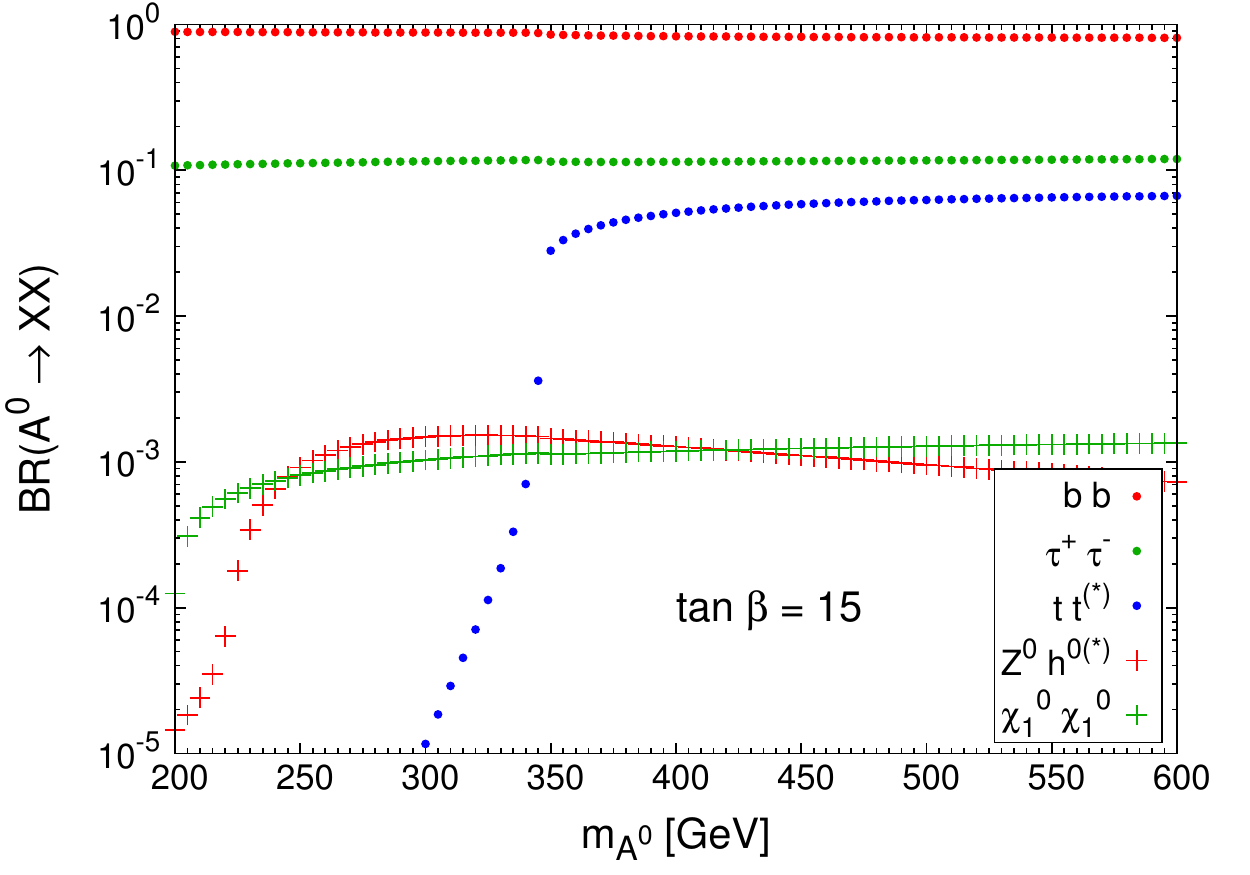}
\end{tabular}
\caption{$A^0$ decay channels in bino LSP scenario for $M_S =$ 40 TeV,
$m_s =$ 7 TeV, $A_t =$ 0, $M_1 =$ 100 GeV, $M_2 = \mu =$ 1 TeV, $M_3 =$ 3 TeV.
Upper left panel: $A^0$ branching ratios as a function of $\tan\beta$ for $m_{A^0} = $ 300 GeV.
Upper right panel: $A^0$ branching ratios as a function of $\tan\beta$ for $m_{A^0} = $ 500 GeV.
Lower left panel: $A^0$ branching ratios as a function of $m_{A^0}$ for $\tan\beta = 7.5$.
Lower right panel: $A^0$ branching ratios as a function of $m_{A^0}$ for $\tan\beta = 15$.
Shaded gray area represents values of $m_{h^0} <$ 124 GeV. The discontinuities around
$m_{A^0} \simeq$ 350 GeV in lower panels are due to the fact that $t \bar t$ channel
starts to be kinematically allowed.}
\label{fig:BRA0decays_scenario1}
\end{center}
\end{figure}

The results of the branching ratios of $H^0$ and $A^0$ as a functions of $\tan\beta$ and $m_{H^0(A^0)}$
within this scenario are contained in Figures~\ref{fig:BRH0decays_scenario1}
and~\ref{fig:BRA0decays_scenario1}, for the parameters: $M_S =$ 40 TeV,
$m_s =$ 7 TeV, $A_t =$ 0, $M_1 =$ 100 GeV, $M_2 = \mu =$ 1 TeV and $M_3 =$ 3 TeV\footnote{On the one hand, we select $M_2 = \mu = $ 1 TeV in order to
generate the maximum gaugino-higgsino mixing allowed in this scenario. On the other hand,
we set $M_3 =$ 3 TeV because we want to maintain the spectrum below 1 TeV scale
as simple as possible. We keep this criterion in the following sections for this kind of plots.}.
We have to be warned about the regions with low values of $\tan\beta$,
since in most of the cases $m_{h^0}$ does not reach the lower limit of 124 GeV (shaded gray area in the plots).
From these plots we find the following salient features:

\begin{itemize}
\item For values of $\tan\beta$ $\gtrsim$ 10, $H^0 \to b \, \bar{b}$ and $A^0 \to b \, \bar{b}$
      are the dominant modes, with BR $\simeq 0.9$.  

\item For values of $\tan\beta$ $\lesssim$ 10, $t \, \bar{t}$ channel can compete
      with $b \, \bar{b}$ if it is kinematically allowed ($m_{H^0(A^0)} \gtrsim$ 350 GeV) or even be the dominant one
      for values of $\tan\beta$ $\lesssim$ 6. Notice that in the plots of branching ratios as a function
      of $m_{H^0 (A^0)}$ there is a discontinuity around 350 GeV because of the opening of
      $t \bar t$ channel.
      
\item The one-loop decays $H^0 \to g g$ and $H^0 \to \gamma \gamma$ have very low BR
      for the allowed $h^0$ mass region, below $10^{-3}$ and $10^{-5}$, respectively.

\item $H^0 \to W^+ W^-$ channel reaches its largest value, BR $\simeq$ 0.1, for values
      of $\tan\beta$ $\lesssim$ 7.5 and when the $t \bar t$ channel is not kinematically allowed.

\item The maximum value of BR($H^0 \to Z^0 Z^0$) is approximately 0.03, occurring for
      $\tan\beta$ $\lesssim$ 7.5 and with the $t \bar t$ channel closed.
        
\item $H^0 \to h^0 h^0$ mode, when is kinematically allowed ($m_{H^0} \gtrsim$ 250 GeV), is slightly larger than $W^+ W^-$ channel and
      reaches its maximum value BR $\simeq$ 0.2 for $\tan\beta$ $\lesssim$ 7.5
      with $t \bar t$ channel closed too.
       
\item The largest value of BR($A^0 \to Z^0 h^0$) is 0.03, requiring the same conditions as
      $H^0 \to W^+ W^-$, $Z^0 Z^0$, $h^0 h^0$ modes.

\item In this case the branching ratios of the invisible channels, $H^0 \to \tilde{\chi}^0_1 \tilde{\chi}^0_1$
      and $A^0 \to \tilde{\chi}^0_1 \tilde{\chi}^0_1$, reach at the most values of ${\cal O}(10^{-3})$.
     
\end{itemize}

\begin{figure}[t!]
\begin{center}
\begin{tabular}{cc}
\includegraphics[width=75mm]{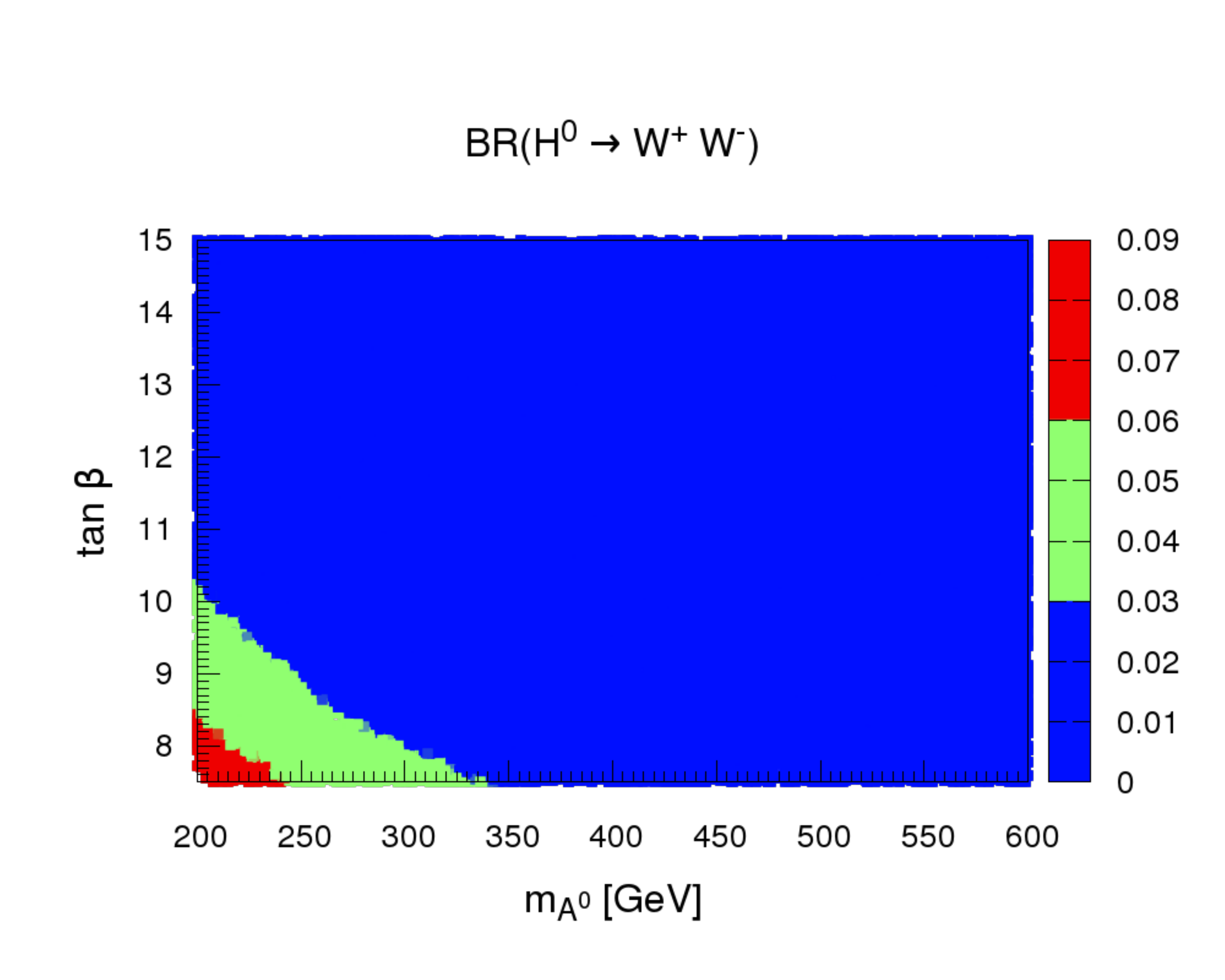} &
\includegraphics[width=75mm]{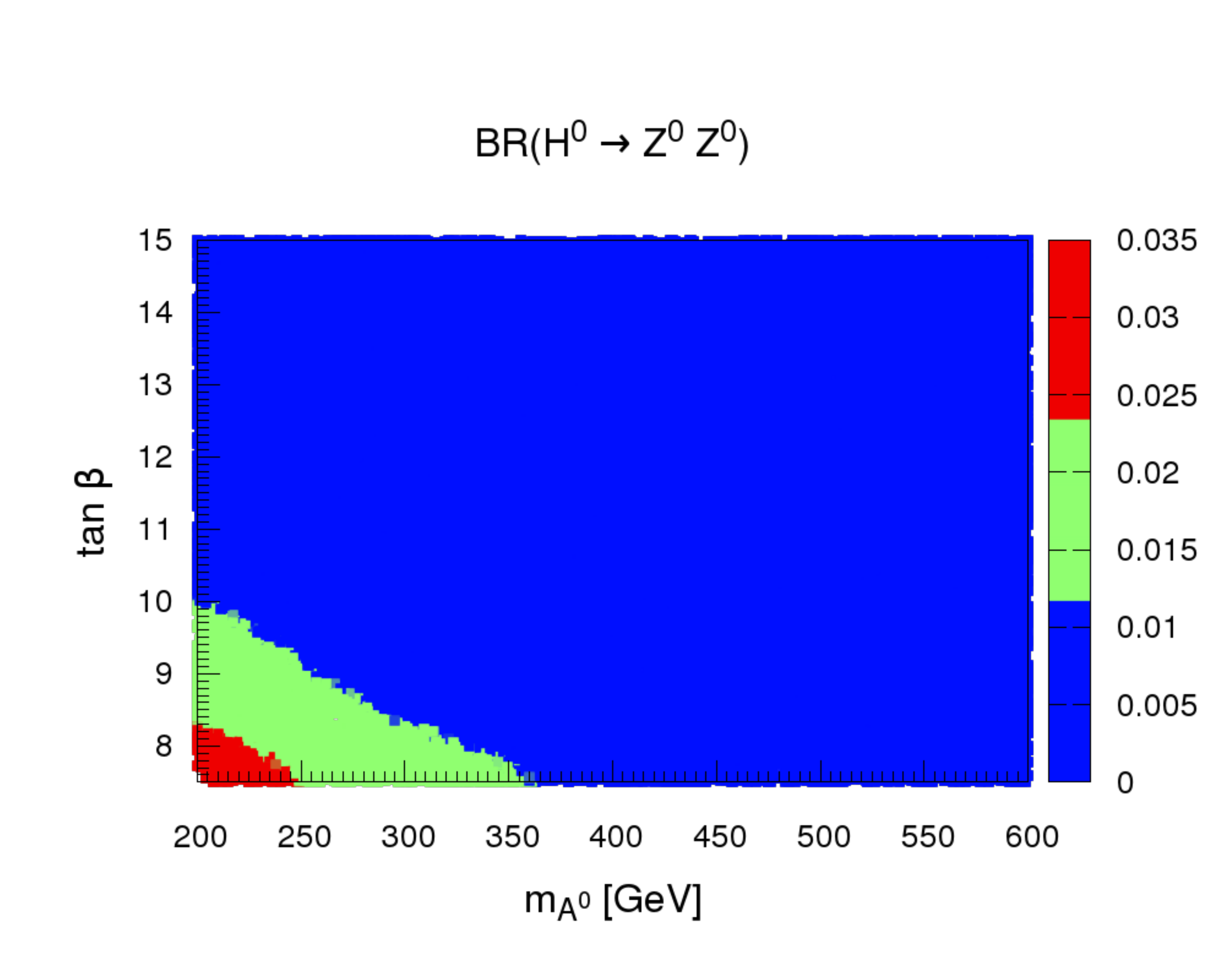} \\
\includegraphics[width=75mm]{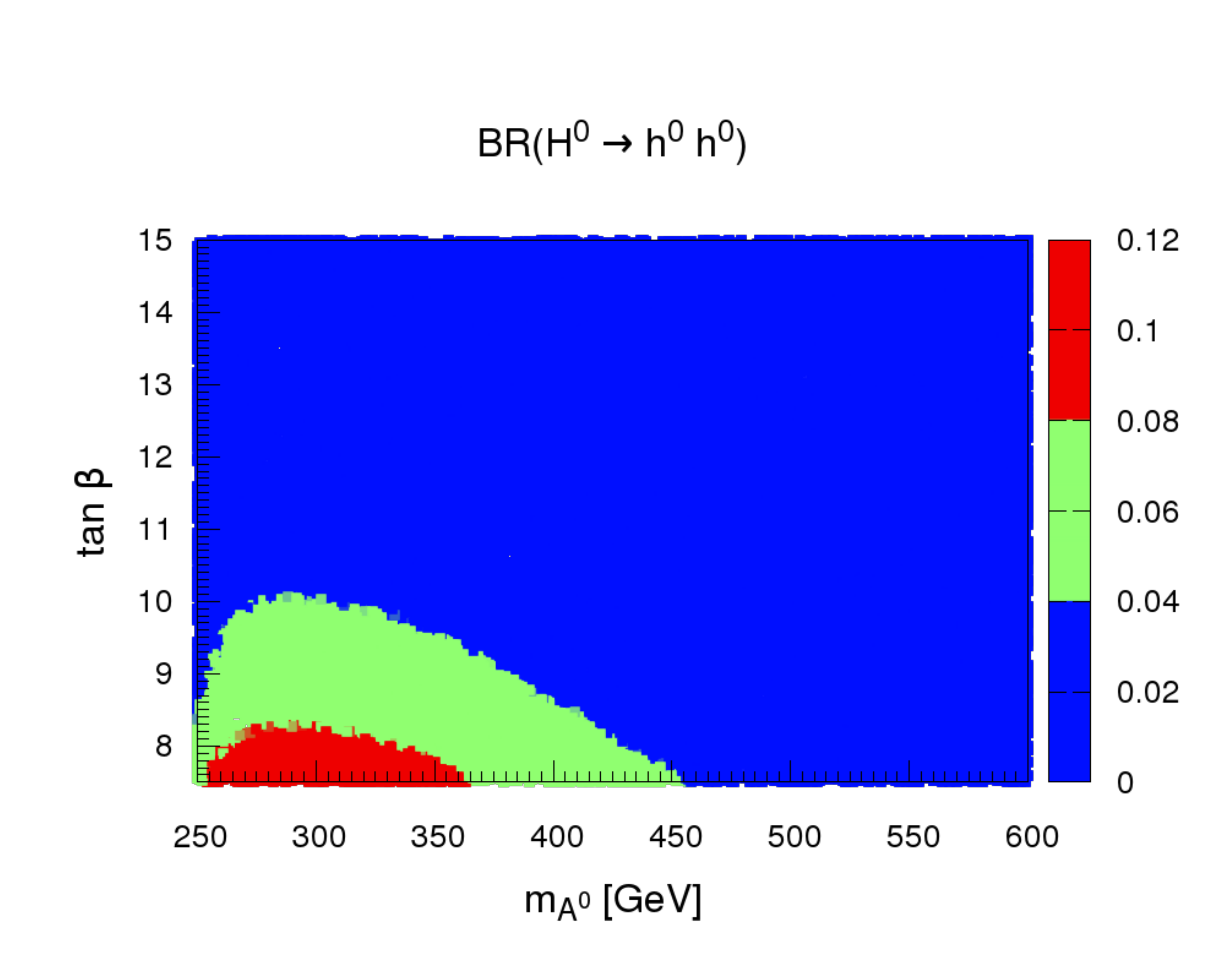} &
\includegraphics[width=75mm]{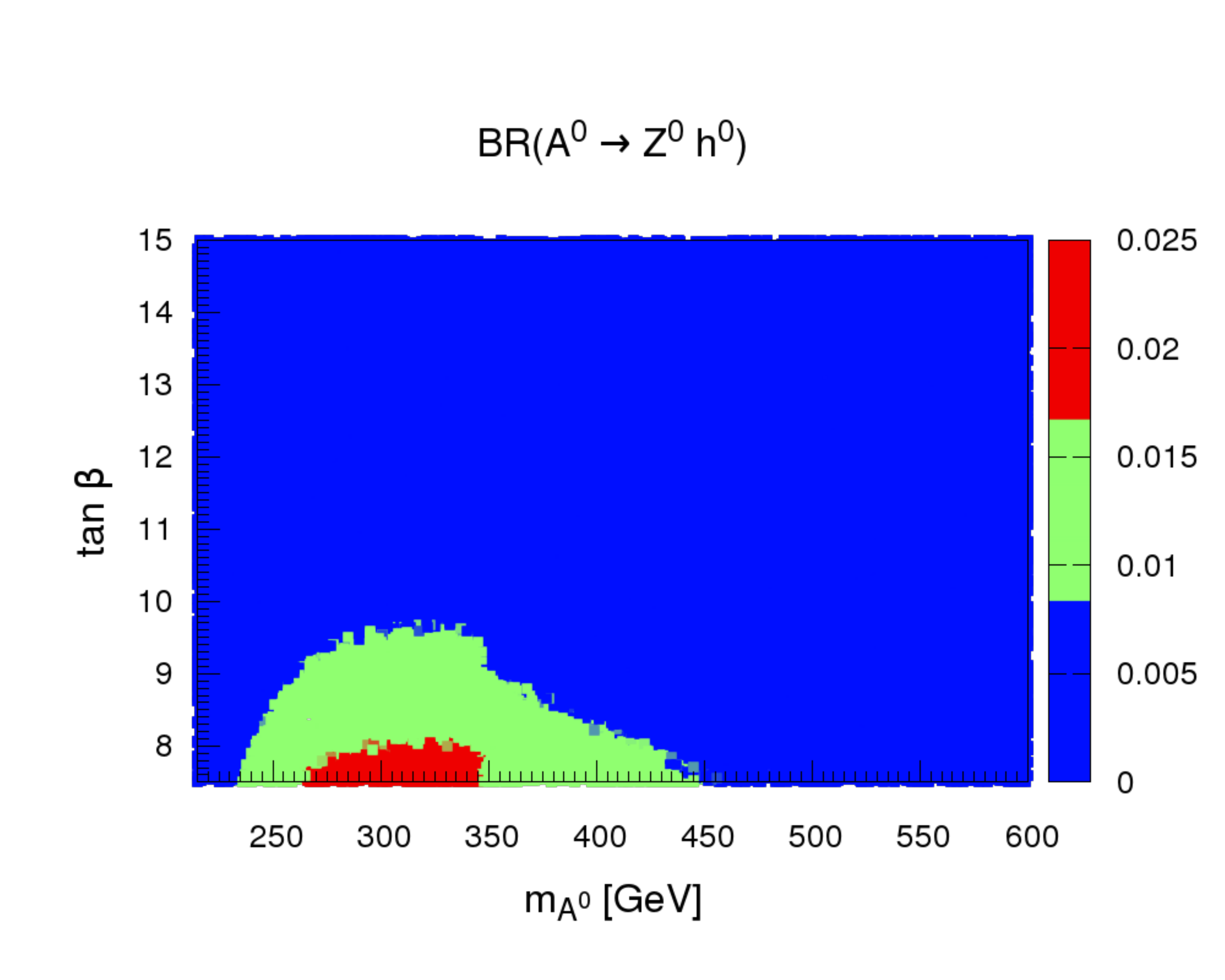}
\end{tabular}
\caption{$H^0$ and $A^0$ branching ratios decays in the plane $m_{A^0}$$-$$\tan\beta$.
Upper left panel: BR$(H^0 \to W^+ W^-)$.
Upper right panel: BR$(H^0 \to Z^0 Z^0)$.
Lower left panel: BR$(H^0 \to h^0 h^0)$.
Lower right panel: BR$(A^0 \to Z^0 h^0)$.
The scan is done in bino LSP scenario with $M_S =$ 40 TeV,
$m_s =$ 7 TeV and $A_t =$ 0 over the following ranges:
60 GeV $< |M_1| <$ 150 GeV, 1 TeV $< |M_2| \,, |\mu| \,, M_3 <$ 3 TeV.}
\label{fig:HAcontour_scenario1}
\end{center}
\end{figure}

Thus, in the most of the regions of the parameter space studied in this bino LSP scenario (for $\tan\beta \gtrsim$ 10), the dominant channels
are $(H^0 ,\, A^0) \to b \, \bar{b}$, followed by $(H^0 ,\, A^0) \to \tau^+ \, \tau^-$, whose couplings increase
with $\tan\beta$. These channels are very difficult to detect since the signals have a large
SM background and/or low detection efficiency. We notice that the most promising decay modes in this scenario are
$H^0\to Z^0 Z^0$ and $H^0 \to h^0 h^0$, which have a sizable BR and a possibly clean signature, while
for $A^0$ we have the mode $A^0 \to Z ^0 h^0$, which has also a possibly clean signature, but not such
a large BR as in the case of $H^0 \to h^0 h^0$. The decay $H^0\to W^+ W^-$ is also a very interesting channel, with a BR larger than $H^0\to Z^0 Z^0$ mode,
but a not such a clean signature.
To further analyze the significance of these modes, we shall now discuss
the contour plots in the plane $m_{A^0}$$-$$\tan\beta$, as shown in Figure~\ref{fig:HAcontour_scenario1}.

\begin{itemize}
\item The decay modes $H^0 \to Z^0 Z^0$ $(W^+ W^-)$ can have  BR $\gtrsim$ 0.01 (0.03) only for 
      the green and red strips that lie within  7.5 $\lesssim \tan\beta \lesssim$ 10 and
      200 GeV $\lesssim m_{A^0} \leq$ 350 GeV.
      
\item Similarly, the decay mode $H^0 \to h^0 h^0$ can reach a BR above 0.04 for 
      $\tan\beta \lesssim$ 10 and 250 GeV $\lesssim m_{A^0} \lesssim$ 450 GeV.

\item The decay mode $A^0 \to Z^0 h^0$ have a similar behavior, but now  BR $\gtrsim$ 0.01 for 
      slightly lower $\tan\beta \lesssim$ 10.
\end{itemize}

\begin{figure}[t!]
\begin{center}
\begin{tabular}{cc}
\includegraphics[width=75mm]{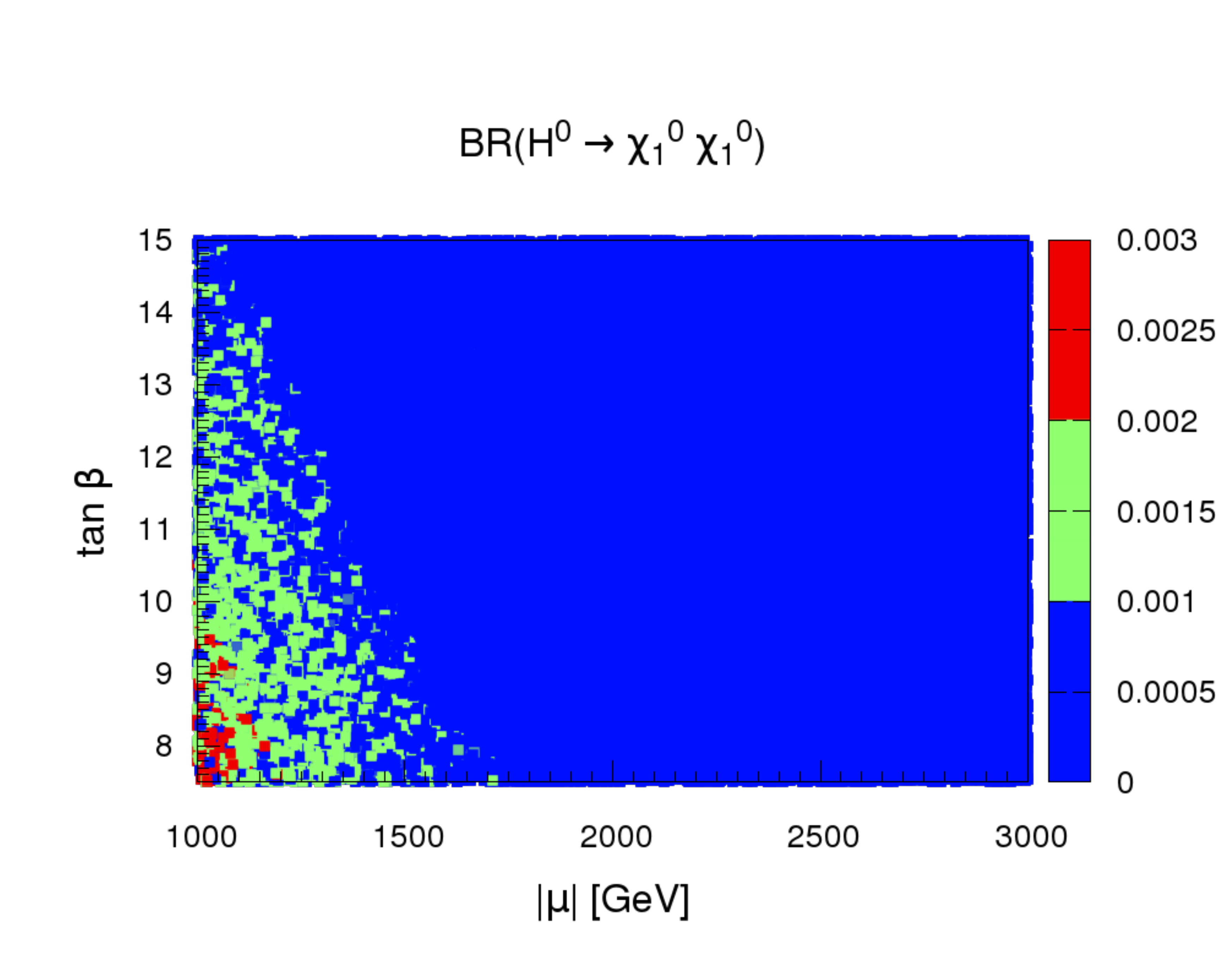} &
\includegraphics[width=75mm]{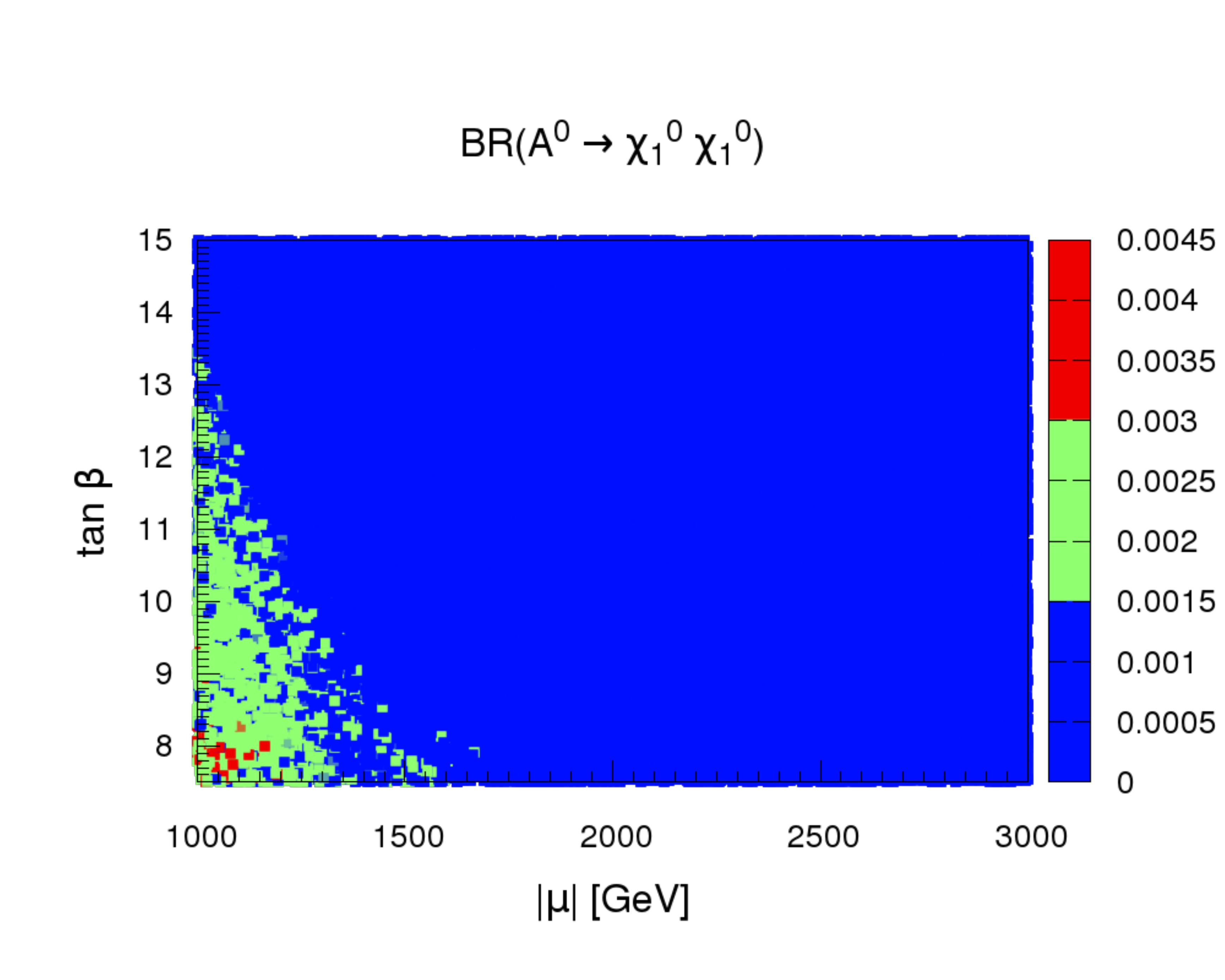} 
\end{tabular}
\caption{$H^0$ and $A^0$ branching ratios of invisible channels in the plane $|\mu|$$-$$\tan\beta$.
Left panel: BR$(H^0 \to \tilde \chi_1^0 \tilde \chi_1^0)$.
Right panel: BR$(A^0 \to \tilde \chi_1^0 \tilde \chi_1^0)$.
The scan is done in bino LSP scenario with $M_S =$ 40 TeV,
$m_s =$ 7 TeV and $A_t =$ 0 over the following ranges:
60 GeV $< |M_1| <$ 150 GeV, 200 GeV $< m_{A^0} <$ 600 GeV,
1 TeV $< |M_2| \,, |\mu| \,, M_3 <$ 3 TeV.}
\label{fig:HAinvisible_scenario1}
\end{center}
\end{figure}
 
On the other hand, the invisible decays of $H^0$ and $A^0$ are presented in Figure~\ref{fig:HAinvisible_scenario1}
in the plane $|\mu|$$-$$\tan\beta$, due to the fact that in this scenario $\mu$ governs the higgsino component
of the bino-like $\tilde \chi_1^0$, necessary to have non-negligible Higgs$-\tilde \chi_1^0-\tilde \chi_1^0$
couplings.
There, we notice that only for $|\mu| \simeq$ 1000$-$1500 GeV
and $\tan\beta \lesssim$ 14 it is possible to obtain BR $\gtrsim$ 0.001, which seems quite difficult to detect at the LHC.

\section{Wino LSP scenario}
\label{winoLSP}

The content of particles of the wino LSP scenario below 1 TeV scale
is the following: one wino-like neutralino $\tilde \chi_1^0$
(100 GeV $\lesssim m_{\tilde \chi_1^0} \lesssim$ 150 GeV),
one wino-like chargino $\tilde \chi_1^\pm$ ($m_{\tilde \chi_1^\pm} \simeq m_{\tilde \chi_1^0}$),
one light Higgs boson $h^0$ (124 GeV $< m_{h^0} <$ 127 GeV), one heavy Higgs boson $H^0$,
one pseudoscalar Higgs boson $A^0$ and one charged Higgs pair $H^\pm$
(200 GeV $\lesssim m_{H^0}, \, m_{A^0}, \, m_{H^\pm} \lesssim$ 600 GeV).
The minimum value of $m_{\tilde \chi_1^\pm}$, governed by the wino soft mass $M_2$ in this scenario, is chosen for respecting the present lower bound~\cite{Beringer:1900zz}.
In order to have one wino-like neutralino, one needs to require $|M_2| \ll |M_1|$, $|\mu|$,
appearing a new particle in the spectrum at the EW scale, a wino-like chargino, very degenerate in mass
with the neutralino.

For the study of the decays into charginos, we have to attend to the interactions of charginos $\tilde \chi_1^\pm$
with $H^0$ and $A^0$ Higgs bosons, whose couplings are~\cite{Djouadi:2001fa}:
\begin{eqnarray}
G^{L,R}_{\tilde \chi^-_1 \tilde \chi^+_1 H^0} &=& \frac{1}{\sqrt{2} \sin\theta_W} \left[ \cos\alpha \, V_{11} \, U_{12} + \sin\alpha \, V_{12} \, U_{11}
\right] \,, \label{Hchi1chi1} \\
G^{L}_{\tilde \chi^-_1 \tilde \chi^+_1 A^0} &=& - \frac{1}{\sqrt{2} \sin\theta_W} \left[ \sin\beta \, V_{11} \, U_{12} + \cos\beta \, V_{12} \, U_{11}
\right] \,, \label{Achi1chi1L}\\
G^{R}_{\tilde \chi^-_1 \tilde \chi^+_1 A^0} &=& - G^{L}_{\tilde \chi^-_1 \tilde \chi^+_1 A^0} \,, \label{Achi1chi1R}
\end{eqnarray}
where $U_{ij}$ and $V_{ij}$ are the entries of the two real matrices $U$ and $V$ that diagonalize
the chargino mass matrix in the wino-higgsino basis. As in the case of the invisible decays of Higgs bosons,
the ($H^0$, $A^0$) $\to \tilde \chi^+_1 \tilde \chi^-_1$ decay modes have sources from Higgs-higgsino-gaugino
couplings and if $\tilde \chi^\pm_1$ was pure wino (or pure higgsino), the width of these modes of $H^0/A^0$ will vanish.

\begin{figure}[t!]
\begin{center}
\begin{tabular}{cc}
\includegraphics[width=75mm]{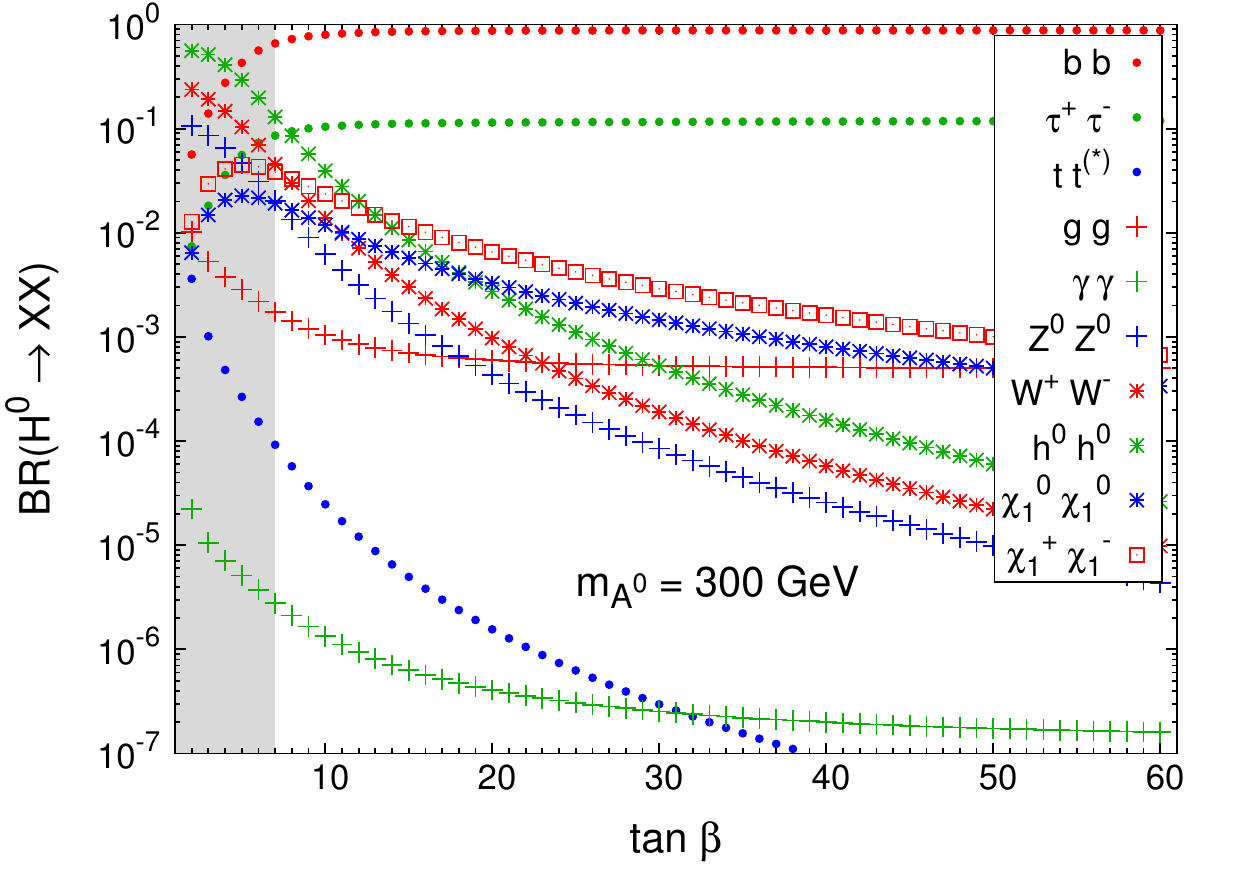} &
\includegraphics[width=75mm]{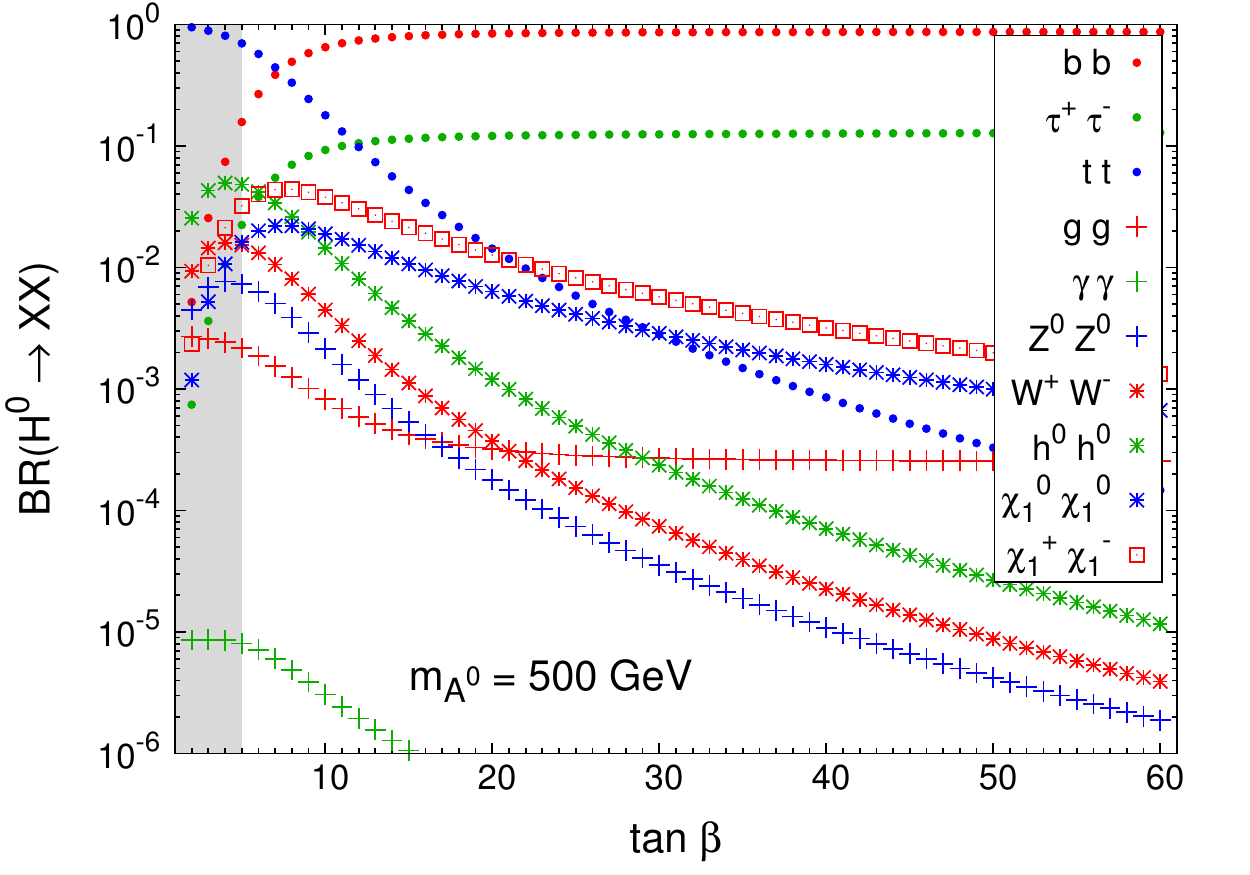} \\
\includegraphics[width=75mm]{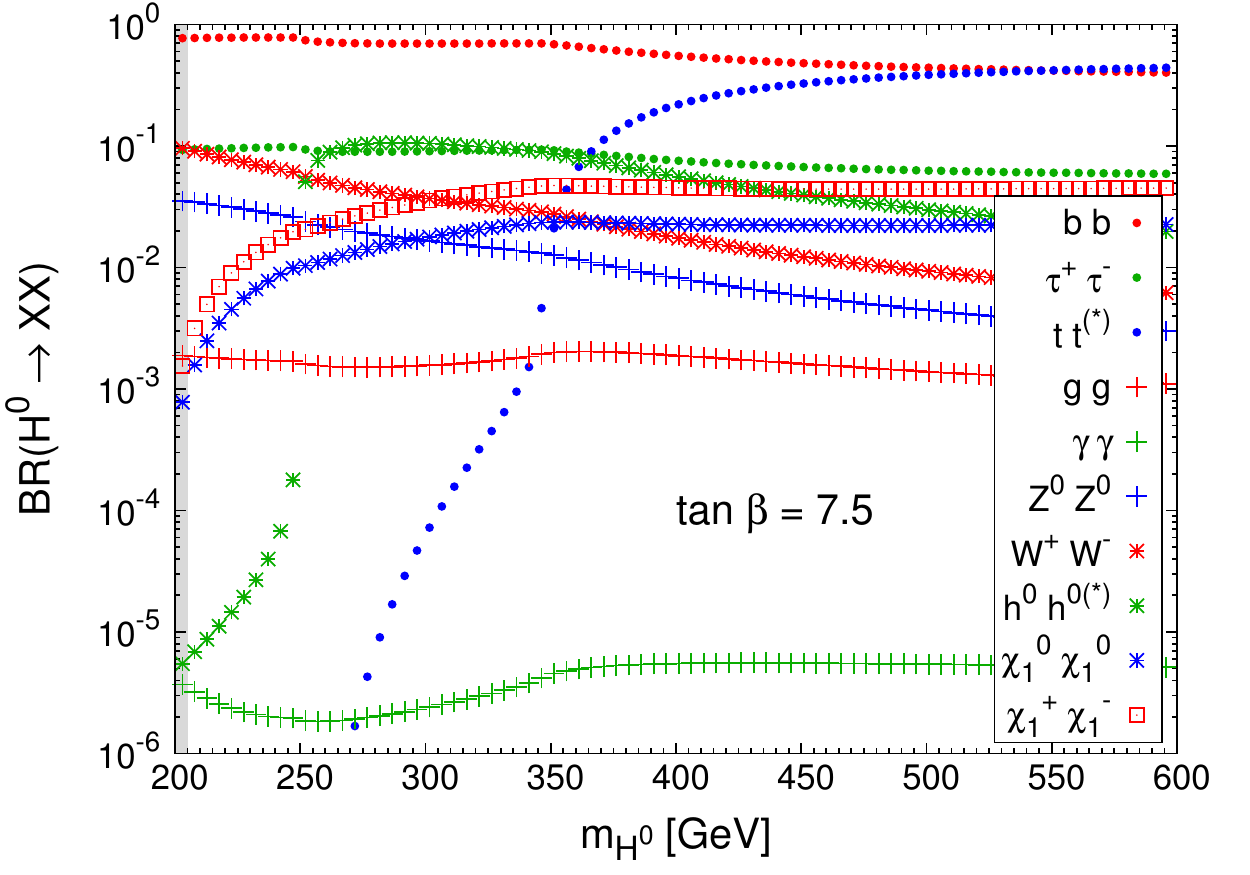} &
\includegraphics[width=75mm]{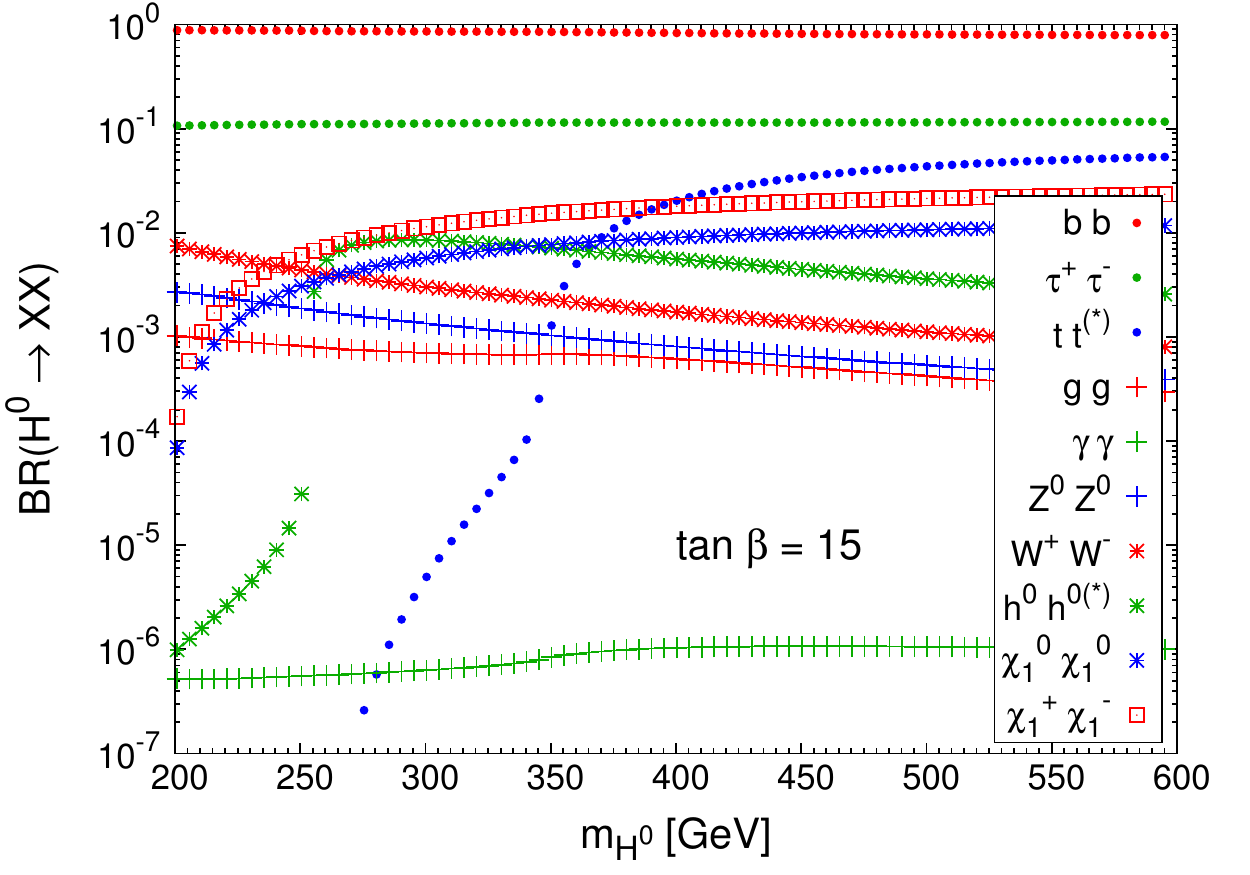}
\end{tabular}
\caption{$H^0$ decay channels in wino LSP scenario for $M_S =$ 40 TeV,
$m_s =$ 7 TeV, $A_t =$ 0, $M_2 =$ 100 GeV, $M_1 = \mu =$ 1 TeV, $M_3 =$ 3 TeV.
Upper left panel: $H^0$ branching ratios as a function of $\tan\beta$ for $m_{A^0} = $ 300 GeV.
Upper right panel: $H^0$ branching ratios as a function of $\tan\beta$ for $m_{A^0} = $ 500 GeV.
Lower left panel: $H^0$ branching ratios as a function of $m_{H^0}$ for $\tan\beta = 7.5$.
Lower right panel: $H^0$ branching ratios as a function of $m_{H^0}$ for $\tan\beta = 15$.
Shaded gray area represents values of $m_{h^0} <$ 124 GeV. The discontinuities around
$m_{H^0} \simeq$ 250, 350 GeV in lower panels are due to the fact that $h^0 h^0$ and $t \bar t$ channels
start to be kinematically allowed.}
\label{fig:BRH0decays_scenario2}
\end{center}
\end{figure}

\begin{figure}[t!]
\begin{center}
\begin{tabular}{cc}
\includegraphics[width=75mm]{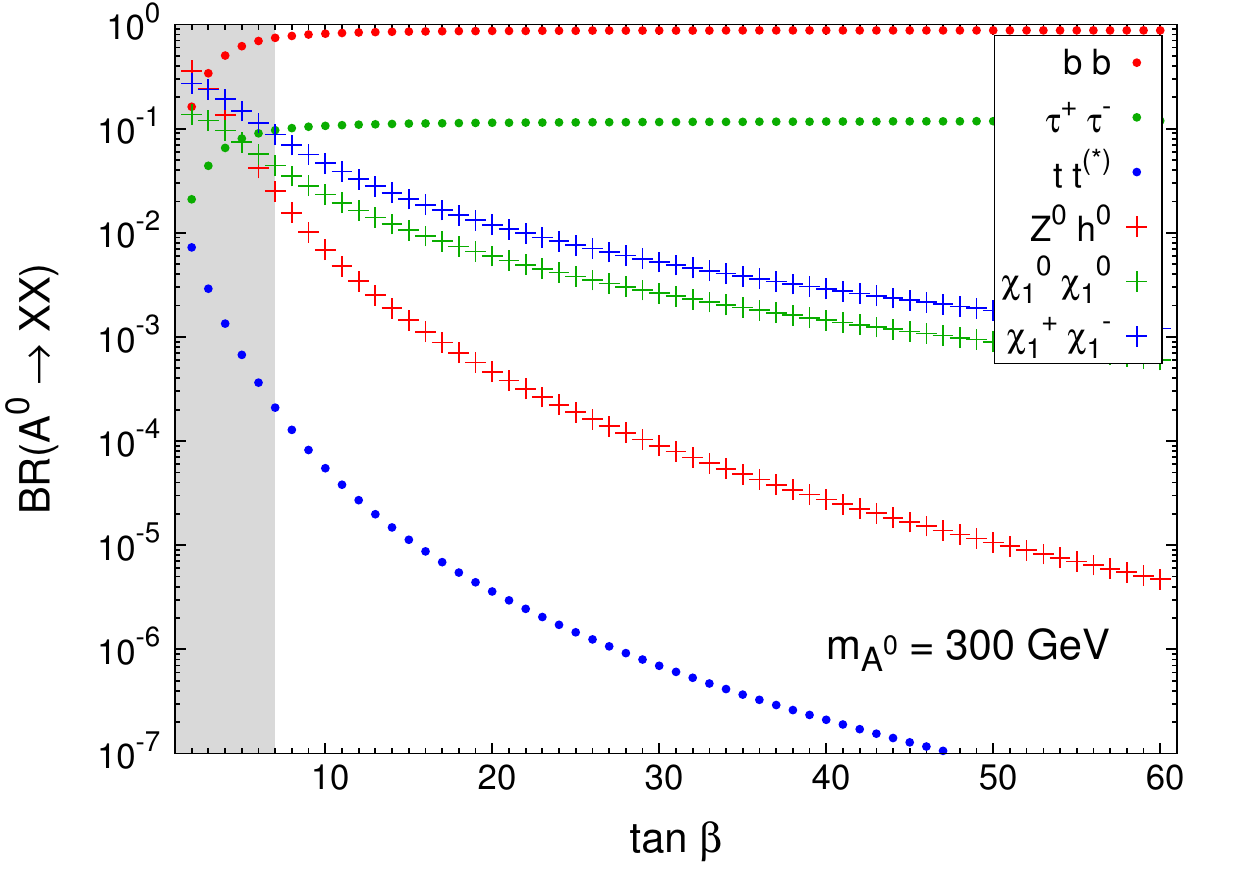} &
\includegraphics[width=75mm]{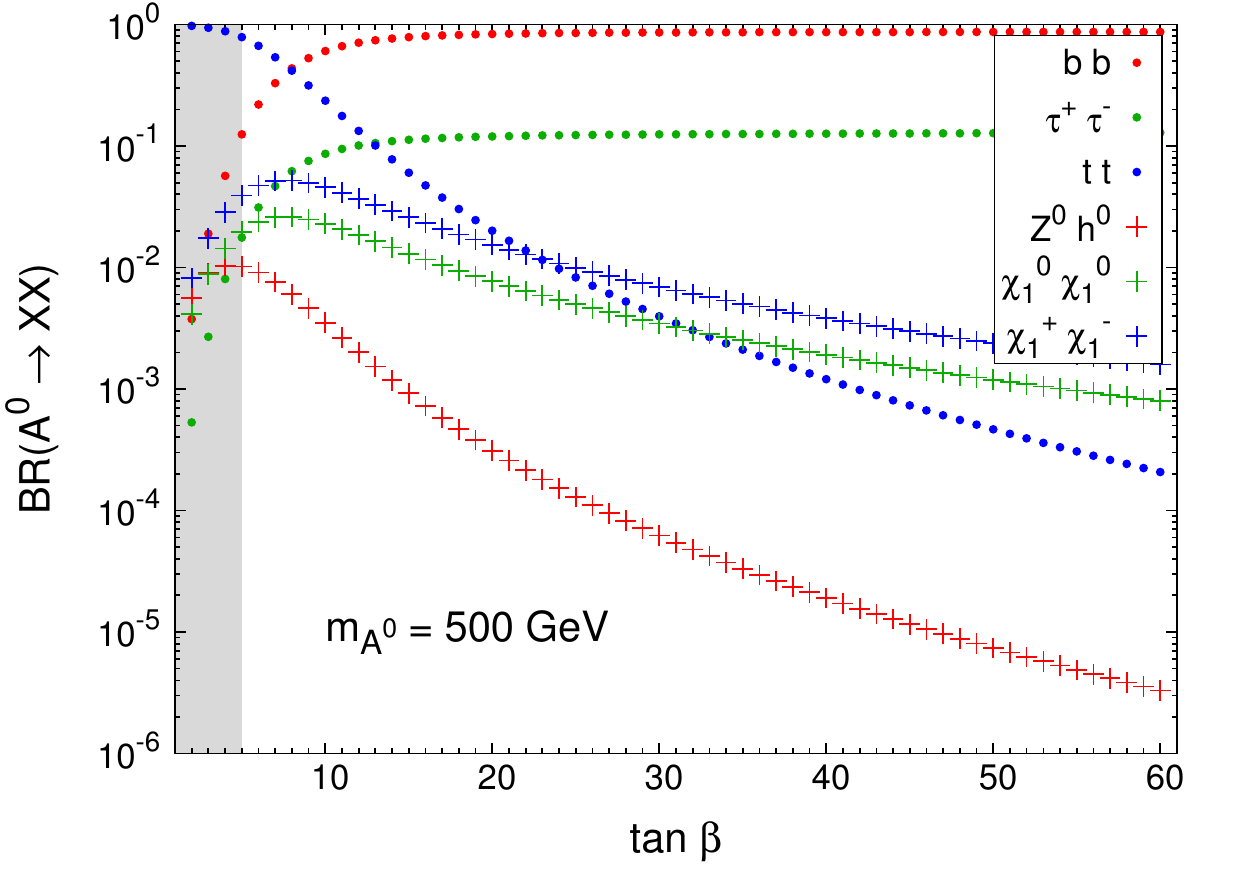} \\
\includegraphics[width=75mm]{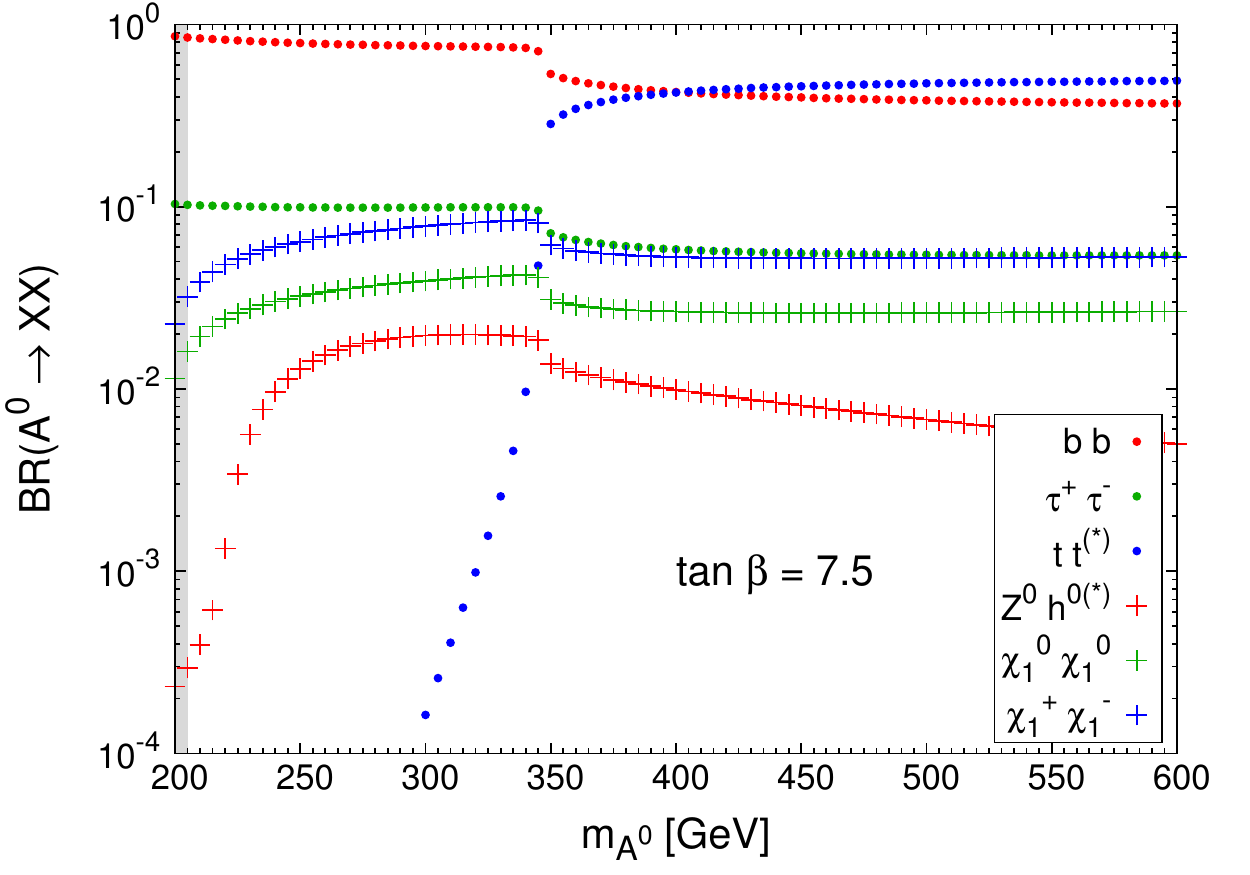} &
\includegraphics[width=75mm]{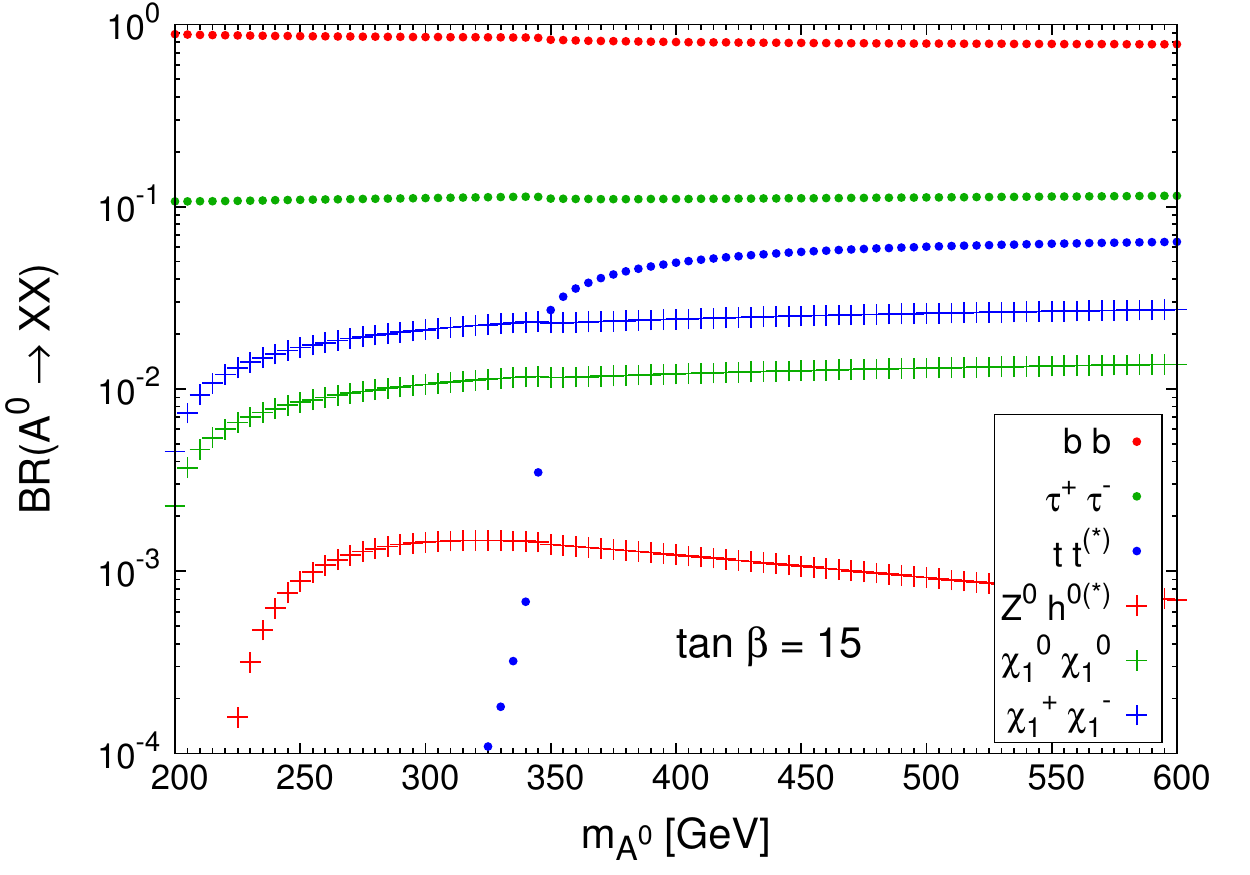}
\end{tabular}
\caption{$A^0$ decay channels in wino LSP scenario for $M_S =$ 40 TeV,
$m_s =$ 7 TeV, $A_t =$ 0, $M_2 =$ 100 GeV, $M_1 = \mu =$ 1 TeV, $M_3 =$ 3 TeV.
Upper left panel: $A^0$ branching ratios as a function of $\tan\beta$ for $m_{A^0} = $ 300 GeV.
Upper right panel: $A^0$ branching ratios as a function of $\tan\beta$ for $m_{A^0} = $ 500 GeV.
Lower left panel: $A^0$ branching ratios as a function of $m_{A^0}$ for $\tan\beta = 7.5$.
Lower right panel: $A^0$ branching ratios as a function of $m_{A^0}$ for $\tan\beta = 15$.
Shaded gray area represents values of $m_{h^0} <$ 124 GeV. The discontinuities around
$m_{A^0} \simeq$ 350 GeV in lower panels are due to the fact that $t \bar t$ channel
starts to be kinematically allowed.}
\label{fig:BRA0decays_scenario2}
\end{center}
\end{figure}

The results of the branching ratios of $H^0$ and $A^0$, within this second scenario,
are contained in Figures~\ref{fig:BRH0decays_scenario2} and~\ref{fig:BRA0decays_scenario2}
for the parameters $M_S =$ 40 TeV, $m_s$ = 7 TeV, $A_t =$ 0,
$M_2 =$ 100 GeV, $M_1 = \mu =$ 1 TeV, $M_3 =$ 3 TeV. We present the results of branching ratios, like in the previous section,
as a function of $\tan\beta$ and $m_{H^0 (A^0)}$ for both $H^0$ and $A^0$ bosons.
From these plots we find the following salient features:

\begin{itemize}
\item For $H^0$ and $A^0$ the dominant mode is again the decay into $b \bar{b}$
      with BR $\simeq$ 0.9, for values of $\tan\beta \gtrsim$ 10.
      If $t \bar t$ channel is kinematically allowed, for values of
      $\tan\beta \lesssim$ 10 it becomes the dominant one.
       
\item $H^0 \to Z^0 Z^0$, $W^+ W^-$, $h^0 h^0$ channels exhibit the same behavior as
      in bino LSP scenario, reaching similar maximum values of branching ratios (0.03, 0.1 and 0.15, respectively).

\item A very interesting feature of this scenario is the increase of the branching ratios
      of the invisible decays, which grows up around an order of magnitude respect the branching ratios of bino LSP scenario,
      reaching the largest values of BR($H^0 \to \tilde{\chi}^0_1 \tilde{\chi}^0_1$) $\simeq$ 2 $\times$ $10^{-2}$
      and BR($A^0 \to \tilde{\chi}^0_1 \tilde{\chi}^0_1$) $\simeq$ 4 $\times$ $10^{-2}$.
       
\item A  new feature of this scenario is the appearance of the decays of $H^0$ and $A^0$ into charginos $\chi^+_1 \chi^-_1$.
      Indeed, the branching ratios of this modes are even larger than the invisible ones, reaching values of almost 0.05 and 0.1, respectively.
     
\end{itemize}

\begin{figure}[t!]
\begin{center}
\begin{tabular}{cc}
\includegraphics[width=75mm]{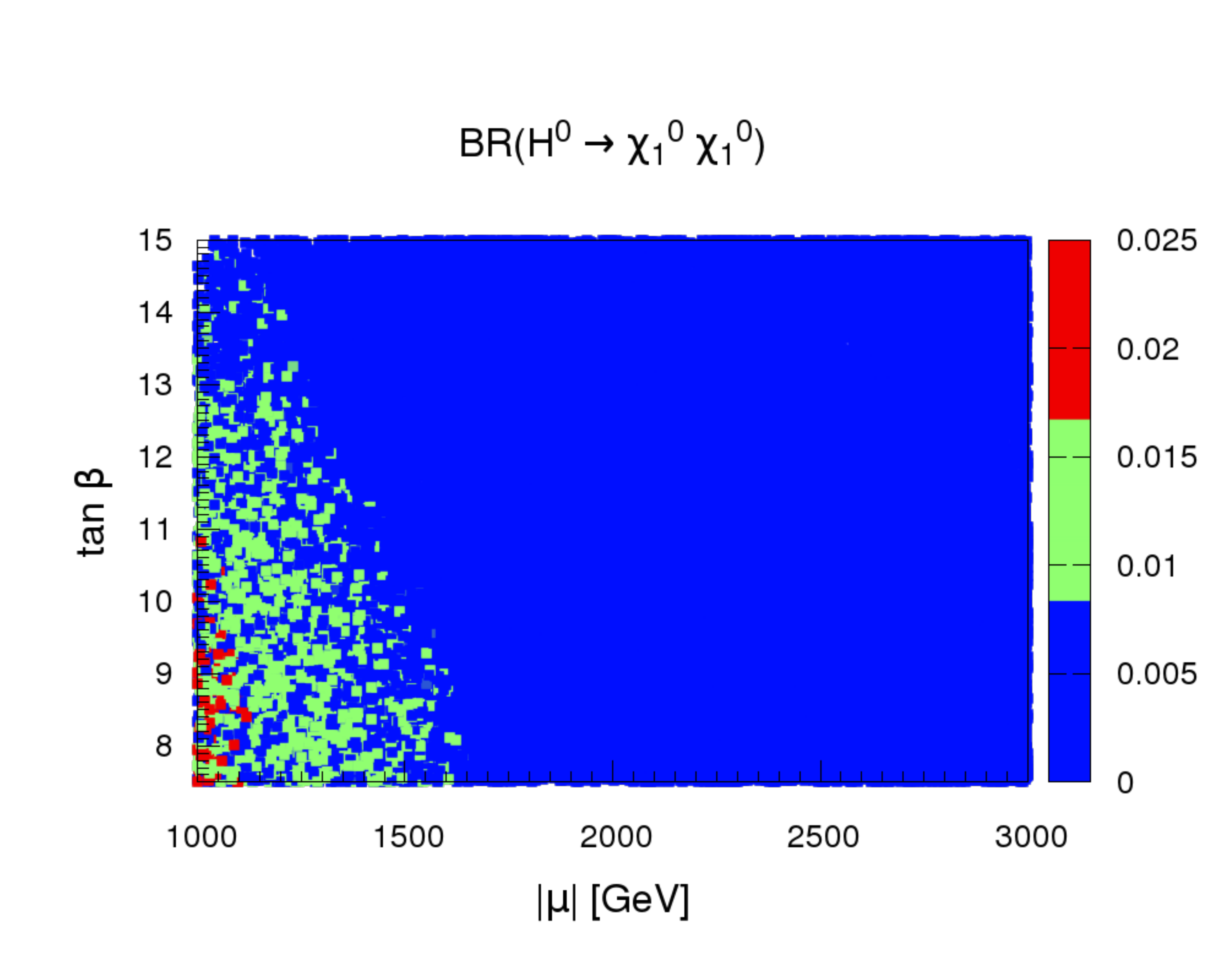} &
\includegraphics[width=75mm]{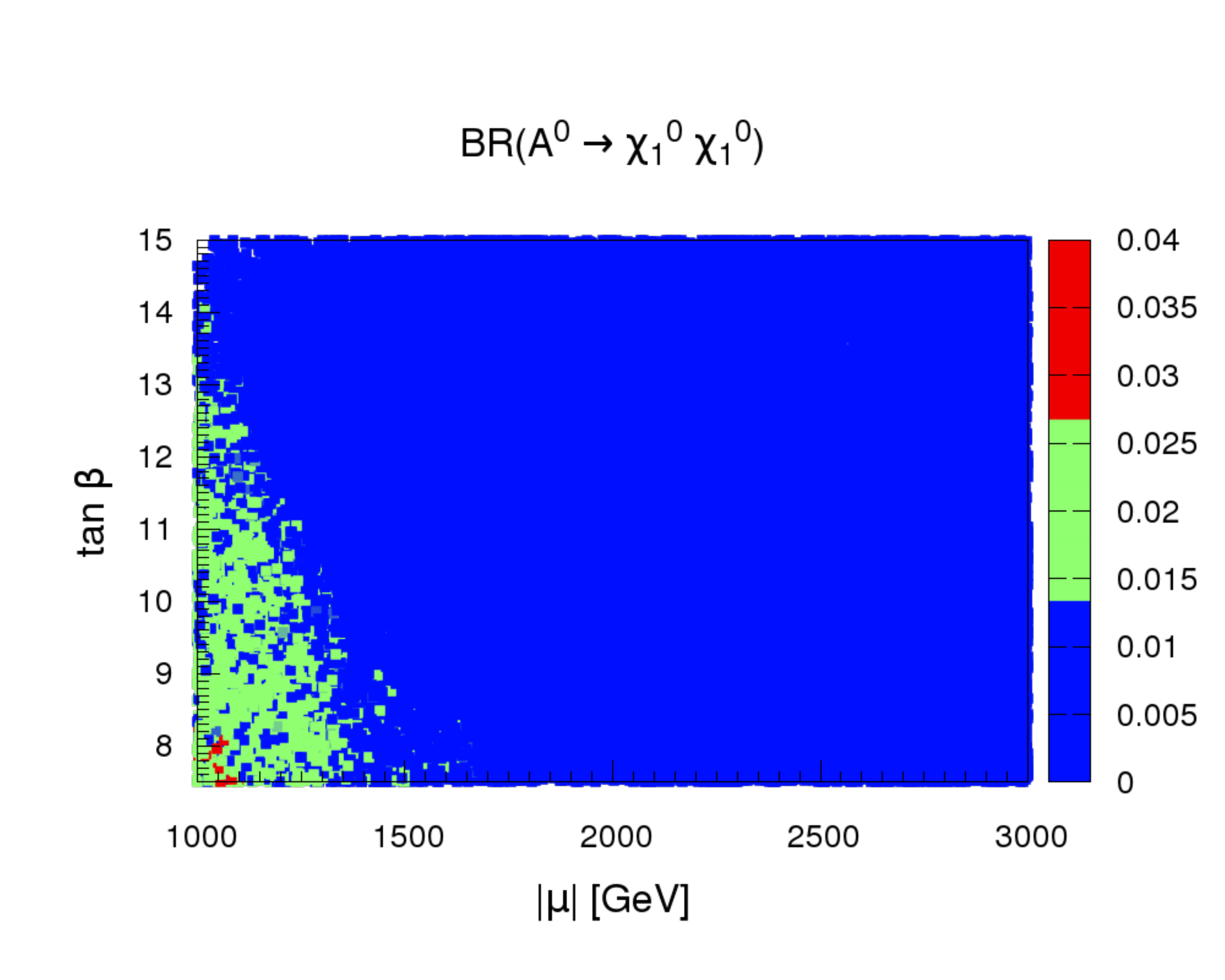} \\
\includegraphics[width=75mm]{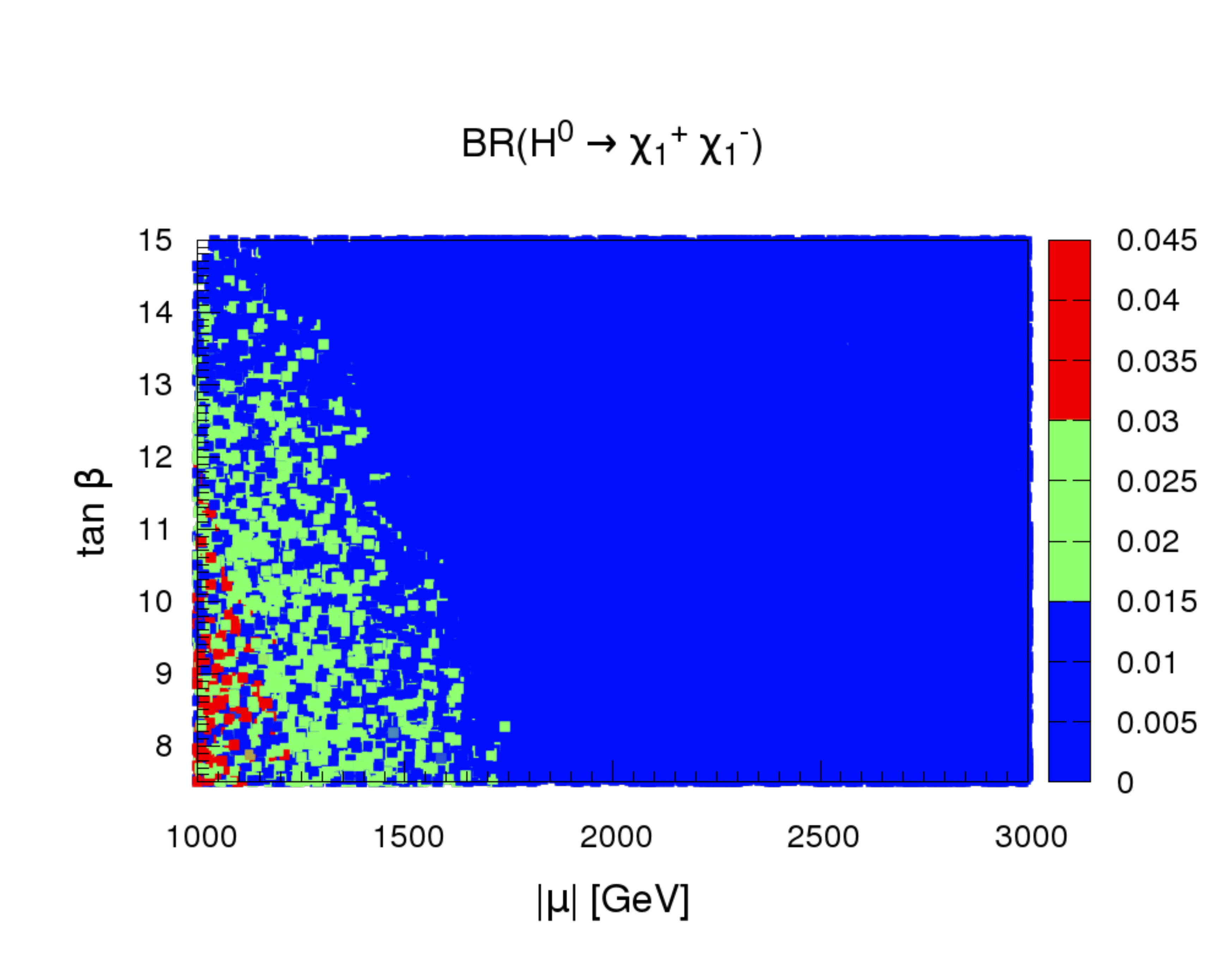} &
\includegraphics[width=75mm]{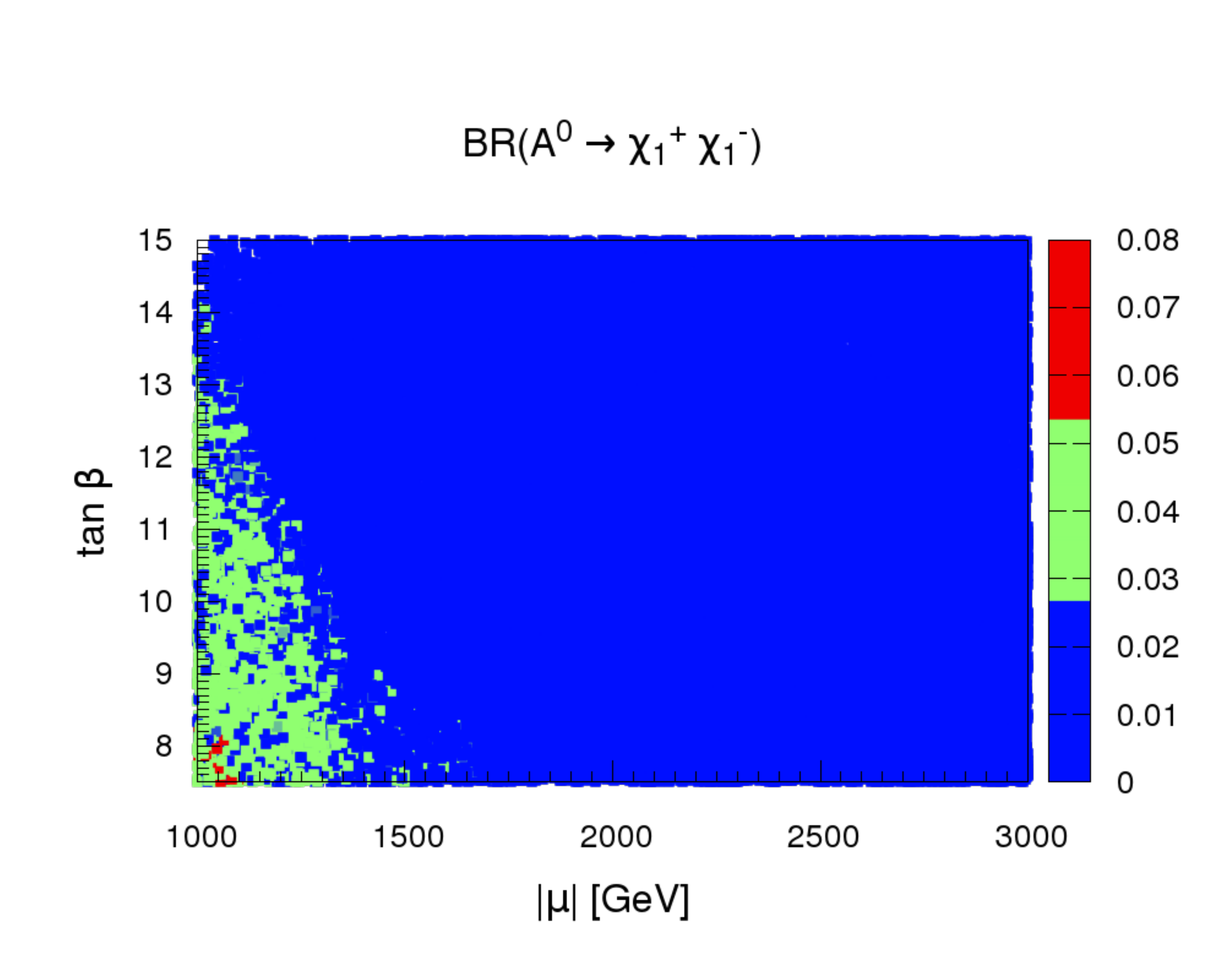}
\end{tabular}
\caption{$H^0$ and $A^0$ branching ratios of neutralino and chargino
channels in the plane $|\mu|$$-$$\tan\beta$.
Upper left panel: BR$(H^0 \to \tilde \chi_1^0 \tilde \chi_1^0)$.
Upper right panel: BR$(A^0 \to \tilde \chi_1^0 \tilde \chi_1^0)$.
Lower left panel: BR$(H^0 \to \tilde \chi_1^+ \tilde \chi_1^-)$.
Lower right panel: BR$(A^0 \to \tilde \chi_1^+ \tilde \chi_1^-)$.
The scan is done in wino LSP scenario with $M_S =$ 40 TeV,
$m_s =$ 7 TeV and $A_t =$ 0 over the following ranges:
100 GeV $< |M_2| <$ 150 GeV, 200 GeV $< m_{A^0} <$ 600 GeV,
1 TeV $< |M_1| \,, |\mu| \,, M_3 <$ 3 TeV.}
\label{fig:HAinvisible_scenario2}
\end{center}
\end{figure}

As in the previous scenario, we notice that the most promising decay modes in this scenario are
$H^0 \to Z^0 Z^0$ and $H^0 \to h^0 h^0$, which have a sizable branching ratio and a possibly clean signature, while
for $A^0$ we have the mode $A^0 \to Z^0 h^0$, which has a possibly clean signature, but not such a
large BR. The behavior of these branching ratios in the plane $m_{A^0}$$-$$\tan\beta$ is in fact similar to the
previous scenario too.
The new feature of this scenario is the moderate enhancement of the invisible decays as well
as the opening of the decays into charginos. To further analyze the significance of these modes, we shall now discuss
the contour plots in the plane $|\mu|$$-$$\tan\beta$, displayed in Figure~\ref{fig:HAinvisible_scenario2}.

\begin{itemize}
\item The branching ratios for the invisible decays of $H^0$ and $A^0$ are presented
      in the upper panels of Figure~\ref{fig:HAinvisible_scenario2}. 
      There, we notice that only for $|\mu| \lesssim$ 1200 GeV and $\tan\beta \lesssim$ 11
      it is possible to obtain BR$(H^0 \to \tilde \chi_1^0 \tilde \chi_1^0)$ $\gtrsim$ 1.5 $\times$ $10^{-2}$.
      If we want BR$(A^0 \to \tilde \chi_1^0 \tilde \chi_1^0)$ $\gtrsim$ 1.5 $\times$ $10^{-2}$,
      we need values of $|\mu| \lesssim$ 1400 GeV and $\tan\beta \lesssim$ 13.

\item The branching ratios for the decay mode into charginos can be slightly larger, as shown
      in the lower panels of Figure~\ref{fig:HAinvisible_scenario2}. In the case of
      BR$(H^0 \to \tilde \chi_1^+ \tilde \chi_1^-)$, we can obtain values around 3$-$4.5 $\times$ $10^{-2}$ 
      with $|\mu| \lesssim$ 1250 GeV and $\tan\beta \lesssim$ 11. BR$(A^0 \to \tilde \chi_1^+ \tilde \chi_1^-)$
      can reach values from 3 $\times$ $10^{-2}$ up to 8 $\times$ $10^{-2}$ for 
      $|\mu| \lesssim$ 1500 GeV and $\tan\beta \lesssim$ 13.
            
\end{itemize}

The most important feature of this scenario is the existence of pure supersymmetric
signatures which could depict an evidence of physics beyond the SM different from a 2HDM.
On the one hand, we have non-negligible invisible decay branching ratios (up to 2.5\% for
$H^0 \to \tilde \chi_1^0 \tilde \chi_1^0$ and up to 4\% for $A^0 \to \tilde \chi_1^0 \tilde \chi_1^0$)
that could produce a big amount of missing transverse energy ($E_T^\text{miss}$) in the detectors.
On the other hand, the decays into charginos have even larger branching ratios than the invisible ones.
They could leave a clear new physics signal consisting of very energetic tracks
in the calorimeters, due to the fact that in this scenario charginos are stable particles
at detector level since they are very degenerate in mass with the neutralino
($m_{\tilde \chi_1^+}$ $-$ $m_{\tilde \chi_1^0} <$ 5 MeV) and
their decay length is at least $10^8$ m.

\section{Higgsino LSP scenario}
\label{higgsinoLSP}

The content of particles of the higgsino LSP scenario below 1 TeV scale
is the following: two higgsino-like neutralinos $\tilde \chi_1^0$ and $\tilde \chi_2^0$,
one higgsino-like chargino pair $\tilde \chi_1^\pm$
(100 GeV $\lesssim m_{\tilde \chi_1^0}$, $m_{\tilde \chi_2^0}$, $m_{\tilde \chi_1^\pm} \lesssim$ 150 GeV),
one light Higgs boson $h^0$ (124 GeV $< m_{h^0} <$ 127 GeV), a couple of heavy Higgs bosons, one scalar $H^0$
and one pseudoscalar $A^0$, as well as one charged Higgs pair $H^\pm$
(200 GeV $\lesssim m_{H^0}, \, m_{A^0}, \, m_{H^\pm} \lesssim$ 600 GeV).
The minimum value of $m_{\tilde \chi_1^\pm}$, governed by $\mu$ in this scenario, is chosen in order to respect the present lower bound~\cite{Beringer:1900zz}.

This scenario with a higgsino-like neutralino as LSP is realized with
$|\mu| \ll |M_1|$, $|M_2|$. Therefore, two additional particles turn up
at low energies near the EW scale: a second higgsino-like neutralino and
one higgsino-like chargino.
The couplings of the lightest neutralino and the chargino pair to neutral Higgs bosons
in this scenario are also given by Eqs. (\ref{Hchi01chi01}$-$\ref{Achi1chi1R}) and the couplings for the second neutralino
can be obtained in~\cite{Djouadi:2001fa}.

\begin{figure}[t!]
\begin{center}
\begin{tabular}{cc}
\includegraphics[width=75mm]{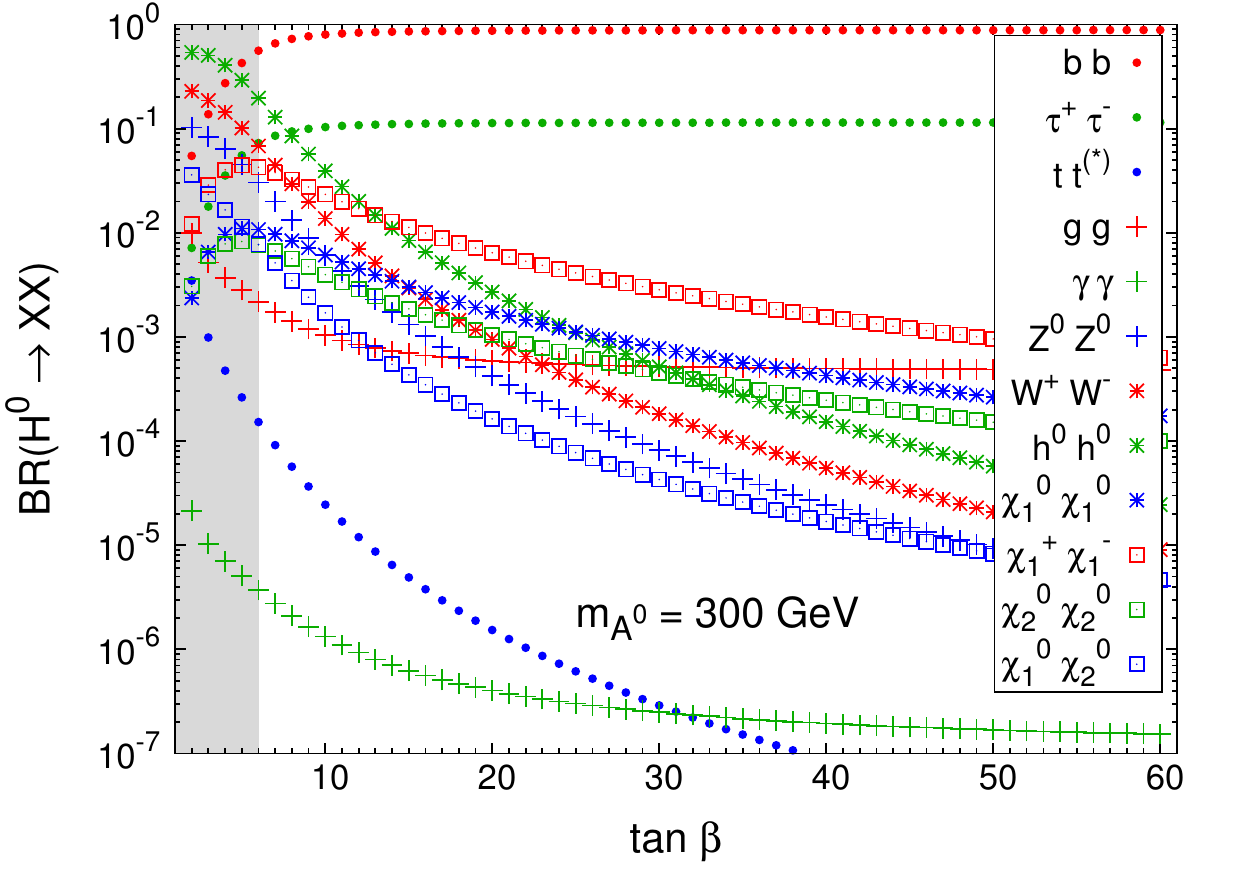} &
\includegraphics[width=75mm]{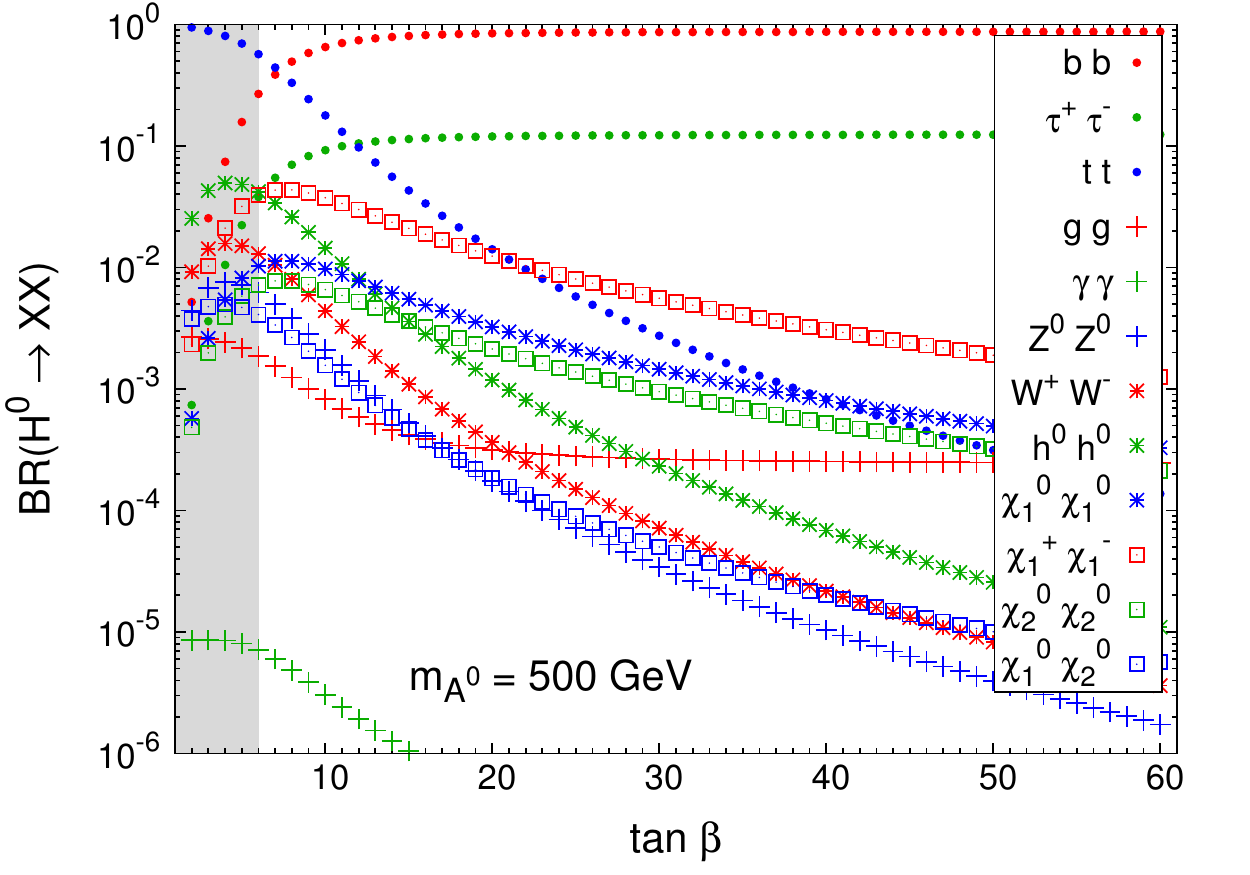} \\
\includegraphics[width=75mm]{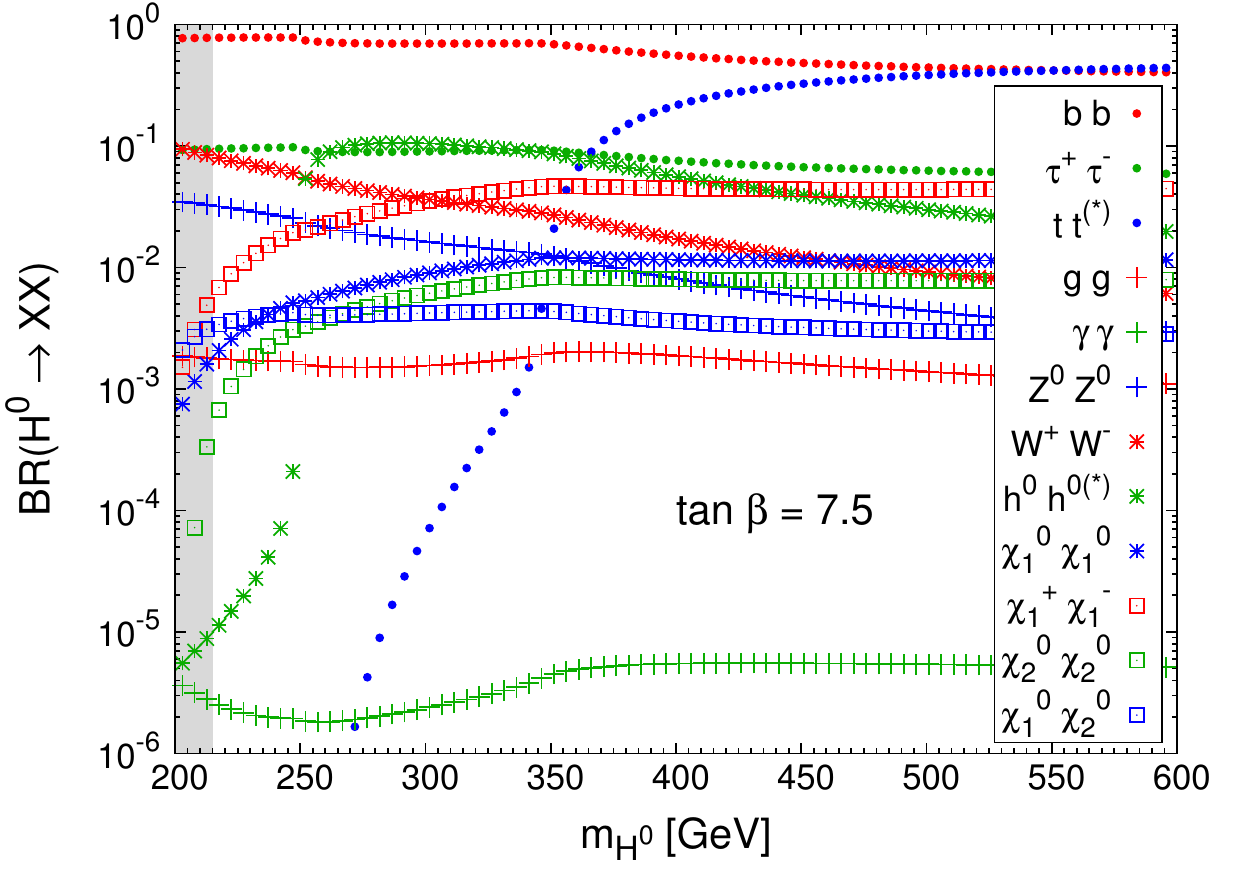} &
\includegraphics[width=75mm]{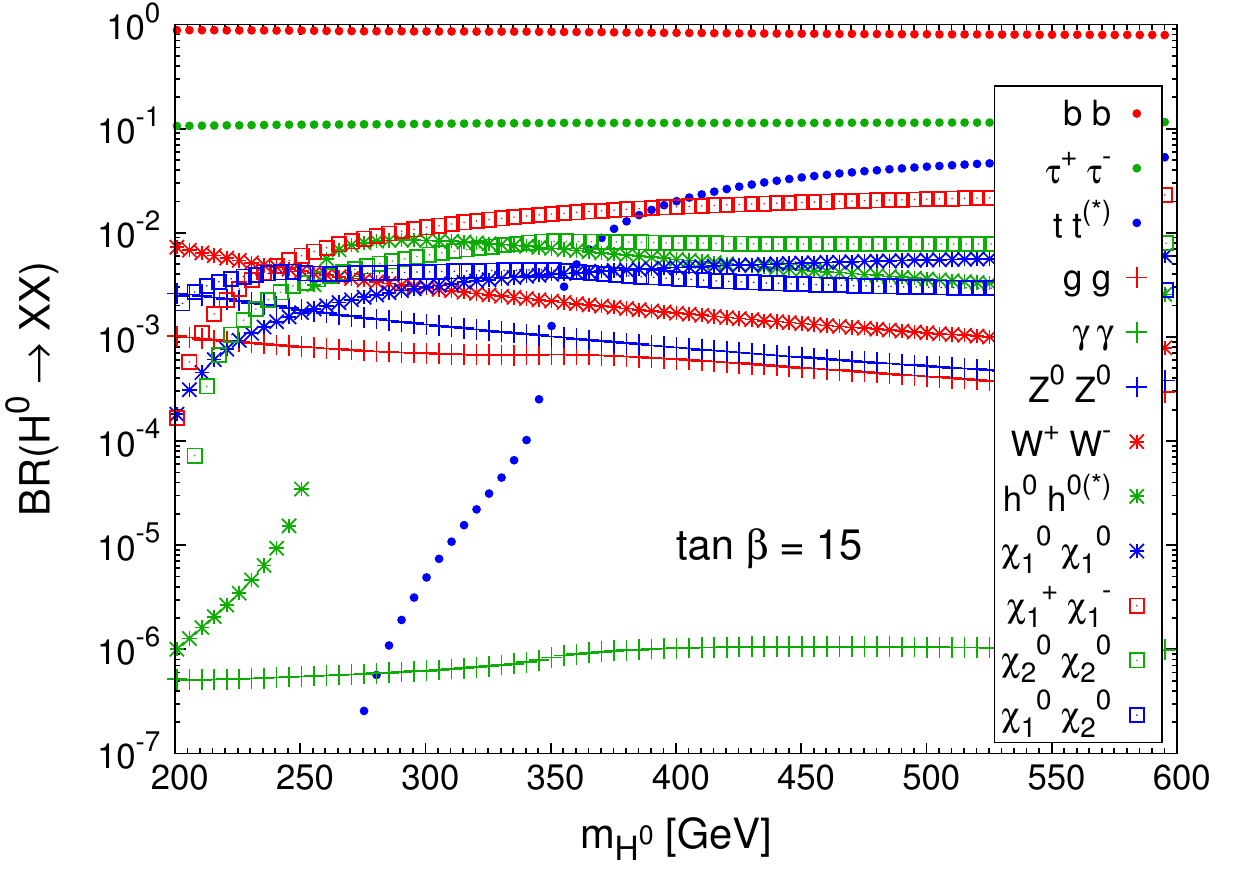}
\end{tabular}
\caption{$H^0$ decay channels in higgsino LSP scenario for $M_S =$ 40 TeV,
$m_s =$ 7 TeV, $A_t =$ 0, $\mu =$ 100 GeV, $M_1 = M_2 =$ 1 TeV, $M_3 =$ 3 TeV.
Upper left panel: $H^0$ branching ratios as a function of $\tan\beta$ for $m_{A^0} = $ 300 GeV.
Upper right panel: $H^0$ branching ratios as a function of $\tan\beta$ for $m_{A^0} = $ 500 GeV.
Lower left panel: $H^0$ branching ratios as a function of $m_{H^0}$ for $\tan\beta = 7.5$.
Lower right panel: $H^0$ branching ratios as a function of $m_{H^0}$ for $\tan\beta = 15$.
Shaded gray area represents values of $m_{h^0} <$ 124 GeV. The discontinuities around
$m_{H^0} \simeq$ 250, 350 GeV in lower panels are due to the fact that $h^0 h^0$ and $t \bar t$ channels
start to be kinematically allowed, respectively.}
\label{fig:BRH0decays_scenario3a}
\end{center}
\end{figure}

\begin{figure}[t!]
\begin{center}
\begin{tabular}{cc}
\includegraphics[width=75mm]{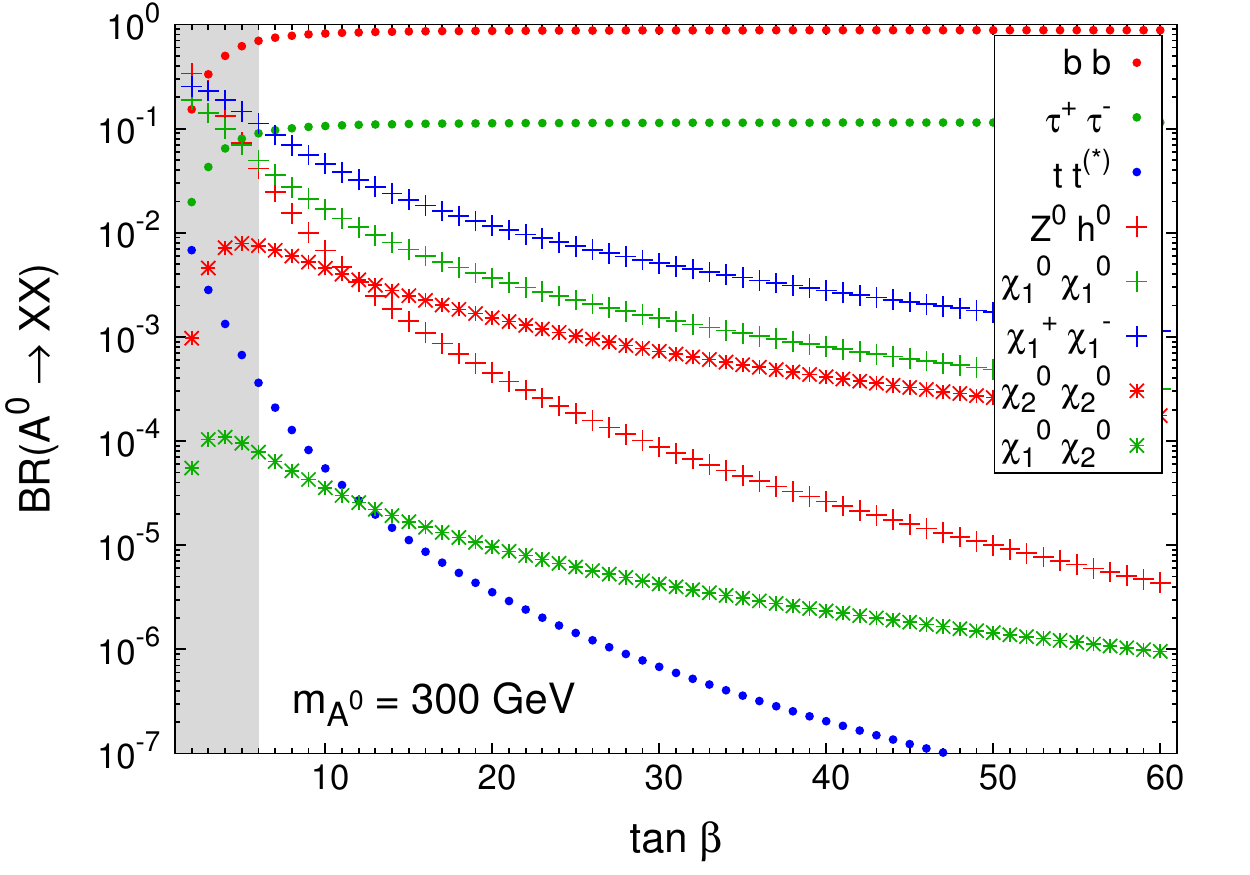} &
\includegraphics[width=75mm]{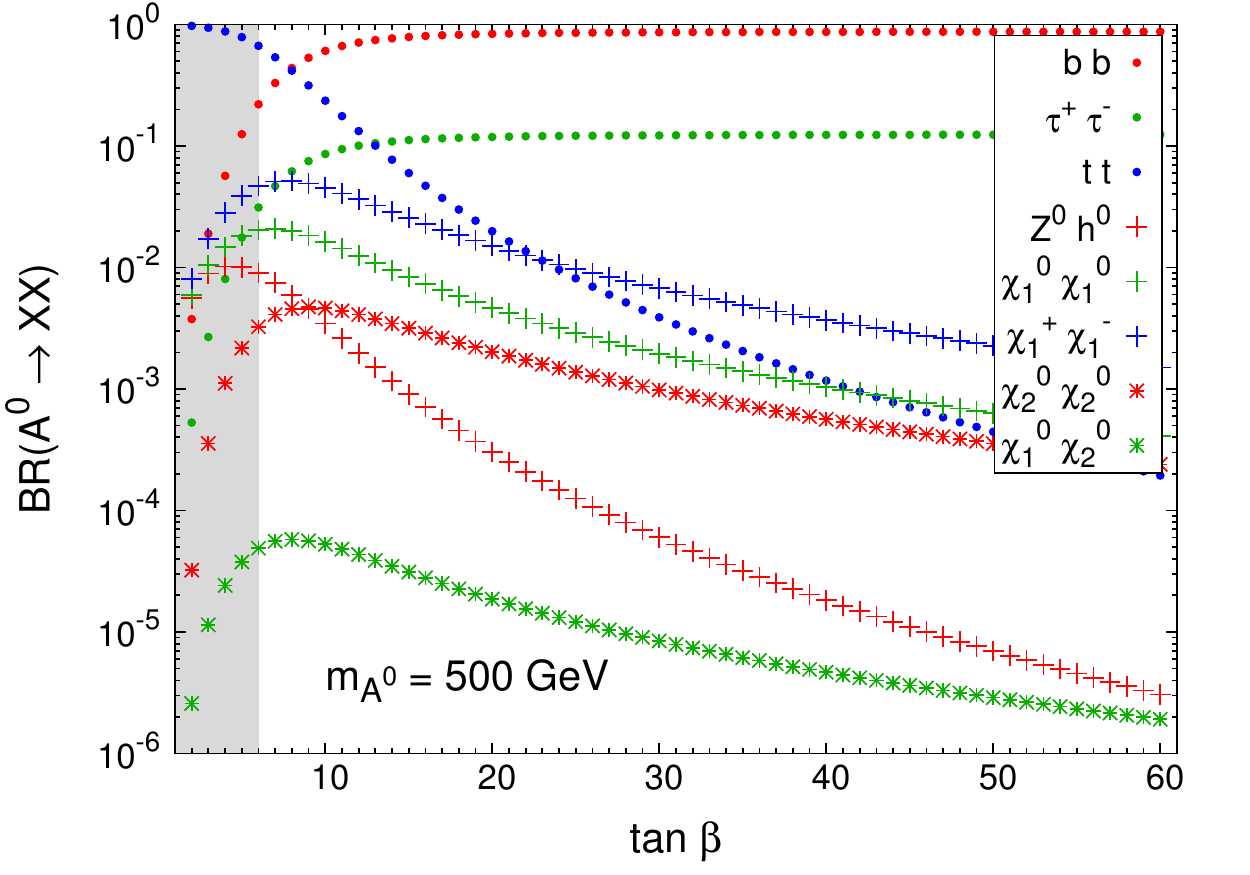} \\
\includegraphics[width=75mm]{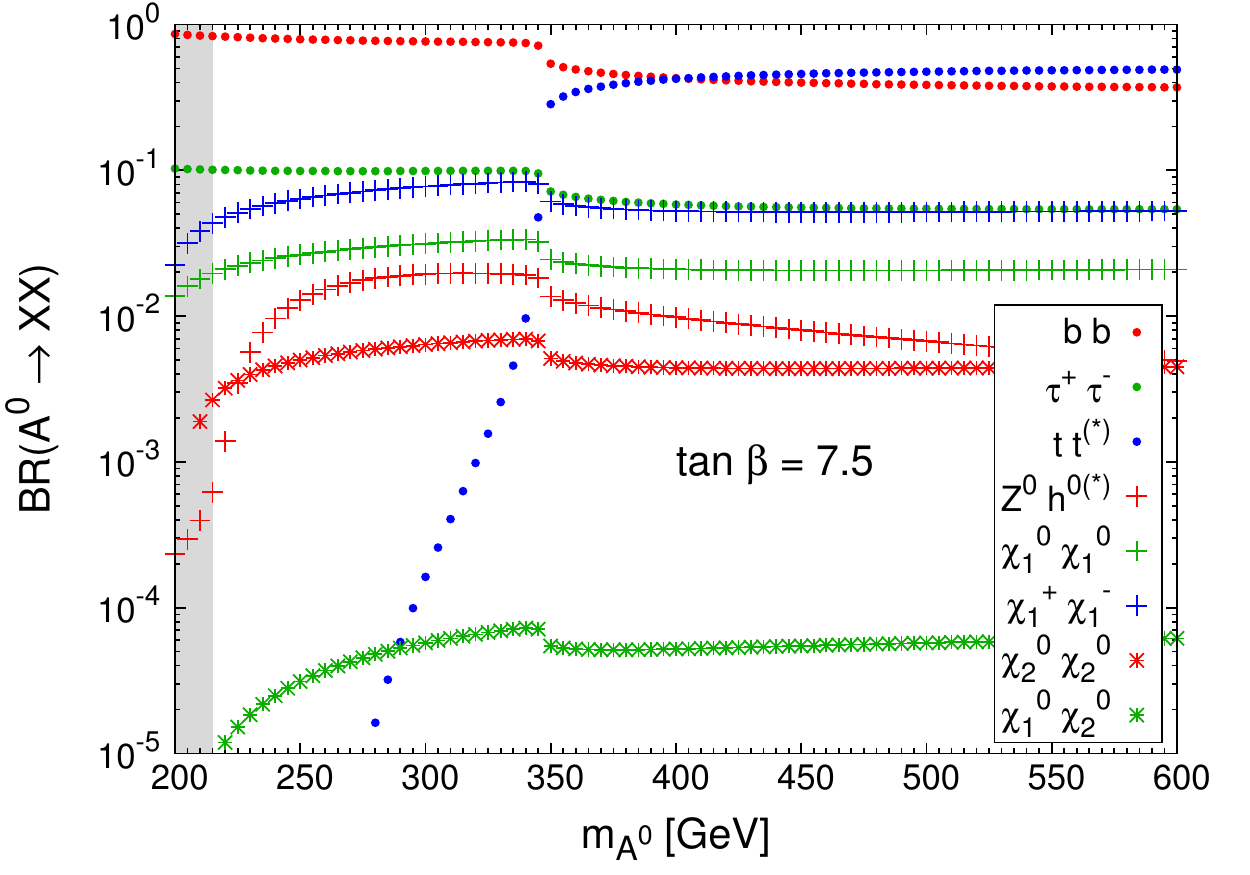} &
\includegraphics[width=75mm]{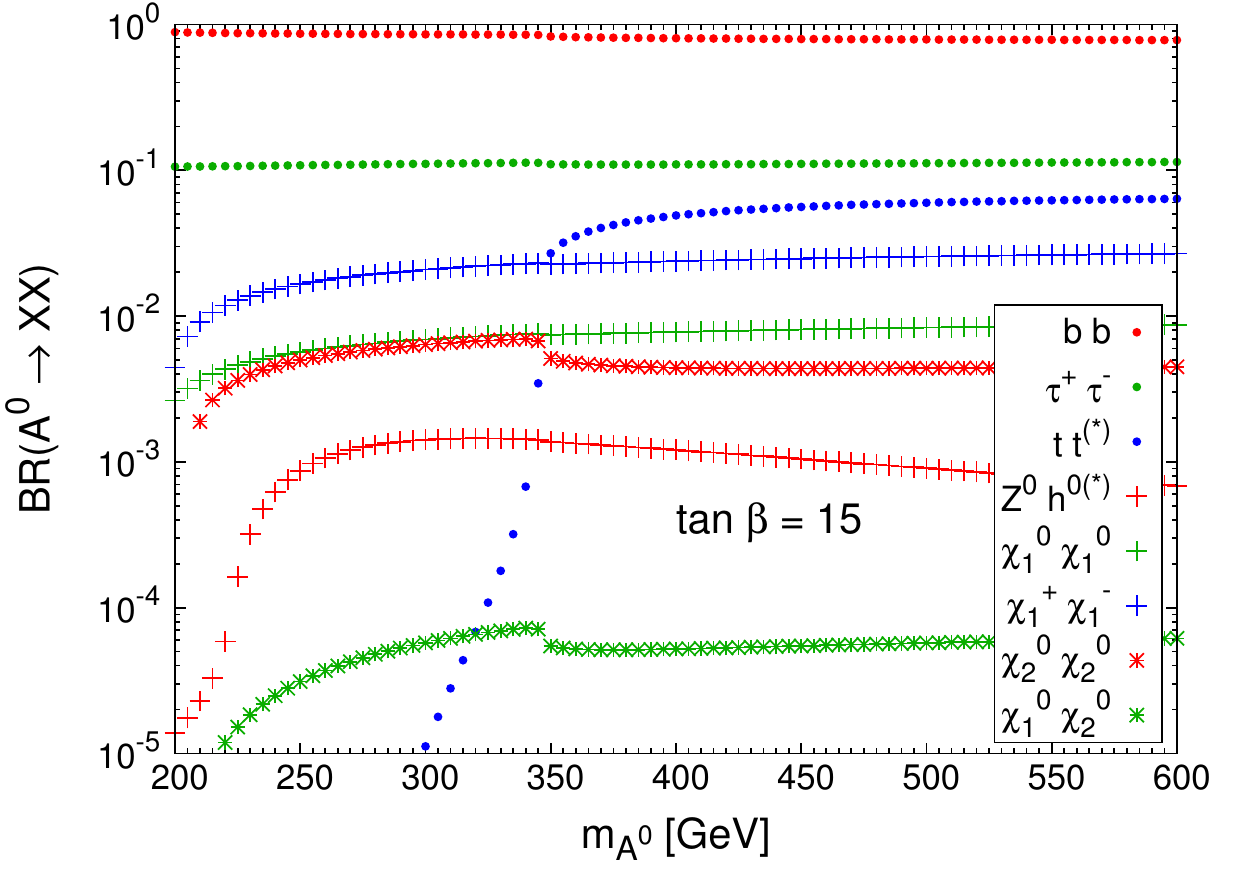}
\end{tabular}
\caption{$A^0$ decay channels in higgsino LSP scenario for $M_S =$ 40 TeV,
$m_s =$ 7 TeV, $A_t =$ 0, $\mu =$ 100 GeV, $M_1 = M_2 =$ 1 TeV, $M_3 =$ 3 TeV.
Upper left panel: $A^0$ branching ratios as a function of $\tan\beta$ for $m_{A^0} = $ 300 GeV.
Upper right panel: $A^0$ branching ratios as a function of $\tan\beta$ for $m_{A^0} = $ 500 GeV.
Lower left panel: $A^0$ branching ratios as a function of $m_{A^0}$ for $\tan\beta = 7.5$.
Lower right panel: $A^0$ branching ratios as a function of $m_{A^0}$ for $\tan\beta = 15$.
Shaded gray area represents values of $m_{h^0} <$ 124 GeV.  The discontinuities around
$m_{A^0} \simeq$ 350 GeV in lower panels are due to the fact that $t \bar t$ channel
starts to be kinematically allowed.}
\label{fig:BRA0decays_scenario3a}
\end{center}
\end{figure}

The results of the branching ratios of  $H^0$ and $A^0$, within this scenario are contained in
Figures~\ref{fig:BRH0decays_scenario3a} and~\ref{fig:BRA0decays_scenario3a}, 
for the parameters: 
$M_S =$ 40 TeV, $m_s = $ 7 TeV, $A_t =$ 0,
$\mu =$ 100 GeV, $M_1 = M_2 =$ 1 TeV and $M_3 =$ 3 TeV.
We present the results of branching ratios, as in the two previous sections,
as a function of $\tan\beta$ and $m_{H^0 (A^0)}$ for both $H^0$ and $A^0$ bosons.
From these plots we find the following salient features:

\begin{itemize}

\item From the plots of branching ratios as function of $\tan\beta$, we notice that
      the decays ($H^0$, $A^0$) $\to b \bar{b}$ are dominant unless $\tan\beta \simeq$ 7 (which is near its lowest 
      allowed value) where the decays into $t \bar{t}$ dominate whenever it is kinematically
      allowed.

\item We also observe that the decays modes $H^0 \to Z^0 Z^0$, $W^+ W^-$, $h^0 h^0$ exhibit similar behavior as
      in the previous scenario, namely they reach maximum values of branching ratios for the lowest allowed value of $\tan\beta$,
      which are of order 0.03, 0.07 and 0.2, respectively, for $\tan\beta \simeq$ 6 with the $t \bar t$ channel closed.
      Similarly, the mode $A^0 \to Z^0 h^0$ reaches a value of BR $\simeq$ 5 $\times$ $10^{-2}$.
      
\item Despite the fact that in this scenario it is possible to have new modes into neutralinos,
      namely, ($H^0$, $A^0$) $\to \tilde{\chi}^0_1 \tilde{\chi}^0_2$, $\tilde{\chi}^0_2 \tilde{\chi}^0_2$, 
      we notice that these modes remain bellow the branching ratios of the invisible decays,
      because of the gaugino component of the neutralino $\tilde{\chi}^0_2$ is too low and produces
      a severe reduction of the couplings to Higgs bosons.
      These invisible modes reach maximum values of order BR($H^0 \to \tilde{\chi}^0_1 \tilde{\chi}^0_1$) $\simeq$ 1 $\times$ $10^{-2}$
      and BR($A^0 \to \tilde{\chi}^0_1 \tilde{\chi}^0_1$) $\simeq$ 5 $\times$ $10^{-2}$ for $\tan\beta \simeq$ 6.
       
\item In this scenario the branching ratios of the decays into charginos ($H^0$, $A^0$) $\to \tilde \chi^+_1 \tilde \chi^-_1$ 
      are even larger than the invisible ones, reaching values of order 0.05 and 0.1, 
      respectively.

\end{itemize}

\begin{figure}[t!]
\begin{center}
\begin{tabular}{cc}
\includegraphics[width=75mm]{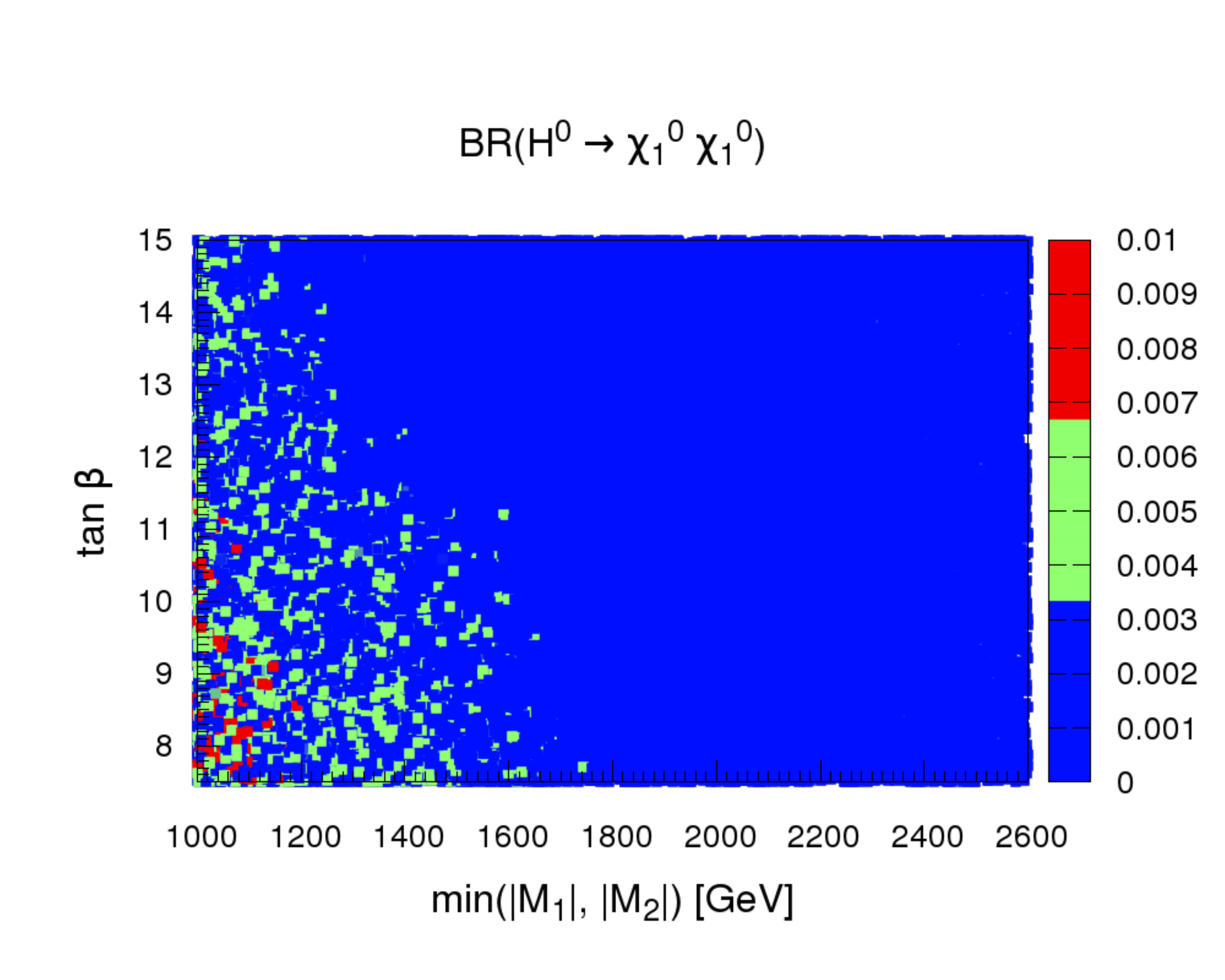} &
\includegraphics[width=75mm]{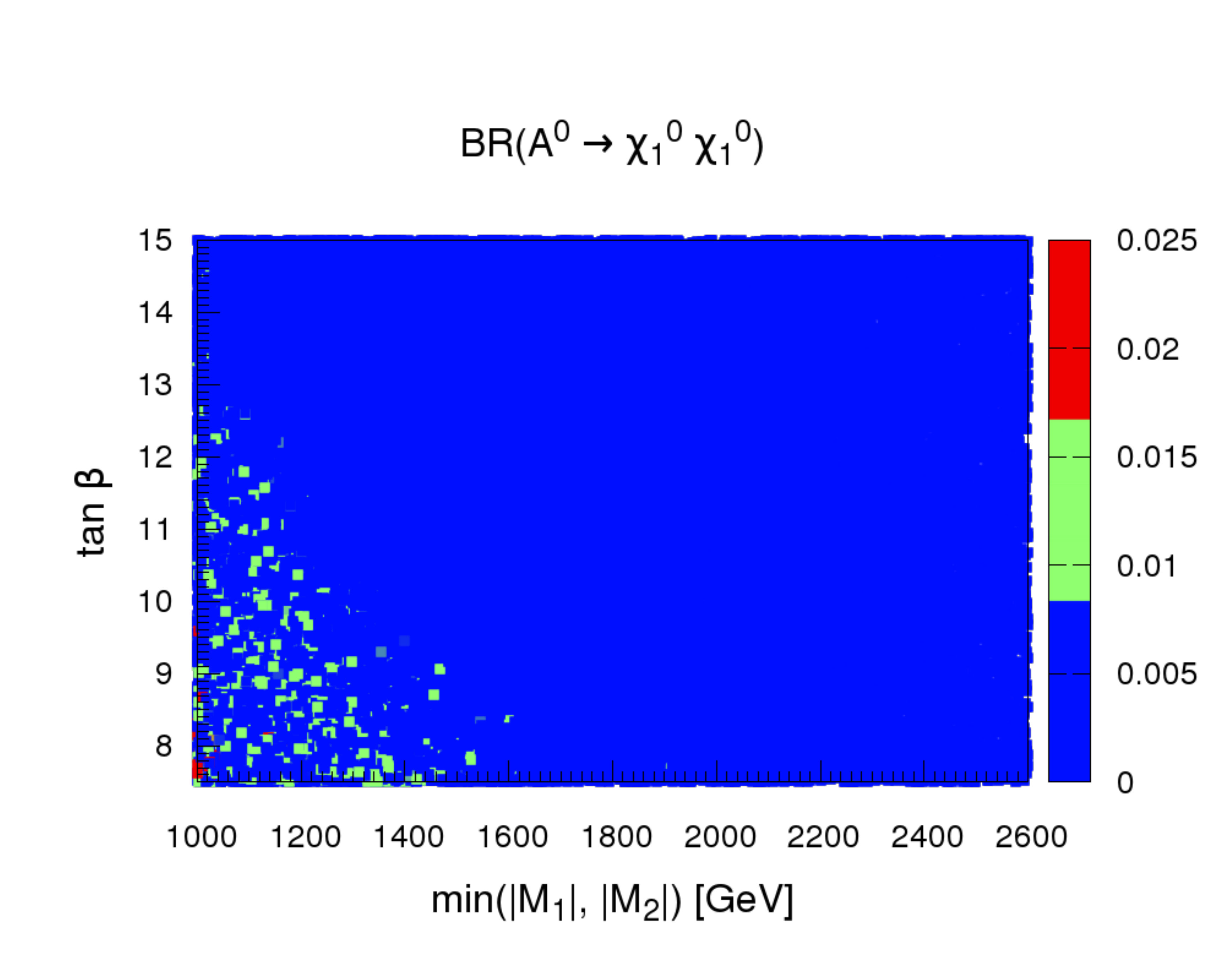} \\
\includegraphics[width=75mm]{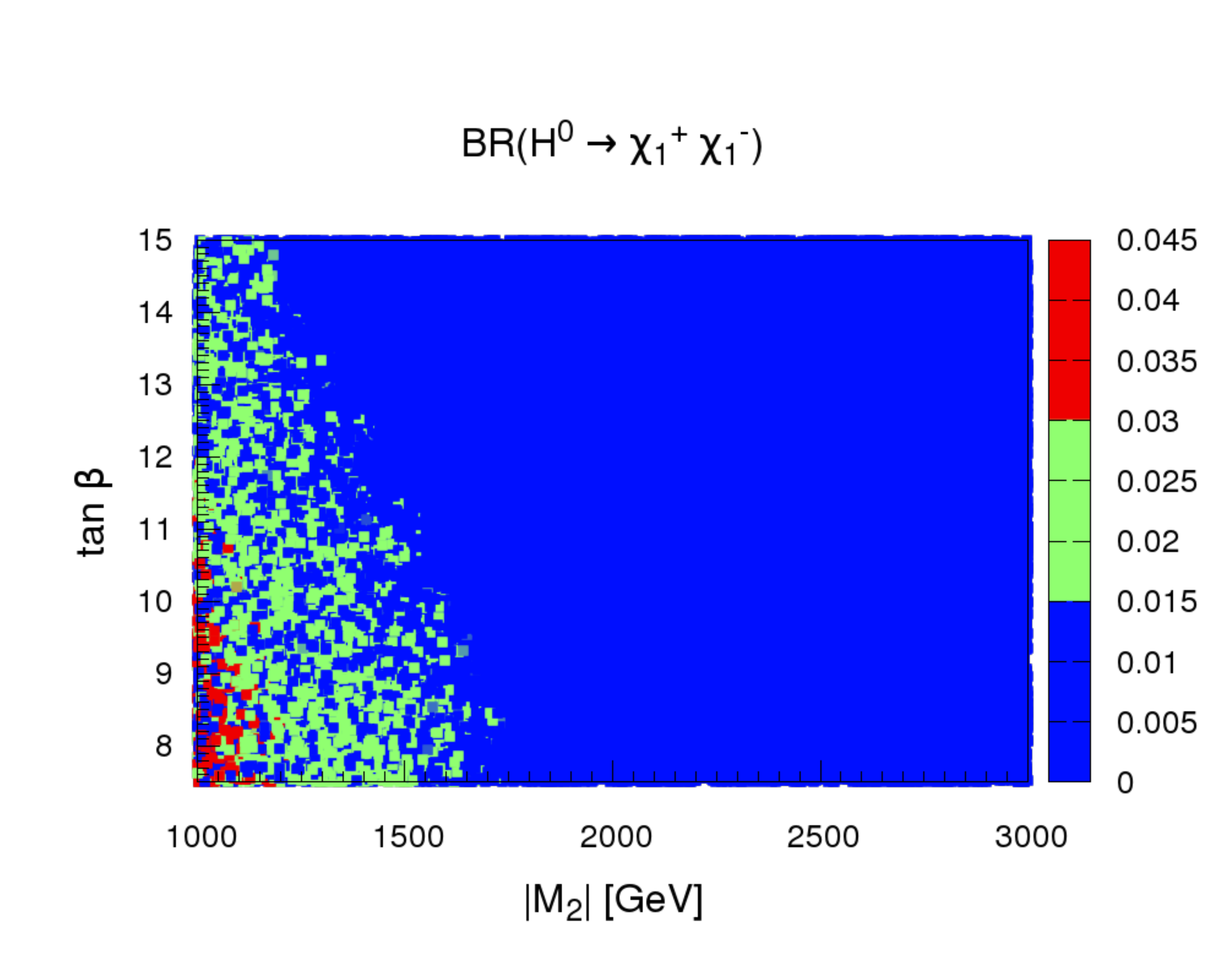} &
\includegraphics[width=75mm]{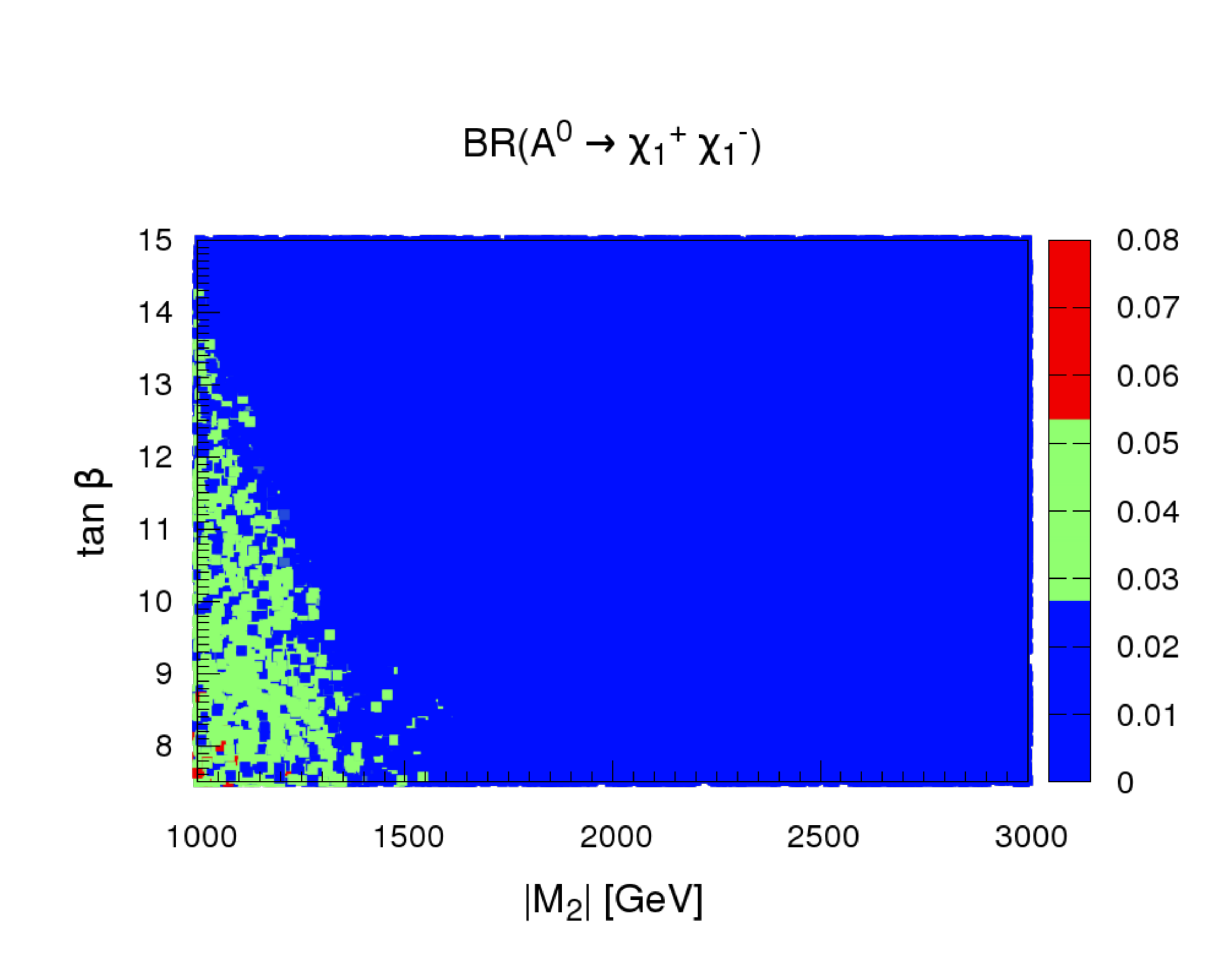}
\end{tabular}
\caption{$H^0$ and $A^0$ branching ratios of neutralino and chargino channels.
Upper left panel: BR$(H^0 \to \tilde \chi_1^0 \tilde \chi_1^0)$ in the plane min($|M_1|$, $|M_2|$)$-\tan\beta$.
Upper right panel: BR$(A^0 \to \tilde \chi_1^0 \tilde \chi_1^0)$ in the plane min($|M_1|$, $|M_2|$)$-\tan\beta$.
Lower left panel: BR$(H^0 \to \tilde \chi_1^+ \tilde \chi_1^-)$ in the plane $|M_2|$$-$$\tan\beta$.
Lower right panel: BR$(A^0 \to \tilde \chi_1^+ \tilde \chi_1^-)$ in the plane $|M_2|$$-$$\tan\beta$.
The scan is done in higgsino LSP scenario with $M_S =$ 40 TeV,
$m_s =$ 7 TeV and $A_t =$ 0 over the following ranges:
100 GeV $< |\mu| <$ 150 GeV, 200 GeV $< m_{A^0} <$ 600 GeV,
1 TeV $< |M_1| \,, |M_2| \,, M_3 <$ 3 TeV.}
\label{fig:HAinvisible_scenario3a}
\end{center}
\end{figure}

As in the previous scenario, we notice that the most promising decay modes in this scenario are
$H^0 \to W^+ W^-$, $Z^0 Z^0$ and $H^0 \to h^0 h^0$, which have a sizable BR and a possibly clean signature, while
for $A^0$ we have the mode $A^0 \to Z^0 h^0$, which has a possibly clean signature but smaller BR. 
The new feature of this scenario is the significance of the invisible decay and
the decays into charginos. To further analyze the significance of these modes, we shall now discuss
the scatter plots in the plane $M_i$$-$$\tan\beta$, displayed in Figure~\ref{fig:HAinvisible_scenario3a}, where $M_i =$ min($|M_1|$, $|M_2|$) for
the invisible modes and $M_i = |M_2|$ for the decays into charginos.

\begin{itemize}
\item For $H^0$ the invisible decay becomes significant, i.e. with a BR close to 0.01 for values of
      $M_i$ near 1000$-$1200 GeV and for $\tan\beta \lesssim 11$. 

\item For $A^0$ we find that the invisible decay has branching ratios in the range 0.01$-$0.015 for similar values of
      $M_i$ (near 1000$-$1200 GeV) and $\tan\beta$ $\lesssim 11$. 
      
\item On the other hand, for $H^0$ the decay mode into charginos has larger branching ratios, within the range  
      0.03$-$0.045 for values of $M_i\simeq 1000$ GeV and for $\tan\beta \lesssim 11$. 

\item For $A^0$ mode into chargino pair we find branching ratios in the range 0.025$-$0.055 for similar values of
      $M_i$ (near 1000$-$1200 GeV) and $\tan\beta \lesssim 13$. 
\end{itemize}

The most important feature of this scenario is, as in the wino LSP scenario,
the existence of pure supersymmetric signatures which could represent an evidence of physics beyond the SM different from a 2HDM.
We have again non-negligible invisible decay branching ratios, although smaller than in the previous scenario
(up to 1\% for $H^0 \to \tilde \chi_1^0 \tilde \chi_1^0$ and up to 2.5\% for $A^0 \to \tilde \chi_1^0 \tilde \chi_1^0$).
The decay into charginos has larger branching ratio than the invisible ones, but in this case
the charginos are not stable at detector level, decaying mainly into a neutralino $\tilde \chi_1^0$ and two quarks,
what produces a signal different from wino LSP scenario, consisting of large $E_T^\text{miss}$ plus jets.

\section{Conclusions}
\label{conclusions}

We have discussed in this paper the results of current searches for Higgs and SUSY at the LHC. They seem to 
favor some Split-inspired SUSY scenarios, which assume that all scalars (except the light 
Higgs boson $h^0$) are much heavier than the fermions. However, we have argued here that the remaining Higgs bosons 
of the MSSM ($H^0$, $A^0$, $H^{\pm}$) do not have to be as heavy as the sfermions, and having them with masses 
near the EW scale  does not pose any conflict with current MSSM constraints. 
On the theoretical side, we notice that there could be constraints on our scenario coming from
RGEs and correct EW symmetry breaking~\cite{Ibarra:2005vb,Bernal:2007uv,Craig:2012di},
but in this paper we have focused more on the phenomenological motivation, rather than on the construction
of a realistic SUSY breaking scheme, which we will treat in a future publication. We have then discussed some SUSY scenarios with heavy sfermions which contain the full 
Higgs sector with masses near the EW scale, and identify 
distinctive signals from these scenarios that could be searched at the LHC.

From our study of the heavy neutral Higgs boson decay channels, we list the following important remarks:

\begin{enumerate}

\item Within the bino LSP scenario, in which only one neutralino has a mass near the EW scale, we found that 
 for low values of $\tan\beta$ and below the threshold for the decay into $t\bar{t}$, the decays
 $H^0 \to W^+ W^-$, $Z^0 Z^0$ and $h^0 h^0$ have branching ratios of order 0.1, 0.03 and 0.1, respectively. Similar results hold for the decay mode 
 $A^0 \to Z ^0 h^0$, whose BR reaches values of order 0.03.
 
\item Within the wino LSP scenario, in which one neutralino and one chargino pair have a mass near the EW scale, 
  we found that the decays $H^0 \to W^+ W^-$, $Z^0 Z^0$ and $h^0 h^0$, as well as $A^0 \to Z ^0 h^0$, have similar BR as in the 
  bino LSP scenario. In this scenario the invisible decay modes ($H^0$, $A^0$) $\to \tilde \chi_1^0 \tilde \chi_1^0$ can reach
  branching ratios of order 0.02 and 0.04, respectively. The decay mode into charginos ($H^0$, $A^0$)  $\to \tilde \chi_1^+ \tilde \chi_1^-$ 
  can reach branching ratios of order 0.05 and 0.1, respectively.
   
\item Within the higgsino LSP scenario, in which two neutralino and one chargino pair have a mass near the EW scale, 
  we found that the decays $H^0 \to W^+ W^-$, $Z^0 Z^0$ and $h^0 h^0$, as well as $A^0 \to Z ^0 h^0$, have similar branching ratios as in the 
  previous scenarios. In this scenario the invisible decay mode ($H^0$, $A^0$) $\to \tilde \chi_1^0 \tilde \chi_1^0$ can reach
  branching ratios of order 0.01 and 0.03, respectively. The decay mode into charginos ($H^0$, $A^0$)  $\to \tilde \chi_1^+ \tilde \chi_1^-$ 
  can reach branching ratios of order 0.05 and 0.08, respectively.
  
\item The bino LSP scenario is very difficult to distinguish from a 2HDM, since its only pure SUSY signatures are
the $H^0$ and $A^0$ invisible decays, which are very suppressed. In contrast, in both wino LSP and higgsino LSP
scenarios the decays into neutralinos and charginos have measurable branching ratios. The main difference between
both scenarios is the chargino lifetime, what leads to very different signals at the detectors. In the wino LSP scenario,
the chargino is nearly stable and leaves an energetic track in the calorimeter. On the other hand, in the higgsino LSP scenario,
the chargino decays mainly into a neutralino and a pair of quarks, with a characteristic signal consisting of large $E_T^\text{miss}$ plus jets.
  
\item On the other hand, we notice that for most of the regions of the parameter space in the three scenarios we have considered
the dominant decay mode of $H^0$ and $A^0$ is into $b\bar{b}$, which seems to be detectable only when these bosons are produced
in association with a $W^\pm$ gauge boson. However, in this case we can take advantage of the enhanced couplings of the heavy Higgs bosons
with the $b\bar{b}$ pair and look for the associated production of $H^0$ and $A^0$ with $b\bar{b}$. Detectability of this
signal at LHC14 was shown to be possible~\cite{Balazs:1998nt,DiazCruz:1998qc}, while more recent studies have
appeared in \cite{Carena:2012rw}.

\item Finally, we estimate the number of events expected for the most interesting $H^0$ and $A^0$ decay modes at the future phase of the LHC, with a center-of-mass energy of $\sqrt{s} =$ 14 TeV and a total integrated luminosity of ${\cal L} =$ 100 fb$^{-1}$. For example, for $\tan\beta =$ 7.5 and $m_{A^0} =$ 300 GeV, the $H^0$ and $A^0$ production cross sections (via gluon fusion and associated with $b \bar b$) reach values around 1 pb. In all the three scenarios considered here, we obtain BR($H^0 \to h^0 h^0$) $\sim 10^{-1}$ and BR($A^0 \to Z^0 h^0$) $\sim 10^{-2}$ and, therefore, we would expect to get around $10^4$ and $10^3$ events, respectively. Regarding the invisible decay modes of $H^0$ and $A^0$, for BR($H^0, A^0 \to \tilde \chi_1^0 \tilde \chi_1^0$) $\sim 2 \times 10^{-2}$ one would achieve around $2 \times 10^3$ events, which would be reduced after considering $b$-tagging process. These results can be used for an indicative of the strength of the the signals at the LHC, but for an estimation of the backgrounds, we focus only on $H^0 \to h^0 h^0 \to 4 b$ channel and follow~\cite{Balazs:1998nt}. The SM cross section of the process $pp \to b \bar b h_\text{SM}$ at LHC14 is $\sigma(pp \to b \bar b h_\text{SM}) \simeq$ 0.5 pb and, thus, with BR($h_\text{SM} \to b \bar b$) $\simeq$ 0.65 and ${\cal L} =$ 100 fb$^{-1}$ we will expect $3.25 \times 10^4$ events for $pp \to b \bar b h_\text{SM} \to b \bar b b \bar b$. From~\cite{Balazs:1998nt} we infer that, for these values of the Higgs mass and luminosity, an enhancement factor of 6 is required. Therefore, one would need 36 $\times$ SM number of events in order to detect $H^0 \to h^0 h^0 \to 4b$ signal. In our case, $\sigma(pp \to h^0 h^0 \to b \bar b b \bar b) \simeq$ 100 fb and, with ${\cal L} =$ 100 fb$^{-1}$, we would obtain around $10^4$ events. We can conclude from Table I of~\cite{Balazs:1998nt} that the mass invariant cut reduces around one order of magnitude the background of $Z b \bar b$ and almost two orders of magnitude $b \bar b b \bar b$ and $b \bar b j j$ backgrounds. Consequently, after these cuts, we would reach an enhancement factor close to 33 $\times$ SM number of events and the signal could be detectable. 
\end{enumerate}

The results presented in this work on the Split-inspired SUSY scenarios with non-universal Higgs masses have concentrated on 
the analysis of the mass spectrum and the branching ratios of the decay modes of $H^0$ and $A^0$ Higgs bosons, which is certainly a first stage. Nevertheless, we have performed 
an exhaustive analysis within the allowed parameter space of these scenarios. In fact,  these results already suggest that 
it can be worthwhile to perform a detailed  simulation study for each one of those promising decay modes, in order to 
determine whether they could be detectable  at the current or future phases of LHC with the expected luminosity.

An analysis of the EWPT in our scenarios and the modifications 
they could induce is also an important issue to be developed in future works. Similarly, an study of
the most general scenario, in which the four neutralinos and the two pair of charginos are
near the EW scale, will be needed in order to analyze in detail all the possible decay modes
of heavy Higgs bosons and the consecutive SUSY decay chains.

\section*{Acknowledgments}

The authors are indebted to Ben Allanach, Sven Heinemeyer and Margarete Muhlleitner for valuable information about
{\tt SOFTSUSY}, {\tt FeynHiggs} and {\tt SUSY-HIT} codes, respectively. E.A. and A.S. thank Ken Kiers for fruitful
discussions and a careful reading of the manuscript. E.A. is financially supported by a MICINN postdoctoral
fellowship (Spain), under grant no. FI-2010-0041, and thanks IFLP-CONICET for hospitality and support.
J.L. Diaz-Cruz acknowledges support from CONACYT-SNI (Mexico).
This work has been partially supported by ANPCyT (Argentina) under
grant no. PICT-PRH 2009-0054 and by CONICET (Argentina) PIP-2011 (E.A., A.S.).

\bibliographystyle{unsrt}

\end{document}